\documentclass[prd,twocolumn,twoside,preprintnumbers,superscriptaddress,nofootinbib]{revtex4}

\usepackage{amsmath,slashed}
\usepackage{graphicx,graphics,color}
\usepackage{dcolumn}
\usepackage[hyperfootnotes=false]{hyperref}
\usepackage{xspace}
\usepackage{enumerate}
\usepackage{varwidth}
\usepackage{comment}
\usepackage{physics}

\usepackage{xcolor}

\usepackage{amssymb}
\usepackage{mathtools}
\usepackage{dsfont}
\usepackage{multirow}
\usepackage{bbold}
\usepackage{xfrac}

\usepackage{cancel}
\usepackage{orcidlink}

\newcommand{\GeV}{\,\text{GeV}}
\newcommand{\MeV}{\,\text{MeV}}

\newcommand{\fm}{\,\text{fm}}

\newcommand{\mpi}{M_\pi}

\newcommand{\mN}{m_N}

\newcommand{\rr}{\mathbf{r}}
\newcommand{\pp}{\mathbf{p}}

\newcommand{\LL}{\mathbf{L}}

\newcommand{\beq}{\begin{equation}}
\newcommand{\eeq}{\end{equation}}

\newcommand{\bsigma}{\boldsymbol{\sigma}}

\newcommand{\vx}{\mathbf{x}}
\newcommand{\vp}{\mathbf{p}}
\newcommand{\vq}{\mathbf{q}}
\newcommand{\bea}{\begin{eqnarray}}
\newcommand{\eea}{\end{eqnarray}}
\newcommand{\be}{\begin{equation}}
\newcommand{\ee}{\end{equation}}
\newcommand{\bma}{\begin{pmatrix}}
\newcommand{\ema}{\end{pmatrix}}
\newcommand{\nn}{\nonumber}

\newcommand{\mupi}{\mu_\pi}

\newcommand{\intem}{1.8/2.0~(EM)}
\newcommand{\intsat}{$\rm NNLO_{sat}$}

\newcommand{\inner}[2]{\langle #1 | #2 \rangle}

\newcommand{\sixjj}[6]{\begingroup\setlength\arraycolsep{1pt}\begin{Bmatrix} #1 & #2 & #3 \\ #4 & #5 & #6 \end{Bmatrix}\endgroup}
\newcommand{\ninejj}[9]{\begingroup\setlength\arraycolsep{1pt}\begin{Bmatrix} #1 & #2 & #3 \\ #4 & #5 & #6 \\ #7 & #8 & #9 \end{Bmatrix}\endgroup}

\allowdisplaybreaks[1]

\begin{document}

\preprint{INT-PUB-26-043
}

\title{Ab initio calculations of \texorpdfstring{$\boldsymbol{\bar\delta_\text{NS}}$}{} and extraction of \texorpdfstring{$\boldsymbol{V_{ud}}$}{} using effective field theory}

\author{Samuel Novario\,\orcidlink{0000-0001-7836-1269}}
\affiliation{Department of Physics, Washington University in Saint Louis, Saint Louis, MO 63130, USA}
\affiliation{McDonnell Center for the Space Sciences at Washington University in St. Louis, MO 63130, USA}

\author{Graham Chambers-Wall\,\orcidlink{0000-0002-9403-8450}}
\affiliation{Department of Physics, Washington University in Saint Louis, Saint Louis, MO 63130, USA}

\author{Vincenzo Cirigliano\,\orcidlink{0000-0002-9056-754X}}
\affiliation{Institute for Nuclear Theory, University of Washington, Seattle WA 91195-1550, USA}

\author{Wouter Dekens\,\orcidlink{0000-0002-7850-5901}}
\affiliation{Institute for Nuclear Theory, University of Washington, Seattle WA 91195-1550, USA}

\author{Jordy~de~Vries\,\orcidlink{0000-0001-5037-5108}}
\affiliation{Institute for Theoretical Physics Amsterdam and Delta Institute for Theoretical
Physics, University of Amsterdam, Science Park 904, 1098 XH Amsterdam, The Netherlands}
\affiliation{Nikhef, Theory Group, Science Park 105, 1098 XG, Amsterdam, The Netherlands}

\author{Stefano Gandolfi\,\orcidlink{0000-0002-0430-9035}}
\affiliation{Theoretical Division, Los Alamos National Laboratory, Los Alamos, NM 87545, USA}

\author{Martin Hoferichter\,\orcidlink{0000-0003-1113-9377}}
\affiliation{Albert Einstein Center for Fundamental Physics, Institute for Theoretical Physics, University of Bern, Sidlerstrasse 5, 3012 Bern, Switzerland}

\author{Emanuele Mereghetti\,\orcidlink{0000-0002-8623-5796}}
\affiliation{Theoretical Division, Los Alamos National Laboratory, Los Alamos, NM 87545, USA}

\begin{abstract}
Radiative corrections limit the precision of $V_{ud}$ extractions from superallowed $\beta$ decays. In particular, the model dependence in the nuclear matrix elements (NMEs) is difficult to assess without a clear organizing principle, which is provided by effective field theory (EFT). We report substantial progress in this program based on ab initio nuclear calculations using the coupled cluster method. Benchmarking results against Quantum Monte Carlo calculations for light nuclei, we evaluate all two-body NMEs that enter the correction $\bar\delta_\text{NS}$ across 12 superallowed transitions relevant for $V_{ud}$, including an estimate of truncation and Hamiltonian uncertainties. We find that certain three-body NMEs at ${\mathcal O}(\alpha^2)$ play an important role, due to a combination of large numerical prefactors and coherent scaling with $Z^2$. We develop a formalism to estimate such NMEs based on nuclear densities to complete the EFT evaluation of $\bar\delta_\text{NS}$, validating the corresponding two-body NMEs against ab initio calculations. We further identify the corrections in the traditional phase-space factor that are captured by the EFT and remove them consistently, which renders the individual transitions mutually consistent. Profiting from a recent calculation of ${\mathcal O}(\alpha^2 Z)$ ultrasoft photons, we find that the scale uncertainty in the EFT is now negligible, while the remaining uncertainty is dominated by contact terms required to renormalize the magnetic Gamow--Teller NMEs. One combination of low-energy constants can be determined from a global fit to the superallowed data, but the second one proves strongly correlated with $V_{ud}$, necessitating additional input to resolve the nuclear-structure effects. We argue that such short-distance contributions currently imply an uncertainty in $V_{ud}$ at the level of $\delta V_{ud}\simeq 6\times 10^{-4}$.
\end{abstract}

\maketitle

\section{Introduction}

Superallowed $\beta$ decays have traditionally been the primary source of information on $V_{ud}$~\cite{Hardy:2020qwl,Gorchtein:2023naa,Hayen:2017pwg}, with a precision at the level of $\delta V_{ud}\simeq 3\times 10^{-4}$, e.g., Ref.~\cite{Hardy:2020qwl} quotes
\begin{equation}
\label{Vud_HT}
V_{ud}=0.97373(31)\,.
\end{equation}
At this level of precision, exquisite control over radiative corrections is required, among them the nucleus-dependent correction factor $\bar\delta_\text{NS}$~\cite{Seng:2018qru,Gorchtein:2018fxl,Seng:2022cnq}, which quantifies a set of nonuniversal corrections that need to be evaluated for each decay separately. In general, a key limitation that arises from the nuclear matrix elements (NMEs) involved concerns uncertainty quantification, since the model dependence is very hard to assess without a clear organizing principle, leading to concerns whether the uncertainty quoted in Eq.~\eqref{Vud_HT}, despite a long history of increasingly refined calculations~\cite{Kinoshita:1958ru,Sirlin:1967zz,Sirlin:1967zza,Abers:1968zz,Jaus:1970tah,Sirlin:1977sv,Sirlin:1981ie,Wilkinson:1982hu,Sirlin:1986cc,Towner:1992xm,
Wilkinson:1993hxz,Wilkinson:1993fva,Czarnecki:2004cw,Marciano:2005ec,Towner:2010zz,Hardy:2014qxa}, provides a realistic estimate of the nuclear uncertainties.
To address this concern, a new approach based on effective field theory (EFT) was put forward in Refs.~\cite{Cirigliano:2024rfk,Cirigliano:2024msg}, in which, in principle, uncertainty estimates can be derived from the EFT expansion at each step in the calculation.
Here, we provide a detailed account of the progress in this program as presented in Ref.~\cite{letter}, to which we also refer for the comparison to a dispersive approach to $\bar\delta_\text{NS}$~\cite{Seng:2018qru,Gorchtein:2018fxl,Seng:2022cnq,Gennari:2024sbn}, consequences for a deficit in first-row CKM unitarity that could point towards physics beyond the Standard Model~\cite{Belfatto:2019swo,Coutinho:2019aiy,Cheung:2020vqm,Belfatto:2021jhf,Branco:2021vhs,Crivellin:2021bkd,Crivellin:2020ebi,Kirk:2020wdk,Crivellin:2021njn,Crivellin:2020lzu,Crivellin:2020klg,Capdevila:2020rrl,Crivellin:2021sff,Crivellin:2020oup,Marzocca:2021azj,Alok:2021ydy, Cirigliano:2022qdm,Cirigliano:2023nol,Dawid:2024wmp,Gialidi:2026tao}, and
a roadmap for future improvements.

Starting point in the EFT is the Wilson coefficient of the semileptonic operator mediating $\beta$ decay, evolved down to the nuclear scale. The perturbative evolution is now known at next-to-leading-logarithmic (NLL) order~\cite{Cirigliano:2023fnz} including $\alpha_s$ corrections~\cite{Moretti:2025ghw}, and if the NMEs were known at sufficient precision, the theoretical uncertainty could be pushed below the level of $1\times 10^{-4}$. This was demonstrated in Ref.~\cite{Cirigliano:2026ios} for pion $\beta$ decay based on the lattice-QCD calculations from Refs.~\cite{Feng:2020zdc,Yoo:2023gln}, strongly motivating the experimental efforts at PIONEER to improve the experimental branching fraction accordingly~\cite{PIONEER:2022yag,PIONEER:2025idw}.

Already at the single-nucleon level  control over the matrix elements becomes substantially more difficult, as reflected by nonperturbative effects traditionally described by the ``$\gamma W$ box'' contribution~\cite{Marciano:2005ec,Seng:2018yzq,Seng:2018qru,Czarnecki:2019mwq,Seng:2020wjq,Hayen:2020cxh,Shiells:2020fqp,Cirigliano:2022yyo,Ma:2023kfr}, which in the EFT approach defines the effective Wilson coefficient $g_V$ at the low scale. Apart from the nonperturbative input, also perturbative corrections are essential to control the scale dependence in the EFT~\cite{Hill:2023acw,Hill:2023bfh,Borah:2024ghn,VanderGriend:2025mdc,Cao:2025zxs,Crosas:2025xyv}. At present, the precision of $V_{ud}$ extracted from neutron decay remains limited by the experimental inputs for lifetime and decay asymmetry, assuming the current best individual measurements~\cite{UCNt:2021pcg,Markisch:2018ndu} at $\delta V_{ud}\simeq 4.2\times 10^{-4}$, while the uncertainty from $g_V$ alone would imply $\delta V_{ud}\simeq 1.2\times 10^{-4}$. Including scale factors, the current uncertainty of $V_{ud}$ determined from neutron decay increases to $\delta V_{ud}\simeq 8.8\times 10^{-4}$~\cite{ParticleDataGroup:2026mpi}.

This uncertainty in $g_V$ automatically propagates to superallowed $\beta$ decays as well, but the error in Eq.~\eqref{Vud_HT} is dominated by the estimate of nuclear corrections from Ref.~\cite{Gorchtein:2018fxl}. To assess the robustness of this estimate, and eventually improve the precision, an EFT formalism was proposed in Refs.~\cite{Cirigliano:2024rfk,Cirigliano:2024msg}, whose salient features we summarize in Sec.~\ref{sec:EFT}. The required NMEs in the EFT approach were already evaluated for the $^{14}$O $\rightarrow$ $^{14}$N transition in Ref.~\cite{Cirigliano:2024msg} using Quantum Monte Carlo (QMC) techniques, and subsequently in Ref.~\cite{King:2025fph} for $^{10}$C $\rightarrow$ $^{10}$B.
In this paper we extend these calculations to medium mass nuclei, using the coupled cluster (CC) method.

In the meantime, it has become apparent that, in addition to the NMEs identified in Refs.~\cite{Cirigliano:2024rfk,Cirigliano:2024msg}, also a particular class of three-body NMEs at ${\mathcal O}(\alpha^2)$ needs to be considered, due to a double enhancement by numerical factors and coherent scaling  with $Z^2$. To obtain an estimate of these NMEs, we developed a formalism using nuclear densities, as described in Sec.~\ref{sec:density}, subsequently validated in the two-body sector against the ab initio calculations presented in Sec.~\ref{sec:NMEs} based on the CC method. These results, ranging from $^{10}$C $\rightarrow$ $^{10}$B to $^{54}$Co $\rightarrow$ $^{54}$Fe, constitute the main outcome of this work, allowing for a first global ab initio evaluation of $\bar\delta_\text{NS}$ for all isotopes relevant for superallowed $\beta$ decays. We performed detailed studies of the CC truncation and Hamiltonian uncertainties, the latter estimated by comparing the Entem--Machleidt potential \intem~\cite{Entem:2003ft,Hebeler:2010xb} (derived from chiral potentials at next-to-next-to-next-to-leading order by renormalization group (RG) methods) and \intsat~\cite{Ekstrom:2015rta} (based on next-to-next-to-leading order (NNLO) chiral interactions optimized including selected isotopes up to oxygen, to ensure consistency with the empirical saturation point of symmetric nuclear matter). While neither interaction provides a fully satisfactory description of the properties of all nuclei of interest, the benchmarking against charge radii shows that the experimental results are encompassed by the span, suggesting that the comparison should provide a realistic assessment. Finally, for a full EFT evaluation also the Coulomb corrections $\bar\delta_C$ need to be evaluated~\cite{Miller:2008my,Miller:2009cg,Seng:2023cvt,Seng:2026hwk}, and the phase-space factor made consistent with the EFT approach. At present, ab initio results for $\bar\delta_C$ are not yet available, see Refs.~\cite{Farren:2024spl,Piarulli:2026abq} for first steps, so that we continue to use $\bar\delta_C$ from Ref.~\cite{Hardy:2020qwl}, while the phase space is treated as described in Sec.~\ref{sec:phase_space} and aligned as much as possible to the EFT approach. Fits to the 12  superallowed transitions studied in this work are presented in Sec.~\ref{sec:fits}, before we conclude with a summary of our main results in Sec.~\ref{sec:summary}.

\section{Factorization of the half-life in the EFT formalism}
\label{sec:EFT}

Using tools from EFT, Refs.~\cite{Cirigliano:2024rfk,Cirigliano:2024msg} provided a factorization formula for the superallowed half-life, expressed as
\begin{align}
    \frac{1}{t}  &=    \frac{G_F^2 |V_{ud}|^2  m_e^5}{\pi^3 \log 2}\,  
    \Big[C_\text{eff}^{(g_V)}(\mu)\Big]^2  \notag\\
    &\quad\times[1 + \bar \delta_R^\prime(\mu)]\,  (1 +  \bar \delta_\text{NS}) \,  ( 1 - \bar  \delta_C)  \,  \bar f(\mu)\,.
    \label{eq:master2}
\end{align}
This expression superficially resembles the standard formula for the superallowed half-life, given, e.g., in Ref.~\cite{Hardy:2020qwl}, but maximally exploits the scale separation in EFT.  
We denote with a bar objects defined in the EFT, which have a counterpart in the standard expression as detailed in Ref.~\cite{Cirigliano:2024msg}. 
Equation~\eqref{eq:master2}
separates the contributions from photons with different virtualities, and allows for the resummation of potentially large logarithms of the ratios of the energy scales in the problem.

We briefly review the main ingredients in this expression.
The coefficient $C_{\text{eff}}^{(g_V)}$ receives 
contributions from hard photons, i.e., photons that have virtualities larger than the typical nuclear scale $\mu \simeq k_F$, with Fermi momentum $k_F$. 
In particular, the Wilson coefficient $C_{\text{eff}}^{(g_V)}$ 
contains perturbative  corrections at the electroweak (EW) scale, the RG evolution from the EW to the QCD scale, nonperturbative contributions encoded in the $\gamma W$ box, and the perturbative QED evolution between the scale $\mu \simeq m_N$ and the nuclear physics scale $\mu \simeq k_F$. 

$\bar\delta_{\text{NS}}$
and $\bar\delta_C$ contain the contributions of potential and soft photons, sensitive to the nuclear scale. Potential photons give the leading contributions to both $\bar\delta_{\text{NS}}$
and $\bar\delta_{C}$. At lowest order in the fine-structure constant $\alpha=e^2/(4\pi)$, the two contributions can be identified with electromagnetic (EM) two-body corrections to the EW current and with isospin-breaking corrections to the nuclear wave function, respectively.
Finally, $\bar f$ and $\bar\delta_R^\prime$  encode contributions from ultrasoft photons, sensitive to low-energy scales as the $\mathcal Q_\text{EC}$ value of the reaction. At lowest order, these corrections include the Fermi and Sirlin functions, respectively. At higher order, they can be expressed in an expansion in powers of $\alpha$
and electron energy $E_e$ with respect to $k_F$. 
We now review the definitions of these objects and the theoretical ingredients used their evaluation.

\subsection{\texorpdfstring{$\boldsymbol{C_{\text{eff}}^{(g_V)}}$}{}}
\label{sec:CgVeff}

The Wilson coefficient $C_{\text{eff}}^{(g_V)}$ is obtained by integrating out hard and soft photons. 
Hard photons with virtualities greater than the nucleon mass are encoded in the single nucleon vector coupling $g_V$ 
appearing in the one-body effective Lagrangian\footnote{For a more detailed discussion, see Ref.~\cite{Cirigliano:2024msg}.}  
\begin{equation}
{\mathcal L}_\text{eff}^\text{1b} = - \sqrt{2} \,  g_V \, G_F V_{ud} \ \bar e_L \gamma_\mu \nu_{eL} \, \bar N v^\mu \tau^+ N\,, 
\end{equation}
where $N^T = (p, n)$ is the nucleon field, $v^\mu$ is the nucleon four-velocity, and $\tau^a$ are the isospin Pauli matrices. 
In the evaluation of $g_V(m_n)$ at the nucleon scale we employ the following input. 
For the perturbative matching and running we rely on the 
NLL analysis of Ref.~\cite{Moretti:2025ghw} (including $\alpha_s$ corrections, NLL$_s$), which extends earlier results from  
Refs.~\cite{Sirlin:1981ie,Cirigliano:2023fnz}.
For the nonperturbative matching at the nucleon scale, we use 
the subtracted $\gamma W$ box as evaluated in Ref.~\cite{Cirigliano:2023fnz}, 
where various dispersive analyses~\cite{Seng:2018yzq,Seng:2018qru,Czarnecki:2019mwq,Shiells:2020fqp,Seng:2020wjq,Hayen:2020cxh}  
were combined with a new analysis of the elastic term.
We find 
\begin{align}
g_V (m_n) &= 1.01143 (12)\,,
\end{align}
with uncertainty dominated by the nonperturbative box input. 
Our central value for $g_V (m_n)$ differs from the one in Ref.~\cite{Moretti:2025ghw} by $+ 7\times 10^{-5}$.
This is entirely due to the different input for the $\gamma W$ box, for which Ref.~\cite{Moretti:2025ghw} 
uses    the lattice QCD result from Ref.~\cite{Ma:2023kfr}. 
For the  transition-independent radiative correction $\Delta_R^V$, commonly used in 
non-EFT-based approaches~\cite{Hardy:2020qwl} to superallowed $\beta$ decays,    we find\footnote{This value differs from $\Delta_R^V=2.436(16)\%$~\cite{Moretti:2025ghw} again due to the use of lattice QCD~\cite{Ma:2023kfr} for the $\gamma W$ box therein. In both the EFT and the traditional formalism, this change in the $\gamma W$ box would increase $V_{ud}$ by $7\times 10^{-5}$.}  
\be
\label{DeltaRV}
\Delta_R^V  = [g_V(m_n)]^2 \left(1 + \frac{5 \alpha (m_n)}{8 \pi} \right)-1 =2.451(25)\%\,.
\ee
Below the nucleon mass scale we evolve $g_V (\mu)$ to the appropriate  transition-dependent 
scale $\mu_\pi$ as defined in Eq.~\eqref{mupi} below, which is determined by the nuclear radius $R$, using the NLL RG equations~\cite{Cirigliano:2023fnz}.

In addition, $C_{\text{eff}}^{(g_V)}$ receives contributions from 
integrating out soft and potential photons. These start at $\mathcal O(\alpha^2)$, and arise from
 two-body loop and three-body tree-level diagrams. 
The matching 
condition for $C_{\text{eff}}^{(g_V)}$ at the nuclear scale $\mu = \mu_\pi \simeq  k_F$ is given by~\cite{Cirigliano:2024msg}
\begin{align}\label{eq:Ceff}
C_\text{eff}^{(g_V)} &= g_V\left[c_W^{(g_V,0)}+Z c_W^{(g_V,1)}+ Z^2 c_W^{(g_V,2)}\right]\,,
\end{align}
where $Z$ is the charge of the final-state nucleus and 
\begin{align}
    c_W^{(g_V,0)}&=1+(-1\pm 1)\left(\frac{1}{2}C_\delta -C_\delta^\text{3b}\right)\,,\notag\\
 c_W^{(g_V,1)}&= C_\delta - 2 C_\delta^\text{3b}  \pm C_\delta^\text{3b}\,, \qquad  c_W^{(g_V,2)}=C_\delta^\text{3b}\,, \label{eq:bc}
\end{align}
with the upper (lower) signs referring to $\beta^+$ ($\beta^-$) decays.
The coefficients $C_{\delta}$ and $C_{\delta}^{\text{3b}}$ were first computed in Ref.~\cite{Cirigliano:2024msg}. After 
checking the two-body loop diagrams and  reevaluating the Fourier transform of the three-body diagrams, we find 
\begin{align}
    C_{\delta} &= - \alpha^2 \left( \frac{1}{2}\log \frac{\bar\mu^2}{\Lambda^2} - \frac{7}{8}+  \gamma_E\right)\,, \label{eq:Cdelta} \\
    C^{\text{3b}}_{\delta} &= - \alpha^2 \left( \frac{1}{4} \log \frac{\bar\mu^2}{\Lambda^2} - \frac{1}{2} + \frac{ \gamma_E}{2}\right)\,, \label{eq:Cdelta3b}
\end{align}
where $\Lambda$ denotes an arbitrary subtraction scale.
The dependence on $\Lambda$ is canceled by an analogous dependence in the nuclear-structure-dependent correction $\bar\delta_{\text{NS}}$. 
These results are valid in the $\overline{\text{MS}}_\chi$ scheme \cite{Gasser:1984ux}, which is obtained by expressing the dimensional regularization scale $\mu$ in terms of the $\overline{\text{MS}}_\chi$ scale $\bar \mu$ as
\begin{equation}
   \mu = \bar\mu \left(\frac{e^{\gamma_E-1}}{4\pi}\right)^{\frac{1}{2}}\,.
\end{equation}
In the following, we will omit the bar, and all renormalization scales will be understood in the $\overline{\text{MS}}_\chi$ scheme. 
The logarithms in Eqs.~\eqref{eq:Cdelta} and \eqref{eq:Cdelta3b}
agree with the expressions in Ref.~\cite{Cirigliano:2024msg}, but we find different finite pieces.  
In particular, the difference in $C^{\text{3b}}_\delta$ arose from 
the Fourier transform to coordinate space, while the difference in $C_{\delta}$ is due to performing the Dirac algebra in $d$
versus $4$ dimension. 
$C_{\text{eff}}^{(g_V)}$ satisfies the RG equation~\cite{Borah:2024ghn,Hill:2023acw,Cirigliano:2024msg,Broadhurst:1994se,Cirigliano:2023fnz}
\begin{equation}
\label{eq:CeffRG}
\frac{d C^{(g_V)}_\text{eff}(\mu)}{d\log \mu} = \gamma^{(g_V)}C_\text{eff}^{(g_V)}(\mu)\,,
\end{equation}
with
\begin{align}
    \gamma^{(g_V)}&=\frac{\alpha}{\pi}\tilde \gamma_0+\left(\frac{\alpha}{\pi}\right)^2\tilde \gamma_1 + \left[\sqrt{1-\alpha^2 Z(Z\pm1)}-1\right] \nonumber \\
    &+ \frac{\alpha^3}{4\pi}Z^2\left(6-\frac{\pi^2}{3}\right)\,,
\end{align}
and
\begin{equation}
\tilde \gamma_0 = -\frac{3}{4},\quad \tilde \gamma_1 = \frac{5 \tilde{n}}{24} + \frac{5}{32} - \frac{\pi^2}{6}\,, 
\end{equation}
with $\tilde n =1 $ for $\mu\leq \mpi$.

\subsection{Nuclear-structure-dependent corrections}

Next, we consider contributions sensitive to nuclear structure.
The exchange of potential and soft photons can be captured by calculating the NMEs  of two- and three-body transition operators between the wave functions of initial and final states~\cite{Cirigliano:2024msg,Cirigliano:2024rfk}. 
The leading contributions to 
$\bar\delta_\text{NS}$ 
arise at $\mathcal O(\alpha \epsilon_\chi)$,
$\mathcal O(\alpha {\mathcal Q}_\text{EC}/\mpi)$,
and ${\mathcal O}(\alpha^2)$. The first corrections are induced by transition operators that are independent of the lepton energies, but require insertions of vertices from subleading terms in the chiral Lagrangian. The second type of operators, on the other hand, can be built using only leading-order interactions, but are proportional to the leptonic energies and momenta. The third corrections arise from the exchanges of two photons between the electron and the nucleons. 
We can  write
\begin{equation}\label{eq:deltaNS}
    \bar\delta_\text{NS} = \delta_\text{NS}^{(0)} + \overline{\delta_\text{NS}^{E}}\,,
\end{equation}
where $\delta_\text{NS}^{(0)}$ is energy independent, and 
$\overline{\delta_\text{NS}^{E}}$ results from averaging energy-dependent contributions over the electron or positron phase space.

At $\mathcal O(\alpha \epsilon_\chi)$ and $\mathcal O(\alpha^2)$, $\delta_\text{NS}^{(0)}$
can be expressed as
\begin{align}\label{deltaNS0}
   \delta^{(0)}_\text{NS}  &= \frac{2}{g_V(\mupi)M_\text{F}^{(0)}}   \sum_{N = n,p} \bigg[ \alpha  \Big( M^\text{mag}_{\text{GT}, N} + M^\text{mag}_{\text{T}, N} +  M^\text{CT}_{\text{GT}, N}   \nonumber \\ &  + M_{\text{LS}, N}  \Big)      
    + g_V(\mu_\pi)\alpha^2 \Big( M^+_{\text{F}, N} +
    M^{+\, \text{3b}}_{\text{F}, N} \Big)  \bigg]\,.
\end{align}
$M^{(\mathcal X)}_{i, N}$ denote the NMEs of EW potentials $\mathcal V$
\begin{align}\label{eq:densities}
M^{(\mathcal X)}_{i, N} =   \langle f |  \mathcal V^{(\mathcal X)}_{i, N} | i \rangle\,,
\end{align}
where the subscript $i = \{\text{F} ,\text{GT}, \text{T}, \text{LS}\}$
denotes the spin structure of the operators, which can have a 
Gamow--Teller (GT), Fermi (F), Tensor (T), or spin-orbit (LS) component. 
The isospin structure of the operators is accounted for by splitting them into an $N=p$ and $N=n$ component, denoting the coupling of the photon emitted by the electron or positron to a ``spectator'' proton or neutron.
The (optional) superscript $\mathcal X \in \{{\text{mag}}, {\text{CT}}, +, +\,{\text{3b}}\}$
denotes the origin and radial dependence of the operator.
With the exception of $\mathcal X = +\,\text{3b}$, the operators are two-body operators.
For positron emission, the operators are given by~\cite{Cirigliano:2024msg}
\begin{align}
    \mathcal V^{\mathcal X}_{\text{F}, N} &= 
    \sum_{j < k} h_{\text{F},N}^{\mathcal X}(r_{jk}) \mathcal O_{\text{F},N}^{(jk)}\,, \label{eq:OF} \\ 
   \mathcal O_{\text{F},N}^{(jk)} & = 
     \tau^{- (j)}  P^{(k)}_N  + (j \leftrightarrow k) \,,  \notag\\
    \mathcal V^{\mathcal X}_{\text{GT}, N} &= \sum_{j < k} h_{\text{GT},N}^{\mathcal X}(r_{jk}) \mathcal O_{\text{GT},N}^{(jk)}\,,  \label{eq:OGT}\\ 
    \mathcal O_{\text{GT},N}^{(jk)}& =   \bsigma^{(j)} \cdot  \bsigma^{(k)} \Big[ \tau^{- (j)}  P^{(k)}_N  + (j \leftrightarrow k)  \Big]\,,
   \notag
    \\
    \mathcal V^{\mathcal X}_{\text{T}, N} & =  \sum_{j < k} h_{\text{T},N}^{\mathcal X}(r_{jk}) \mathcal O_{\text{T}, N}^{(jk)}\,, \label{eq:OT} \\ 
    \mathcal O_{\text{T}, N}^{(jk)}& =   S^{(jk)}({\bf \hat r}) \Big[ \tau^{- (j)}  P^{(k)}_N  + (j \leftrightarrow k)  \Big]\,, \notag
    \\
        \mathcal V^{}_{\text{LS}, N} & =  \sum_{j < k} h_{\text{LS}, N}(r_{j k}) \mathcal O_{\text{LS}, N}^{(jk)}\,, \label{eq:Oso}\\ 
        \mathcal O_{\text{LS}, N}^{(jk)}& =  \tau^{- (j)} P_N^{(k)}  (\LL_{jk} - \LL^{\text{CM}}_{jk}) \cdot \boldsymbol{\sigma}^{(j)} + (j \leftrightarrow k)\,, \notag
\end{align}
with $\rr_{jk} = \rr_j - \rr_k$,
$r_{jk}= |\rr_{jk}|$, $
S^{(jk)}(\hat \rr)=  3 \hat\rr \cdot \bsigma^{(j)} \, \hat\rr  \cdot \bsigma^{(k)} - \bsigma^{(j)} \cdot \bsigma^{(k)}$,  $
\LL_{jk} = - \frac{i}{2} \rr_{jk}\times \left({\boldsymbol{\nabla}}_j - {\boldsymbol{\nabla}}_k\right)$,
and 
$
\LL^{\text{CM}}_{jk} = - \frac{i}{2} \rr_{jk}\times \left({\boldsymbol{\nabla}}_j + {\boldsymbol{\nabla}}_k\right)$
. Notice that when exchanging $j \leftrightarrow k$ $\LL$ remains unchanged, while $\LL^{\text{CM}} \rightarrow - \LL^{\text{CM}}$.
$P_{p,n}$ denote proton and neutron projectors $P_{p,n} = (1\pm \tau_3)/2$.
The radial functions are
\begin{align}
    h^\text{mag}_{\text{GT}, p}(r) 
    &= 4 h^\text{mag}_{T, p}(r)
    = \frac{g_A}{3 \mN}   \frac{1+\kappa_p}{r}\,, \notag\\ 
    h^\text{mag}_{\text{GT}, n}(r) &=4 h^\text{mag}_{\text{T}, n}(r) 
     = \frac{g_A}{3 \mN}   \frac{\kappa_n}{r}\,, \notag\\
     h^\text{CT}_{\text{GT}, p}(r) &= - \frac{4\pi}{3} (g^{N\!N}_{V1} + g^{N\!N}_{V2}) \delta_{\Lambda_{\text{EFT}}}(r)\,,\notag \\  
     h^\text{CT}_{\text{GT}, n}(r) &= - \frac{4\pi}{3} (g^{N\!N}_{V1} - g^{N\!N}_{V2}) \delta_{\Lambda_{\text{EFT}}}(r)\,,\notag \\
     h_{\text{LS}, p}(r) &= + \frac{g_A}{2 m_N} \frac{1}{r}\,,\notag  \\ 
     h_{\text{LS}, n}(r) &= 0\,,\notag \\
     h^+_{{\rm F}, p}(r,\Lambda) &= - \log(r \Lambda)\,, \notag\\
    h^+_{{\rm F}, n}(r,\Lambda) &= 0\,. 
    \label{eq:hplus}
\end{align}
Here $g_A \simeq 1.27$ is the axial charge, $\kappa_p$ and $\kappa_n$
are the proton and neutron anomalous magnetic moments, $\kappa_p \simeq 1.79$ and $\kappa_n \simeq -1.91$.
$\delta_{\Lambda_{\text{EFT}}}(r)$ denotes a regularization of the $\delta$ function, to be chosen compatibly with the strong interaction potentials, including even possible nonlocal regularizations of contact interactions, as the one used in the EM potential~\cite{Entem:2003ft}. The regularization adopted in this paper is discussed in App.~\ref{app:regulator}.
$g_{V1}^{N\!N}$
and $g_{V2}^{N\!N}$
are two-nucleon low-energy constants (LECs).
We introduce the combinations
\begin{align}
\label{LECs_linear_combinations}
    g_{Vp}^{N\!N} = g_{V1}^{N\!N} + g_{V2}^{N\!N}\,,\notag \\
    g_{Vn}^{N\!N} = g_{V1}^{N\!N} - g_{V2}^{N\!N}\,,
\end{align}
and, in order to provide dimensionless matrix elements, define
\begin{equation}
\left \{   g^{N\!N}_{Vp}, g^{N\!N}_{Vn} \right\} \equiv \frac{1}{m_N} \frac{1}{(2 F_\pi)^2} \left\{ \tilde{g}^{N\!N}_{V p}, \tilde{g}^{N\!N}_{Vn} \right\}\,,
\end{equation}
with $F_\pi = 92.3\MeV$ and $m_N = 938.9\MeV$.
The contact matrix elements will be given in units of $\tilde g^{N\!N}_{Vp}$, $\tilde g^{N\!N}_{Vn}$. With this normalization, we expect the LECs to scale as $\tilde g^{N\!N}_{Vi} = \mathcal O(1)$.

Equations~\eqref{eq:OF}--\eqref{eq:Oso} 
agree with the expressions in Ref.~\cite{Cirigliano:2024msg}
with two important exceptions:
\begin{enumerate}
    \item the component proportional to ${\mathbf L}^{\text{CM}}$ was omitted in Ref.~\cite{Cirigliano:2024msg};
    \item the radial function $h_{\text{LS}, p}$ differs by a sign.\footnote{We thank C.-Y.~Seng for pointing out this sign mistake, which originated in taking the Fourier transform of Eq.~(35) in Ref.~\cite{Cirigliano:2024msg}.  }
\end{enumerate}
Because of a large cancellation between the ${\bf L}$ and ${\mathbf L}^{\text{CM}}$ components, 
these differences lead to nonnegligible numerical differences, and the expression given here supersede those from Ref.~\cite{Cirigliano:2024msg}. 

For the three-body matrix element $M^{+\text{3b}}_{F, p}$, the potential is defined as 
\begin{align}
 {\mathcal V}_+^\text{3b}(\rr, \Lambda) &=    \sum_{i \neq j \neq k}  h^{+\,\text{3b}}_{F,p}(r_{ij}, r_{ik},r_{jk}) \tau^{-(i)}  P_p^{(j)} P_p^{(k)} \nonumber \\
    &  =   2  \sum_{i < j < k}   \bigg[ h^{+\,\text{3b}}_{F,p}(r_{ij}, r_{ik}, r_{jk}) \tau^{-(i)}  P_p^{(j)} P_p^{(k)} \nonumber \\ & \quad+ h^{+\,\text{3b}}_{F,p}(r_{ij}, r_{jk}, r_{ik})\tau^{-(j)}  P_p^{(i)} P_p^{(k)} \nonumber  \\ 
    &\quad  + h^{+\,\text{3b}}_{F,p}(r_{jk}, r_{ij}, r_{ik}) \tau^{-(k)}  P_p^{(i)} P_p^{(j)} \bigg]\,,
    \label{eq:V3plusMain}
\end{align}
with
\begin{align}\label{eq:radial3b}
    h^{+\,\text{3b}}_{F,p}(r_{ij},r_{ik},r_{jk},\Lambda) &= - \frac{1}{2} \log \bigg[\frac{\Lambda}{2} \Big(r_{ij} + r_{ik} + r_{jk} \Big) \bigg]\nonumber
\\
& - \frac{1}{4} \frac{r^2_{ij} + r_{ik}^2 - r_{jk} (r_{ij}+ r_{ik})}{r_{ik} r_{ij}}\,.    
\end{align}

In the EFT formalism $\bar\delta_\text{NS}$ also receives an energy-dependent correction at $\mathcal O(\alpha {\mathcal Q}_\text{EC}/\mpi)$.
This correction is given by
\begin{align}\label{deltaNSE}
\overline{\delta^E_\text{NS}} &= \mp \alpha \frac{2}{M_\text{F}^{(0)}} R_A E_0
\tilde{f}_E  M^E_{\text{F}, p}\,,
\end{align}
where the upper sign is for $\beta^+$, the lower sign for $\beta^-$ decays. In Eq.~\eqref{deltaNSE}, we neglected pion-range contributions, which were shown to be small in $^{10}$C and $^{14}$O~\cite{King:2025fph,Cirigliano:2024msg}. We checked that these contributions are negligibly small for all the nuclei of interest.
In Eq.~\eqref{deltaNSE}, we introduced a factor of $R_A = 1.2 A^{1/3}$ fm to make the matrix elements dimensionless.
$E_0$ is the electron endpoint energy, 
\begin{equation}
    E_0 = {\mathcal Q}_\text{EC} - m_e\,,
\end{equation}
in terms of the electron capture 
${\mathcal Q}_\text{EC}$ values given in Ref.~\cite{Hardy:2020qwl}, and summarized in Table~\ref{tab:exp}.  
The quantity $\tilde f_E$
arises from the phase-space average, and is given by
\begin{align}
\label{tildefE}
 \tilde{f}_E &= \frac{1}{E_0}   \left(\frac{4}{3}\langle E_e \rangle + \frac{1}{6} E_0 + \frac{1}{2}\left\langle \frac{m^2_e}{ E_e} \right\rangle \right)\,,
 \end{align}
where the phase-space average is defined in Eq.~\eqref{eq:PSaverage} below.
The matrix element $M_{F,p}^E$ has the same definition as in Eq.~\eqref{eq:OF}, with 
\begin{equation}\label{eq:hE}
    h^E_{\text{F}, p}(r)  = - \frac{r}{2 R_A}\,.
\end{equation}

\subsection{Ultrasoft corrections and the phase-space factor}

Finally, we consider the 
corrections arising from 
ultrasoft photons, sensitive to low-energy scales as the electron momentum. The largest contribution in this regime comes from the Fermi function.
Here we use the calculation of the Fermi function in dimensional regularization \cite{Hill:2023acw,Hill:2023bfh}, in which the traditional dependence on the nuclear radius $R$
is replaced by a dependence on the renormalization scale $\mu$.
After translating the results of Refs.~\cite{Hill:2023acw,Hill:2023bfh} from $\overline{\text{MS}}$ to $\overline{\text{MS}}_\chi$, the Fermi function to all orders in $Z^n\alpha^n$ is given by
 \begin{align}
 \label{eq:FF}
\bar F(\beta,\mu)  &= \frac{4\eta}{(1+\eta)^2}\frac{2(1+\eta)}{\Gamma(2\eta+1)^2}|\Gamma(\eta+iy)|^2 e^{\pi y}\notag\\
&\quad\times \left(\frac{2 |\pp_e|}{\mu}e^{1/2-\gamma_{E}}\right)^{2(\eta-1)}\,,
\end{align}
with $\eta = \sqrt{1-\alpha^2 Z^2}$ and $y =\mp Z\alpha/\beta$.
For the scale choice $\mu = R^{-1} e^{1/2 - \gamma_E}$, Eq.~\eqref{eq:FF}
differs from the traditionally employed Fermi function~\cite{Wilkinson:1982hu} by $4\eta/(1+\eta)^2\approx1-\alpha^4Z^4/16$.
We encode the Fermi function into the  phase-space factor $\bar f$
\begin{equation}\label{eq:PS}
 \bar f(\mu) =  \frac{1}{m_e^5} \int_{m_e}^{E_0} d E_e w_0(E_e)  
    \, \tilde C (E_e) \, 
 \bar F (\beta, \mu)\,,
\end{equation}
where
\begin{equation}
w_0 (E_e) 
=   |\pp_e| E_e (E_0-E_e)^2\,.
\end{equation}
The factor  $\tilde C (E_e)$   encodes corrections due to the 
nuclear EW form factor, nuclear recoil, 
atomic electron screening, and  atomic overlap~\cite{Hardy:2020qwl,Gorchtein:2023naa}, for which we take the expressions of Refs.~\cite{Hayen:2017pwg,Hardy:2004id,Hardy:2008gy}, with the subtractions and modifications described in Sec.~\ref{sec:phase_space} and App.~\ref{app:subtractions}.

At $\mathcal O(\alpha)$, 
$\delta_R' (E_e,\bar\mu)$ is determined by the Sirlin function, as computed in heavy-baryon chiral perturbation theory~\cite{Ando:2004rk,Cirigliano:2022hob}.
We will use here the
$\mathcal O(\alpha^2 Z)$ corrections
computed in the same scheme, which were derived in Refs.~\cite{Crosas:2025xyv,Cao:2025zxs}.
$\delta^\prime_{R}$ and $\delta^E_{\text{NS}}$ are energy dependent. To study corrections to the half-life, we define the phase-space average in terms of the Fermi function, as
\begin{align}
\label{eq:PSaverage}
\bar G(\mu) &\equiv \langle G(\mu) \rangle\\
&\equiv \frac{\int_{m_e}^{E_0} d E_e w_0(E_e)  
    \, \tilde C (E_e) \, 
 \bar F (\beta, \mu) \,  
    \tilde G (E_e,\mu)  }{\int_{m_e}^{E_0} d E_e  w_0 (E_e)\, \tilde C (E_e) \, 
 \bar F (\beta, \mu)  }\,,\notag
 \end{align} 
where $\tilde G$ denotes any energy-dependent function. 
We use the bracket and bar notation interchangeably.

\section{Fermi matrix elements and nuclear charge distributions}
\label{sec:density}

The nuclear-structure-dependent contributions discussed above require input on two- and three-body matrix elements. At first sight, it is surprising that even three-body matrix elements at ${\mathcal O}(\alpha^2)$ need to be considered, which is why such contributions were not evaluated in Ref.~\cite{Cirigliano:2024msg}. However, we found that for increasing $Z$ such effects become more and more essential, as the corresponding NMEs prove to scale with $Z^2$, in contrast to their two-body counterparts, for which a cancellation prevents any coherent enhancement at all. Since, in addition, the matching coefficient~\eqref{eq:Cdelta3b} is enhanced by a factor $\pi^2$ over its naive power-counting estimate, in a similar way as for the Fermi function,\footnote{As argued in Ref.~\cite{Cirigliano:2024msg}, given countless examples in which such numerical factors are not immediately related to a given topology, e.g., in pion--nucleon scattering~\cite{Gasser:2002am,Hoferichter:2009ez,Hoferichter:2009gn}, the multiple scattering series~\cite{Beane:2002wk,Liebig:2010ki,Baru:2010xn,Baru:2011bw,Baru:2012iv}, or the axial charge of the nucleon~\cite{Cirigliano:2022hob,Hall:2025ytt,Gialidi:2026yyk,Tomalak:2026ido}, we do not attempt to capture such factors by a modified power counting, but consider them to be only revealed by explicit calculations.}   
we do need the corresponding NMEs for a phenomenologically viable evaluation of $\bar\delta_\text{NS}$. Further, we do not have a full ab initio calculation for these NMEs available, but we do have access to the nuclear densities, both weak and point-proton, corresponding to the two Hamiltonians employed in the CC calculations. Accordingly, we devised a strategy to estimate Fermi matrix elements in terms of these densities, outlined in this section and further detailed in App.~\ref{app:density}, while App.~\ref{app:FermiGas} presents another consistency check using a Fermi-gas model. In the numerical analysis, we will validate the density treatment in the two-body sector, apply it in the three-body case to evaluate $M_{\text{F},p}^{+\,\text{3b}}$, and show that even higher contributions (not captured by the Fermi function) are sufficiently suppressed, see App.~\ref{app:higher-body}.

Taking ${\mathcal V}_F^E$ or  the ${\mathcal O}(\alpha^2)$ potential ${\mathcal V}_F^+$  as an example, their form in terms of fields can schematically be written as 
$\hat V_\text{2b} = \int_{\vx_i} p_{\vx_1}^\dagger p_{\vx_2}^\dagger V(\vx_1,\vx_2)p_{\vx_2}n_{\vx_1}$. One can rearrange the fields so that they form bilinears, $\hat V_\text{2b} = \int_{\vx_i} (p^\dagger n)_{\vx_1} (p^\dagger p)_{\vx_2} V(\vx_1,\vx_2)$, which are related to the charge and weak density operators. Taking the matrix element of $\hat V_\text{2b}$ and inserting complete sets of states between each of the bilinears, then leads to 
\beq\label{eq:densities2b}
M_\text{2b}= \sum_n \int_{\vx_1,\vx_2} \rho^W_{fn}(\vx_1)\rho_{ni}^\text{EM}(\vx_2)V(\vx_1,\vx_2)\,,
\eeq
where 
$\langle \vp_f|\hat V_\text{2b}|\vp_i\rangle = (2\pi)^3\delta ^3(\vp_i - \vp_f)M_\text{2b}$, while  $\rho^W_{fn}$ and $\rho^\text{EM}_{ni}$ are the transition matrix elements of the weak and charge operators, between the initial/final states and intermediate state $|n\rangle$. Approximating the above expression by restricting the sum to the initial/final states then allows us to estimate the two-body matrix elements with an integral over the weak and charge densities. These densities determine the bulk properties of the respective nuclei, scaling coherently with the number of nucleons, and are thus expected to yield the dominant effect compared to excited-state corrections. Below, we will validate this approximation at the two-body level against explicit ab initio results. 
In the case of ${\mathcal V}_F^E$, Eq.~\eqref{eq:densities2b} is reminiscent of the Zemach moments discussed in Refs.~\cite{Wang:2016rqh,Friar:2003zg,Holstein:1974zza}.

The same approach leads to similar relations for the three-body potential ${\mathcal V}_+^\text{3b}$. In this case, there are additional terms arising from the anticommutation relations of the nucleon fields. These lead to two-body terms, which can similarly be expressed as integrals over densities, see App.\ \ref{app:density} for details. There are several ways to order the bilinears, giving rise to multiple relations
\begin{align}\label{eq:densities3b}
M_\text{3b} &=\sum_{m,n}\int_{\vx_i}\rho^\text{EM}_{fm}(\vx_3)\rho^\text{EM}_{mn}(\vx_2)\rho^{W}_{ni}(\vx_1)V(\vx_1,\vx_2,\vx_3)\nn\\
&-
\sum_n\int_{\vx_i}\Big[\rho^\text{EM}_{fn}(\vx_2)\rho^{W}_{ni}(\vx_1)\nn\\
&\quad+2\rho^{W}_{fn}(\vx_2)\rho^\text{EM}_{ni}(\vx_1)\Big] V(\vx_1,\vx_2,\vx_2)
\nn\\
 &=
\sum_{m,n}\int_{\vx_i}\rho^\text{EM}_{fm}(\vx_3)\rho^W_{mn}(\vx_2)\rho_{ni}^\text{EM}(\vx_1)V(\vx_1,\vx_2,\vx_3)\nn\\
&-
2\sum_n\int_{\vx_i}\rho^{W}_{fn}(\vx_2)\rho^\text{EM}_{ni}(\vx_1)V(\vx_1,\vx_2,\vx_2)
\nn\\
&=
\sum_{m,n}\int_{\vx_i}\rho^W_{fm}(\vx_3)\rho_{mn}^\text{EM}(\vx_2)\rho_{ni}^\text{EM}(\vx_1)V(\vx_1,\vx_2,\vx_3)\nn\\
&-
\sum_n\int_{\vx_i}\rho^{W}_{fn}(\vx_2)\rho^\text{EM}_{ni}(\vx_1)V(\vx_1,\vx_2,\vx_2)\,.
\end{align}
These expressions are in principle equivalent, however, differences can arise when neglecting the excited states. In what follows, we will use Eq.~\eqref{eq:densities3b} to estimate the matrix elements of ${\mathcal V}_+^\text{3b}$ and we will use the comparison of Eq.~\eqref{eq:densities2b} with explicit calculations, as well as the differences between the representations in Eq.~\eqref{eq:densities3b}, to assign an uncertainty.
In addition, we will approximate the radial function of the three-body potential by its logarithmic term, neglecting the second line in Eq.\ \eqref{eq:radial3b}. We checked that the difference falls safely within our uncertainty estimates.

\section{Nuclear Matrix Elements}
\label{sec:NMEs}

In this section, we present the main outcome of this work: 
first ab initio calculation of $\bar\delta_{\text{NS}}$ for $0^+$ transitions ranging from $^{10}$C $\rightarrow$ $^{10}$B to $^{54}$Co $\rightarrow$ $^{54}$Fe. The calculations are performed using CC theory, and are compared to QMC calculations in $^{10}$C and $^{14}$O. In addition, we calculated  the nuclear charge and weak distributions, which we use to constrain three-body matrix elements and corrections to the phase-space factor. 

\subsection{Coupled Cluster method}

The CC calculations begin with the intrinsic nuclear Hamiltonian
\begin{equation}
H =
\sum_{i<j}\left[
\frac{(\mathbf p_i-\mathbf p_j)^2}{2\mN A}
+ V_{N\!N}^{(i,j)}
\right]
+
\sum_{i<j<k} V_{N\!N\!N}^{(i,j,k)}\,,
\label{eq:Hamiltonian}
\end{equation}
where $\mN$ is the nucleon mass and $V_{N\!N}$ and $V_{N\!N\!N}$ are the two- and three-nucleon interactions.
For these calculations, we employ the \intem~\cite{Entem:2003ft, Hebeler:2010xb} and \intsat~\cite{Ekstrom:2015rta} chiral interactions, which provide complementary Hamiltonians for estimating the interaction dependence of the calculated NMEs. Details of the two-body matrix elements in the harmonic-oscillator basis are provided in App.~\ref{app:TBME}.

Calculations with \intem~use a harmonic-oscillator frequency $\hbar\omega=16\MeV$, while those with \intsat~use $\hbar\omega=12\MeV$.
The largest calculations include single-particle states with $N=2n+l\leq N_\text{max}=10$, and residual basis dependence is estimated from the difference between $N_\text{max}=10$ and $N_\text{max}=8$.
Three-nucleon matrix elements are restricted to $N_1+N_2+N_3\leq 16$ for both interactions.

For each superallowed transition, we use the even--even member of the isobaric multiplet as the many-body reference state.
Starting from a spherical oscillator basis, we construct an axially deformed Hartree--Fock state $|\Phi_0\rangle$, with reference energy $E_\text{ref}=\langle\Phi_{0}|H|\Phi_{0}\rangle$, thereby incorporating static correlations that would otherwise appear as high-order particle--hole excitations in a spherical expansion~\cite{Ring:1980, Duguet:2014jja, Qiu:2017, Tsuchimochi:2018, Novario:2020kuf}. The Hamiltonian is then normal ordered with respect to the reference state, and residual three-nucleon terms are neglected in the normal-ordered two-body approximation~\cite{Hagen:2007ew, Roth:2011vt}.

CC theory~\cite{Coester:1958seh, Coester:1960anj,Cizek:1966zzp,Kummel:1978zz, Bartlett:2007zza,Hagen:2013nca} is formulated in terms of the similarity-transformed Hamiltonian
\begin{equation}
    \overline H_\text{N} = e^{-T} H_\text{N} e^T\,,
\label{eq:Hbar}
\end{equation}
where $H_\text{N} = H - E_\text{ref}$ is the normal-ordered Hamiltonian and the cluster operator $T$ creates particle--hole excitations from $|\Phi_0\rangle$. The correlated right ground state is $|\Psi_0\rangle = e^T |\Phi_0\rangle$, with correlation energy $\Delta E_{0} = \langle\Phi_{0}|\overline H_\text{N}|\Phi_{0}\rangle$, and the amplitudes in $T$ are determined by decoupling the reference from excited configurations within the chosen truncation. Since $\overline H_\text{N}$ is non-Hermitian, the corresponding left ground state is written as $\langle\widetilde\Psi_{0}|=\langle\Phi_0|(1+\widetilde\Lambda)e^{-T}$, where $\widetilde\Lambda$ is a deexcitation operator truncated consistently with $T$~\cite{Bartlett:2007zza,Hagen:2013nca}.

The odd--odd partner in the superallowed transition is described with charge-exchange equation-of-motion (EOM) CC theory~\cite{Bartlett:2007zza, Shavitt:2009, Jansen:2011gb, Jansen:2012ey, Hagen:2012fb, Binder:2012mk}.
The right and left EOM states are written as
$|\Psi_k\rangle = R_k|\Psi_0\rangle = R_k e^T|\Phi_0\rangle$
and
$\langle\widetilde\Psi_k|
 = \langle\Phi_0|\widetilde L_k e^{-T}$,
respectively, and satisfy the biorthogonal eigenvalue equations
\begin{align}
    \overline H_\text{N} R_k|\Phi_0\rangle &= \Delta E_k R_k|\Phi_0\rangle \,, \notag \\
    \langle\Phi_0|\widetilde L_k\overline H_\text{N} &= \Delta E_k\langle\Phi_0|\widetilde L_k\,, \notag \\
    \langle\Phi_0|\widetilde L_k R_{k'}|\Phi_0\rangle &= \delta_{kk'}\,.
\label{eq:EOM}
\end{align}
Here $R_k$ and $\widetilde L_k$ are the charge-exchange excitation
and deexcitation operators, respectively, with the quantum numbers
carried by the transition operator.
The eigenvalues are the EOM state correlation energies, $\Delta E_{k} = E_{k} - E_\text{ref}$.

\begin{table*}[t]
\centering
    \renewcommand{\arraystretch}{1.3}
\begin{tabular}{c c rrr rrr rrr rrr}
\toprule
Transition  & Interaction   & \multicolumn{3}{c}{ $M^\text{mag}_{\rm GT, p}$} & \multicolumn{3}{c}{ $M^\text{mag}_{\rm GT, n}$} & 
\multicolumn{3}{c}{ $M^{\rm CT}_{\rm GT, p}$} & \multicolumn{3}{c}{ $M^{\rm CT}_{\rm GT, n}$} 
\\ \cline{3-5}\cline{6-8}\cline{9-11}\cline{12-14}
& & \multicolumn{1}{c}{T-1} & \multicolumn{1}{c}{N10} & \multicolumn{1}{c}{N8} & \multicolumn{1}{c}{T-1} & \multicolumn{1}{c}{N10} & \multicolumn{1}{c}{N8}
  & \multicolumn{1}{c}{T-1} & \multicolumn{1}{c}{N10} & \multicolumn{1}{c}{N8} & \multicolumn{1}{c}{T-1} & \multicolumn{1}{c}{N10} & \multicolumn{1}{c}{N8}
\\\colrule

\multirow{2}{*}{$^{10}$C $\rightarrow$ $^{10}$B} &  \intem   & $-0.375 $&$ -0.337 $&$ -0.340 $&$ 0.087 $&$  0.081 $&$     0.082 $&$ 0.082 $&$     0.078 $&$     0.079$&$ 0.045 $&$     0.046 $&$     0.047$ \\ 
& \intsat &$ -0.331 $&$    -0.293 $&$    -0.294$&$ 0.089 $&$     0.082 $&$     0.082 $&$ 0.066 $&$     0.063 $&$     0.063 $&$ 0.042 $&$     0.043 $&$     0.042 $\\

\multirow{2}{*}{$^{14}$O $\rightarrow$ $^{14}$N}  &$ 
$&$ -0.484 $&$ -0.500 $&$ -0.501 $&$ 0.181 $&$ 0.222 $&$ 0.223 $&$ 0.083 $&$ 0.087 $&$ 0.087 $&$  0.043 $&$ 0.050 $&$ 0.050 
$\\
&&$ -0.434 $&$ -0.457 $&$ -0.455 $&$ 0.180 $&$  0.228 $&$ 0.225 $&$ 0.065 $&$ 0.069$&$ 0.069 $&$ 0.039 $&$ 0.045 $&$ 0.045 $\\

\multirow{2}{*}{$^{26}$Si $\rightarrow$$^{26}$Al} &&$ -0.377 $&$ -0.392 $&$ -0.394 $&$
0.126 $&$ 0.141 $&$ 0.143 $&$
0.104 $&$ 
0.105 $&$ 
0.106 $&$
0.067 $&$ 
0.083 $&$
0.085
$\\
&&$ -0.330 $&$-0.354 $&$ -0.354 $&$ 
0.122 $&$ 0.134 $&$ 0.133 $&$ 
0.077 $&$ 
0.080 $&$
0.080 $&$
0.056 $&$ 0.068 $&$ 0.068
$\\
 
\multirow{2}{*}{$^{34}$Ar $\rightarrow$ $^{34}$Cl} &$ 
$&$ -0.452 $&$ -0.475 $&$ -0.477 $&$ 0.231 $&$ 0.278 $&$ 0.280 $&$ 0.100 $&$ 0.100 $&$ 0.102 $&$ 0.071 $&$ 0.081 $&$ 0.082 $\\
&&$ -0.396 $&$ -0.436 $&$ -0.436 $&$ 0.202 $&$ 0.270 $&$ 0.268 $&$ 0.074 $&$ 0.077 $&$ 0.078 $&$ 0.056 $&$ 0.066 $&$ 0.067 $\\

\multirow{2}{*}{$^{38}$Ca $\rightarrow$ $^{38}$K} &&$
-0.483 $&$ -0.493 $&$ -0.496 $&$  0.229 $&$ 0.270 $&$     0.271 $&$
0.101 $&$ 0.100 $&$ 0.102 $&$
0.071 $&$ 0.082 $&$ 0.083 $\\
 &&$-0.423 $&$  -0.442 $&$ -0.443 $&$ 
0.210 $&$ 0.248 $&$ 0.249 $&$ 
0.075 $&$ 0.075 $&$ 0.077 $&$
0.058 $&$ 0.065 $&$ 0.066 $\\
\colrule
\multirow{2}{*}{$^{26}$Al $\rightarrow$ $^{26}$Mg} &$ 
$&$  -0.188 $&$     -0.208 $&$    -0.210 $&$ 
0.259 $&$     0.270 $&$  0.271 $&$ 
0.068 $&$     0.083 $&$     0.085 $&$
0.106 $&$    0.107 $&$     0.108 $\\
&&$  -0.183 $&$    -0.196 $&$   -0.196 $&$ 
0.228 $&$     0.243 $&$    0.244$&$ 
0.057 $&$     0.068 $&$    0.068 $&$ 
0.080 $&$     0.083 $&$    0.083 $\\
 
\multirow{2}{*}{$^{34}$Cl $\rightarrow$ $^{34}$S} &&$
-0.342 $&$    -0.408 $&$    -0.410 $&$ 
 0.309 $&$     0.324 $&$     0.326 $&$ 
 0.071 $&$     0.080 $&$     0.082 $&$ 
 0.101 $&$     0.101 $&$     0.103 $\\ 
&&$ -0.300 $&$    -0.394$&$    -0.392 $&$
0.272 $&$     0.298 $&$     0.298 $&$ 0.056 $&$     0.065 $&$     0.066 $&$
0.076 $&$     0.079 $&$     0.079 $\\
 
\multirow{2}{*}{$^{38}$K $\rightarrow$ $^{38}$Ar} &&$ 
-0.339 $&$    -0.397 $&$    -0.399 $&$ 
 0.331 $&$     0.338 $&$     0.340 $&$
 0.071 $&$     0.081 $&$     0.082 $&$ 
 0.103 $&$     0.102 $&$     0.103  
$\\
&&$ -0.311 $&$    -0.366 $&$    -0.366 $&$
 0.291 $&$     0.303 $&$     0.304 $&$
 0.057 $&$     0.064 $&$     0.065 $&$
 0.076 $&$     0.077 $&$     0.078  $\\
 
\multirow{2}{*}{$^{42}$Sc $\rightarrow$ $^{42}$Ca} &&$
-0.217 $&$    -0.226 $&$    -0.228 $&$ 
 0.264 $&$     0.234 $&$     0.237 $&$
0.089 $&$     0.093 $&$     0.095 $&$
0.124 $&$     0.111 $&$     0.113
$\\
&&$ 
-0.200 $&$ -0.208 $&$ -0.209$&$
 0.221 $&$  0.205 $&$ 0.207 $&$
0.068 $&$     0.071 $&$     0.071 $&$
0.086 $&$     0.082 $&$     0.083 $\\

\multirow{2}{*}{$^{46}$V $\rightarrow$ $^{46}$Ti}
 &&$ -0.267 $&$ -0.268 $&$    -0.272 $&$
 0.238 $&$     0.232 $&$     0.235 $&$
 0.095 $&$     0.101 $&$     0.103 $&$ 
 0.117 $&$     0.116 $&$     0.119 $\\
&&$  -0.243 $&$    -0.241 $&$    -0.242   $&$
 0.208 $&$     0.207 $&$     0.209 $&$ 
 0.074 $&$     0.077 $&$     0.078 $&$
 0.087 $&$     0.088 $&$     0.088 $\\
 
 \multirow{2}{*}{$^{50}$Mn $\rightarrow$ $^{50}$Cr} 
&&$ -0.276 $&$    -0.274 $&$   -0.278 $&$
    0.245 $&$     0.241 $&$     0.243 $&$ 
    0.101 $&$     0.105 $&$     0.107 $&$ 
    0.122 $&$     0.120 $&$     0.122 
$\\
&&$  
-0.244 $&$   -0.239 $&$    -0.241 $&$ 
0.214 $&$     0.210 $&$     0.212 $&$ 
0.076 $&$     0.078 $&$     0.079 $&$
0.089 $&$     0.087 $&$     0.088 $\\

 \multirow{2}{*}{$^{54}$Co $\rightarrow$ $^{54}$Fe} 
&$ $&$ -0.283 $&$ -0.281 $&$ -0.284 $&$
0.252 $&$     0.251 $&$     0.253 $&$
0.105 $&$     0.110 $&$     0.112 $&$ 
0.127 $&$     0.126 $&$     0.128 $\\
&&$  -0.246 $&$    -0.239 $&$    -0.242 $&$
0.212 $&$     0.215 $&$     0.217 $&$ 
0.078 $&$     0.079 $&$     0.080 $&$ 
0.089 $&$     0.089 $&$     0.090 $ \\
\botrule
  \end{tabular}
\caption{NMEs of the magnetic and contact GT operators, for the \intem\ and the \intsat\ interactions.
For each matrix element, the columns labeled N8 and N10 are computed with singles + doubles CC truncation (CCSD) and $N_{\text{max}}=8$ and $10$, respectively. The columns labeled T-1 include triples (CCSDT-1), by adding the difference 
of $N_{\text{max}}=6$ calculations done with and without triples to the $N_{\text{max}}=10$ CCSD calculation.
}\label{tab:1}
 \end{table*}

In superallowed transitions, the isobaric analog state has $J^{\pi}=0^{+}$. However, because the deformed reference breaks rotational symmetry, we instead target charge-exchange EOM states with $J^{\pi}_{z}=0^{+}$, several of which can carry nonzero Fermi transition strength. We therefore identify the isobaric analog state in the deformed basis by its Fermi matrix element,
\begin{equation}
    \Big\langle \widetilde\Psi_0\Big| \sum_{a=1}^A \tau_a^\mp \Big|\Psi_k\Big\rangle
    \Big\langle \widetilde\Psi_k\Big| \sum_{a=1}^A \tau_a^\pm \Big|\Psi_0\Big\rangle
    \simeq |M_\text{F}^{(0)}|^{2}=2\,,
\label{eq:IAS_diagnostic}
\end{equation}
as expected for a $T=1$ superallowed transition in the isospin-symmetric limit. In most cases, one of the lowest states carries nearly all of the Fermi strength, while the remaining candidates have negligible transition amplitudes. In the rare case that the strength is fragmented between nearby states, both contributions are included.

\begin{figure*}[t!]
    \centering
\includegraphics[width=\linewidth]{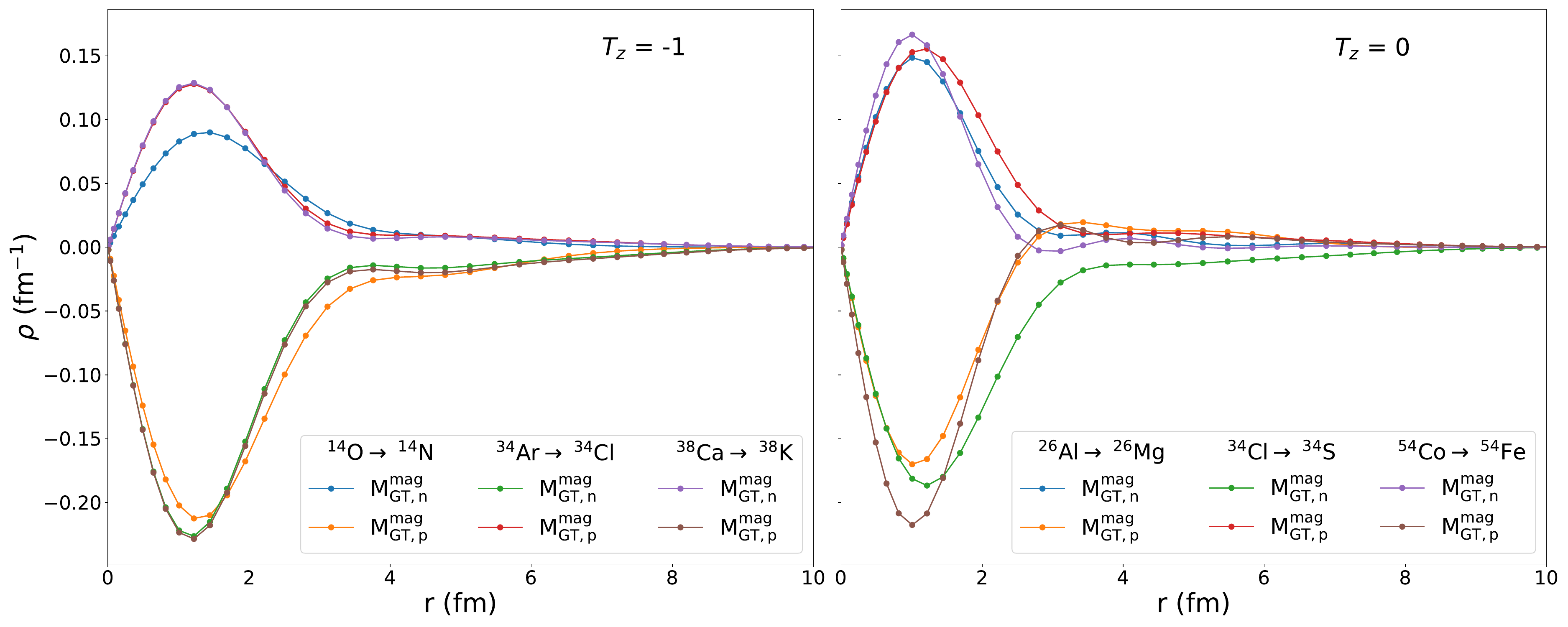}
\caption{GT operator densities for three $T_z = -1$ nuclei (left) and three $T_z = 0$ nuclei (right). The operator densities are calculated with the \intem\, interaction, with CCSD(N8). $T_z=0$ nuclei show a larger cancellation between  $M^{\text{mag}}_{\text{GT}, n}$
and $M^{\text{mag}}_{\text{GT}, p}$.
}\label{fig:GTdensities}
\end{figure*}

The EFT operators, derived in Ref.~\cite{Cirigliano:2024msg} and entering Eq.~\eqref{eq:densities} and Eqs. \eqref{eq:OF}--\eqref{eq:Oso}, are evaluated in their similarity-transformed form,
\begin{equation}
\overline{\mathcal V^{(\mathcal X)}_{i, N}} = e^{-T} \mathcal V^{(\mathcal X)}_{i, N} e^T \,.
\label{eq:Obar}
\end{equation}
The corresponding CC matrix element is
\begin{align}
    \big| M^{(\mathcal X)}_{i, N} \big|^2 =&
    \Big\langle\Phi_0\Big|(1+\widetilde\Lambda) \overline{\mathcal V^{(\mathcal X)\dagger}_{i, N}} R_{k}\Big|\Phi_0\Big\rangle \nonumber\\
    &\Big\langle\Phi_0|\widetilde L_{k} \overline{\mathcal V^{(\mathcal X)}_{i, N}}\Big|\Phi_0\Big\rangle \,.
\label{eq:CC_ME}
\end{align}
The relative sign of each matrix element is taken from the corresponding left transition amplitude, and the remaining overall phase ambiguity is fixed separately for each decay by choosing the sign convention for which the bare Fermi matrix element is positive.  The CC calculations utilize the coordinate-space two-body operators entering $\bar\delta_\text{NS}$, whose harmonic-oscillator matrix elements are evaluated with regulator factors chosen consistently with the underlying chiral interaction, as detailed in Apps.~\ref{app:regulator} and~\ref{app:TBME}.

Our largest calculations are performed at the singles-and-doubles level (CCSD), while leading triples correlations (CCSDT-1)~\cite{watts:1993,watts:1995} are included through a limited-space correction.  Denoting any of the matrix elements $M^{(\mathcal X)}_{i, N}$ by $M$, we compute
\begin{align}
    M &= M_{(N_\text{max}=10)}^\text{CCSD} + \Delta M^{(T)} \,, \notag \\
    \Delta M^{(T)} &=
    M_{(N_\text{max}=6)}^\text{CCSDT-1} - M_{(N_\text{max}=6)}^{\rm CCSD}\,.
\label{eq:triples_correction}
\end{align}
The correction $\Delta M^{(T)}$ is evaluated using the corresponding CCSDT-1 truncation for the ground-state amplitudes, charge-exchange EOM amplitudes, and similarity-transformed operators.
This procedure incorporates the leading effects of triples correlations, while retaining the larger single-particle spaces needed for the final matrix elements.

The dominant CC uncertainties arise from finite-basis and many-body truncations, while additional uncertainties enter through rotational symmetry breaking and the choice of nuclear Hamiltonian and oscillator frequency.
We estimate the first two contributions explicitly, following uncertainty estimates used in related deformed-reference CC calculations of scalar nuclear observables~\cite{Simonis:2019spj, Novario:2020kuf, Hagen:2022tqp}.
For each interaction, the finite-basis uncertainty is taken as the difference between the matrix elements obtained in the $N_\text{max} =10$ and $N_\text{max} =8$ model spaces.
The uncertainty associated with omitted higher-order particle--hole excitations is estimated conservatively from the full CCSDT-1 correction of Eq.~\eqref{eq:triples_correction}.  These two contributions are combined in quadrature,
\begin{align}
    \sigma_M^\text{basis} &= \Big| M_{(N_\text{max}=10)}^\text{CCSD}
    - M_{(N_\text{max}=8)}^\text{CCSD} \Big|\,, \nonumber\\
    \sigma_M^\text{trunc} &= \Big|\Delta M^{(T)}\Big|\,, \nonumber\\ \sigma_M^\text{CC}
    &= \sqrt{ \big(\sigma_M^\text{basis}\big)^2
    + \big(\sigma_M^\text{trunc}\big)^2 }\,.
\label{eq:cc_uncertainty}
\end{align}

Rotational-symmetry-breaking effects are not assigned an additional independent uncertainty, but previous deformed-reference CC studies of scalar observable indicate that they are small compared with the dominant basis and truncation effects in medium-mass nuclei~\cite{Hagen:2022tqp}; for light nuclei, see the discussion of $^{10}$C below.
Combined dependence on the nuclear Hamiltonian and oscillator frequency is probed through the complementary calculations with \intem~at $\hbar\omega=16\MeV$ and \intsat~at $\hbar\omega=12\MeV$. Because the interaction and oscillator frequency are varied together, their resulting spread is kept separate from $\sigma_M^\text{CC}$ in the final analysis.

The CC framework offers favorable computational scaling necessary to access medium-mass nuclei beyond the reach of exact diagonalization and current QMC calculations. Together with charge-exchange EOM techniques and similarity-transformed Hamiltonian and EW operators, this enables a consistent many-body treatment of the superallowed transitions.

\subsection{\texorpdfstring{$\boldsymbol{\mathcal O(\alpha \epsilon_\chi)}$}{} corrections to \texorpdfstring{$\boldsymbol{\delta^{(0)}_{\text{NS}}}$}{}}

We start from discussing the nuclear-structure-dependent and electron-energy-independent corrections that arise at $\mathcal O(\alpha \epsilon_\chi)$. These corrections are most closely related to the traditional formalism of Ref.~\cite{Towner:1992xm}
and to the dispersive results of Refs.~\cite{Seng:2022cnq,Gennari:2024sbn}.

\begin{table*}[t]
\centering
    \renewcommand{\arraystretch}{1.3}
\begin{tabular}{cc rr rr rr}
\toprule
Transition  & Interaction &  \multicolumn{2}{c}{ $M^\text{mag}_{\text{T}, p}$} &  \multicolumn{2}{c}{$M^\text{mag}_{\text{T}, n}$} & 
\multicolumn{2}{c}{ $M_{\text{LS},p}$ }
\\
& & ab initio & adopted
& ab initio & adopted
& ab initio & adopted\\
\colrule
\multirow{2}{*}{$^{10}$C $\rightarrow$ $^{10}$B} & \intem & $0.0024(13)$  & \multirow{2}{*}{$0.0019(20)$}  &  $0.0014(12)$ & \multirow{2}{*}{$0.0016(14)$} & $0.0082(34)$ & \multirow{2}{*}{$0.0086(39)$}\\
& \intsat & 
$0.0014(15)$ & & $0.0018(11)$ & & $0.0091(29)$ & 
\\

\multirow{2}{*}{$^{14}$O $\rightarrow$ $^{14}$N}  & & 
$0.0045(4)$  & \multirow{2}{*}{$0.0048(13)$} & $0.0131(44)$ & \multirow{2}{*}{$0.0130(53)$} & $-0.0143(15)$ & \multirow{2}{*}{$-0.0138(20)$} 
\\
& & $0.0050(11)$ & & $0.0130(52)$ & & $-0.0133(13)$ &
\\
 
\multirow{2}{*}{$^{26}$Si $\rightarrow$ $^{26}$Al}  & &  $0.0050(8)$  & \multirow{2}{*}{$0.0040(19)$}  & $0.0020(24)$ & \multirow{2}{*}{$0.0025(30)$} & $0.0107(37)$
& \multirow{2}{*}{$0.0100(45)$}
\\
&  & $0.0029(8)$ & & $0.0030(25)$ & & $0.0092(35)$ & 
\\
 
\multirow{2}{*}{$^{34}$Ar $\rightarrow$ $^{34}$Cl}  & &  $0.0016(13)$  & \multirow{2}{*}{$0.0017(15)$}  &$0.0098(6)$ & \multirow{2}{*}{$0.0097(17)$} & $-0.0099(20)$
& \multirow{2}{*}{$-0.0095(28)$}
\\
&  & $0.0018(14)$ & & $0.0097(17)$ & & $-0.0092(24)$ & 
\\

\multirow{2}{*}{$^{38}$Ca $\rightarrow$ $^{38}$K}  & & $0.0022(2)$ & \multirow{2}{*}{$0.0020(9)$} & $0.0103(1)$ &
\multirow{2}{*}{$0.0101(16)$}
& $-0.0192(21)$ & \multirow{2}{*}{$-0.0183(29)$}\\
& & 
$0.0018(7)$ & &  $0.0098(14)$ & & $-0.0175(16)$ &
\\
\colrule
\multirow{2}{*}{$^{26}$Al $\rightarrow$ $^{26}$Mg} & & $-0.0028(35)$ & \multirow{2}{*}{$-0.0035(42)$} & $-0.0037(5)$ & \multirow{2}{*}{$-0.0030(11)$} & $-0.0044(33)$ & \multirow{2}{*}{$-0.0042(37)$}\\
& &
$-0.0042(34)$ & & $-0.0024(3)$ & & $-0.0040(35)$ &\\
 
\multirow{2}{*}{$^{34}$Cl $\rightarrow$ $^{34}$S} & & 
$-0.0140(7)$ & \multirow{2}{*}{$-0.0139(24)$} &$-0.0009(7)$ 
& \multirow{2}{*}{$-0.0010(11)$}
& $-0.0000(42)$ & \multirow{2}{*}{$-0.0006(74)$}
\\ 
&  &
$-0.0139(24)$ & & $-0.0012(10)$ & & $-0.0012(68)$ &
\\
 
\multirow{2}{*}{$^{38}$K $\rightarrow$ $^{38}$Ar} & & $-0.0146(1)$ & \multirow{2}{*}{$-0.0143(22)$} & $-0.0015(1)$ & \multirow{2}{*}{$-0.0013(6)$} &  $0.0032(52)$ & \multirow{2}{*}{$0.0022(62)$}
\\
& & 
$-0.0139(18)$ & & $-0.0012(4)$ & & $0.0012(39)$ &
\\
 
\multirow{2}{*}{$^{42}$Sc $\rightarrow$ $^{42}$Ca} & & 
$-0.0047(5)$  & \multirow{2}{*}{$-0.0056(21)$} & $-0.0037(15)$ & \multirow{2}{*}{$-0.0026(26)$} & $-0.0209(31)$ & \multirow{2}{*}{$-0.0217(39)$}\\
& & 
$-0.0066(11)$ & & $-0.0014(7)$ & & $-0.0225(6)$ & 
\\

\multirow{2}{*}{$^{46}$V $\rightarrow$ $^{46}$Ti} & & 
$-0.0040(30)$ & \multirow{2}{*}{$-0.0045(35)$} & $-0.0009(4)$ & \multirow{2}{*}{$-0.0001(13)$} & $-0.0064(14)$
& \multirow{2}{*}{$-0.0081(58)$}
\\
&  &
$-0.0051(29)$ & &  $0.0006(6)$ & & $-0.0098(42)$ &
\\

\multirow{2}{*}{$^{50}$Mn $\rightarrow$ $^{50}$Cr} & & 
$-0.0020(25)$ & \multirow{2}{*}{$-0.0039(68)$}& $-0.0018(9)$ & \multirow{2}{*}{$-0.0008(22)$} & $-0.0074(12)$ & \multirow{2}{*}{$-0.0100(79)$} \\
&  &
$-0.0059(49)$ & & $0.0001(12)$ & & $-0.0125(53)$ & 
\\

\multirow{2}{*}{$^{54}$Co $\rightarrow$ $^{54}$Fe} & & 
$-0.0015(38)$ & \multirow{2}{*}{$-0.0029(63)$}& $-0.0039(18)$ & \multirow{2}{*}{$-0.0026(40)$}& $-0.0090(5)$ & \multirow{2}{*}{$-0.0113(55)$}
\\
& & $-0.0043(49)$ & & $-0.0012(27)$ & & $-0.0137(31)$ & 
\\
\botrule
  \end{tabular}
\caption{NMEs of the tensor and spin-orbit operators. For each type of matrix elements, the second column denotes the values adopted in the fits, with errors treated as statistical.
}\label{tab:LST}
  \end{table*}

The results for the matrix elements of the  GT magnetic and contact operators
for 12 superallowed transitions
are given in Table~\ref{tab:1}. For the first five transitions, the isospin of the initial state is  $T_z = -1$, where here we follow the traditional nomenclature of Ref.~\cite{Hardy:2020qwl}, which assigns $T_z = -1/2$ to the proton.\footnote{We only follow this convention for the labeling of the superallowed transitions to be consistent with the nuclear-physics literature. All calculations are performed in the standard particle-physics isospin conventions.}
The remaining seven transitions have $T_z = 0$. 
The GT magnetic and contact matrix elements are separately cutoff dependent, and thus we expect them to depend on the Hamiltonian. 
In order to assess the theoretical uncertainties arising from the truncations in the many-body method, in Table~\ref{tab:1} 
we show
three results for each matrix element.
The right and middle entries correspond to $M^{\text{CCSD}}_{N_{\text{max}}=8}$ and $M^{\text{CCSD}}_{N_{\text{max}}=10}$, which truncate the CC expansions at the singles and doubles level and differ in $N_{\text{max}}$, $N_{\text{max}}=8$ (N8) and $N_{\text{max}}=10$ (N10), respectively.
The left entry includes the effect of triples as described in Eq.~\eqref{eq:triples_correction}.
To propagate the theoretical errors to the extraction of $V_{ud}$, we will perform separate fits to the CCSDT-1, CCSD(N10), and CCSD(N8) matrix elements, as discussed in Sec.~\ref{sec:fits}.

Figure~\ref{fig:GTdensities} shows the two-body operator densities for the $M_{\text{GT}, N}^{\text{mag}}$ operators, with the density $\rho$ defined so that
\begin{equation}
    M^{\text{mag}}_{\text{GT}, N} = \int d r \rho(r)\,.
\end{equation}
We show the operator densities for three $T_z = -1$ and three $T_z = 0$ nuclei. Due to computational cost, the operator densities are shown for the CCSD(N8) results, and therefore are not fully converged.  The $T_z = 0$ transitions typically show smaller GT matrix elements, because of a larger cancellation between the components proportional to the neutron and proton magnetic moments~\cite{Hardy:2020qwl}. This cancellation appears to be mostly accidental. Rewriting the operators in isospin basis, and stripping the prefactors $(1+\kappa_p \pm \kappa_n)$ from the $T=1$ and $T=2$ components, one can see that for light nuclei the matrix elements of the $T=1$ and $T=2$ operators have similar size, leading to the $T=2$ component to dominate because of the large $1+\kappa_p - \kappa_n \simeq 4.7$ prefactor. 
For medium mass nuclei, the $T=2$ matrix elements become a factor of 5 to 10 smaller than the $T=1$ ones. This suppression is partially compensated by the ratio of the isovector and isoscalar nucleon magnetic moments. For medium mass nuclei, the $T=1$ and $T=2$ GT operators have therefore similar size. As they add and subtract for $T_z=-1$ and $T_z = 0$ nuclei, respectively, the latter nuclei end up with smaller $\bar\delta_{\text{NS}}$.  

\begin{figure*}[t]
    \centering
    \includegraphics[width=\linewidth]{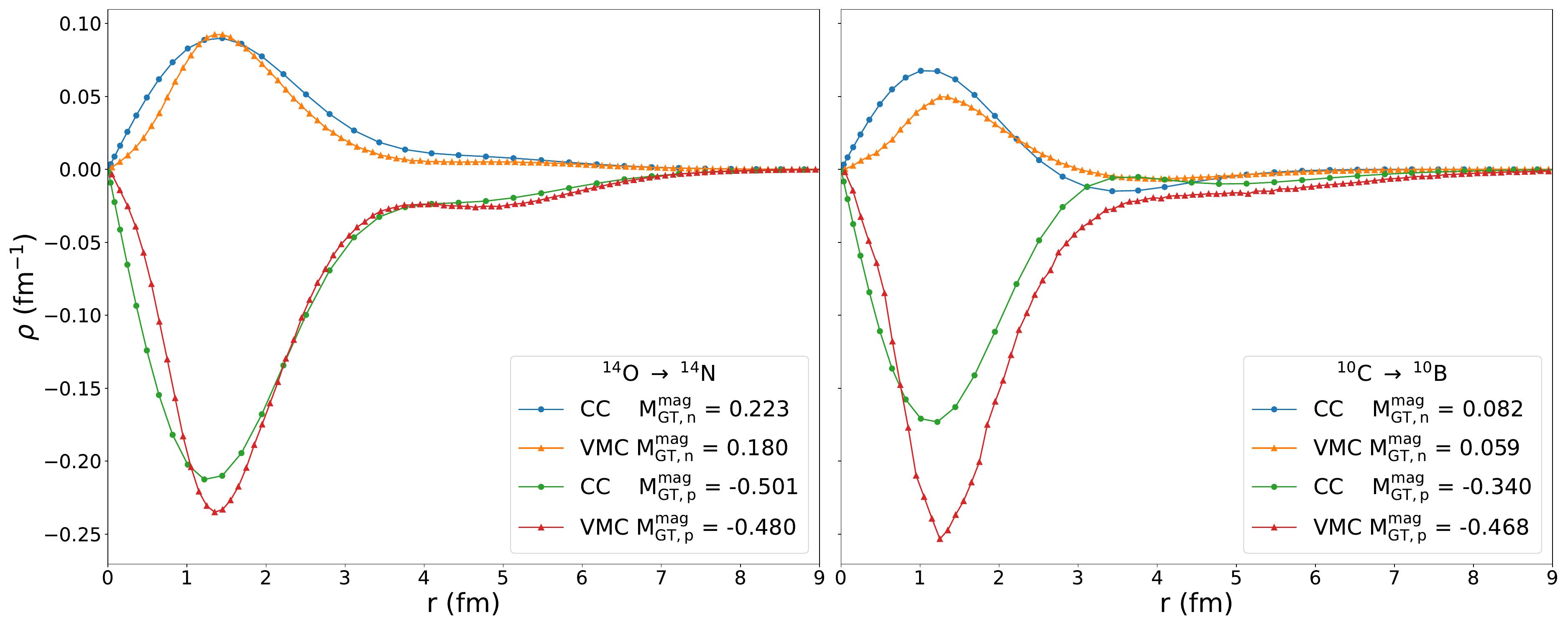}
    \caption{Comparison of the GT operator densities in the present CC calculation to the VMC calculations of Refs.~\cite{Cirigliano:2024msg,King:2025fph}.}
    \label{fig:vmc_vs_cc}
\end{figure*}

The remaining NMEs at $\mathcal O(\alpha \epsilon_\chi)$ are the tensor and spin-orbit matrix elements. These should not depend on the cutoff and on the Hamiltonian, at the claimed accuracy in the EFT expansion. In this case, 
we assign an error to \intem~and \intsat~matrix elements following the prescription described in Eq.~\eqref{eq:cc_uncertainty}. 
We then take the average of the \intem~and \intsat~results as central value, and define the $1\sigma$ error on the matrix element 
to cover the entire range spanned by the \intem~and \intsat~calculations including the CC uncertainties.

The resulting numerical values for the NMEs are given in Table~\ref{tab:LST}. 
We notice that the T and LS matrix elements are typically smaller than $M^{\text{mag}}_{\text{GT}, i}$. 
For the LS matrix elements, 
we observe a large cancellation between the ${\mathbf L}$ and ${\mathbf L}^{\text{CM}}$ components.
This cancellation was overlooked in Ref.~\cite{Cirigliano:2024msg}, which neglected the ${\mathbf L}^{\text{CM}}$ component, but 
has since been observed also in QMC calculations in light nuclei~\cite{King:2025fph}.
Notwithstanding their smallness,
because of the large cancellation between  
$M^{\text{mag}}_{\text{GT}, p}$
and $M^{\text{mag}}_{\text{GT}, n}$ that is observed in most of the $T_z = 0$ transitions, 
the T and LS matrix elements still prove relevant, in contrast to the pion-exchange matrix elements, whose contributions can be safely neglected at the current level of precision.

In Fig.~\ref{fig:vmc_vs_cc}
we compare the results of the present CC calculations with
Variational Monte Carlo (VMC) calculations in $^{14}$O~\cite{Cirigliano:2024msg}
and $^{10}$C~\cite{King:2025fph}. 
We focus the comparison on the GT matrix elements, which are the largest NMEs that enter for the superallowed decays.
Reference~\cite{Cirigliano:2024msg} used the NNLO local chiral interactions of Ref.~\cite{Gezerlis:2014zia}, while Refs.~\cite{King:2025fph}
used interactions in the  Norfolk two- and three-nucleon interaction family (NV2+3)~\cite{Piarulli:2014bda,Piarulli:2016vel,Piarulli:2017dwd,Baroni:2018fdn}.

The figure shows the results obtained with the NV2+3-Ia$^*$ interaction, with other interactions providing similar results \cite{King:2025fph}. We observe a very good agreement between the CC and VMC calculations in $^{14}$O. The proton magnetic matrix elements $M^{\text{mag}}_{\text{GT}, p}$ differ by about 5\%. The difference arises primarily  from the region $r \lesssim 1$ fm, which is expected due to the different regulators used in the CC and VMC calculations. The neutron matrix element is slightly more discrepant, but the discrepancy is reduced after the inclusion of triples correlations. We observe a more significant difference in $^{10}$C. In this case, even after including  triples, the CC $M^{\text{mag}}_{\text{GT}, p}$  calculated with the \intem\, interaction, is 30\% smaller than the QMC result, while, after Green's function MC (GFMC) evolution, the neutron matrix elements are in better agreement, see Eq.~\eqref{eq:QMC10}. Reaction experiments of $^{10}$C (and its analog $^{10}$Be) exhibit features consistent with the formation of clusters~\cite{Curtis:2008zz,Charity:2022kyp,Li:2023msp}. A possible origin of the differences in the matrix elements and operator densities is this clustering, an example of higher-order correlations that are difficult to capture in CC. $\alpha$-like clustering involves genuine four-particle--four-hole correlations, which enter only indirectly at the CCSD and CCSDT-1 levels. In open-shell nuclei, no single configuration dominates. The deformed reference absorbs much of this static correlation into the mean field, but at the cost of broken rotational symmetry, whose restoration requires correlations of all orders, recoverable only by projection and most important in light nuclei~\cite{Hagen:2022tqp}. Similar limitations may affect other open-shell transitions away from shell closures, such as those of $^{26}$Al, $^{46}$V, and $^{50}$Mn. Additionally, the CC and QMC calculations of $^{10}$C use different interactions. It was observed in GFMC predictions of the enhanced $B(E2;2_1^+\to0_1^+)$ in $^{10}$C---a signature of ground state deformation---that the results were sensitive to the underlying nuclear Hamiltonian~\cite{McCutchan:2012tw}. Thus, in the present analysis, we cannot conclusively determine what is the origin of this discrepancy. For future analyses, a more quantitative diagnosis of the discrepancy will certainly be necessary and will require calculations of $\bar{\delta}_{\rm NS}$, as well as other structural quantities, using the same underlying nuclear Hamiltonians.  As we will discuss in Sec.~\ref{sec:fits}, the relatively small 
value (in absolute size) of the CC $M_{\text{GT},p}^{\text{mag}}$ matrix element has an impact on the quality of the fits for the extraction of  $V_{ud}$, and a dedicated benchmark should be pursued in the future.

\begin{table*}[t]
\centering
    \renewcommand{\arraystretch}{1.3}
\begin{tabular}{cc rrr rrr r r r}
\toprule
Transition  & Interaction   &  \multicolumn{3}{c}{ $M^{E}_{\text{F}, p}$} &  \multicolumn{3}{c}{ $M^{+}_{\text{F}, p}$} 
 & $\langle r_{jk}\rangle$ [fm] & $e^{\langle \log r_{jk}\rangle} $ [fm] & $R$ [fm] \\
 & & ab initio & density & adopted & ab initio & density & adopted  & & &\\
\colrule
\multirow{2}{*}{$^{10}$C $\rightarrow$ $^{10}$B} &  \intem   & $-4.84(4)$ & $-4.76$ &  \multirow{3}{*}{ $-4.9(1)$} & $-1.63(4)$  & $-1.41$ &  \multirow{3}{*}{$-1.8(2)$} &  $3.52(3)$ & $3.26(2)$ & \multirow{3}{*}{$3.27(5)$} \\ 
& \intsat  & $-5.05(5)$ & $-5.09$ &  & $-1.92(6)$
& $-1.84$ & & $3.69(4)$ & $3.39(3)$ &\\
& NV2+3-Ia$^*$ &  $-5.24(1)$ & $-4.87$ & & $-2.07(1)$ & $-1.56$ & & $3.83(1)$ & $3.46(1)$ & 
\\

\multirow{2}{*}{$^{14}$O $\rightarrow$ $^{14}$N}  & \intem 
& $-6.24(3)$  & $-6.03$ &  \multirow{3}{*}{$-6.4(3)$} & $-1.59(12)$  & $-1.03$  & \multirow{3}{*}{$-1.9(5)$} & $3.64(2)$ & $3.40(4)$ & \multirow{3}{*}{$3.39(1)$}
\\
& \intsat & $-6.62(8)$ & $-6.52$ & & $-2.20(16)$ & $-1.77$ & & $3.87(5)$ & $3.61(6)$ &  \\
& NNLO & $-6.22(2)$ &  $-5.93$ & &  $-1.52(1)$  & $-0.89$ & & $3.63(1)$ & $3.37(1)$&   \\
 
\multirow{2}{*}{$^{26}$Si $\rightarrow$ $^{26}$Al} & 
& $-10.6(3)$ & $-10.7$ & \multirow{2}{*}{$-10.8(8)$} & $-1.52(8)$  & $-1.04$ & \multirow{2}{*}{$-2.1(8)$} & $4.09(11)$ & $3.86(2)$ & \multirow{2}{*}{$3.99(4)$}\\
& & $-11.1(5)$ & $-11.6$ & & $-2.68(17)$ & $-2.51$ & & $4.30(18)$ & $4.11(4)$ & \\
 
\multirow{2}{*}{$^{34}$Ar $\rightarrow$ $^{34}$Cl} & 
& $-13.7(2)$ & $-13.6$ & \multirow{2}{*}{$-14.1(9)$} & $-1.56(9)$  & $-0.93$ & \multirow{2}{*}{$-2.4(1.0)$} & $4.42(5)$ & $4.15(2)$ & \multirow{2}{*}{$4.29(2)$}\\
& & $-14.4(5)$ & $-14.9$ & & $-3.28(18)$ & $-3.04$ & & $4.67(17)$ & $4.45(3)$ & \\

\multirow{2}{*}{$^{38}$Ca $\rightarrow$ $^{38}$K} & & $-15.1(2)$ & $-15.1$ & \multirow{2}{*}{$-15.6(9)$}&
  $-1.55(12)$  & $-0.76$ & \multirow{2}{*}{$-2.6(1.2)$} & $4.54(6)$ & $4.28(2)$ & \multirow{2}{*}{$4.44(1)$}\\
 & & $-16.2(4)$ & $-16.5$ & & $-3.57(15)$ & $-3.16$ & & $4.86(12)$ & $4.61(3)$ & \\
\colrule
\multirow{2}{*}{$^{26}$Al $\rightarrow$ $^{26}$Mg} & 
& $-9.75(20)$ & $-9.8$  & \multirow{2}{*}{$-10.0(6)$} & $-1.39(11)$ & $-0.85$ & \multirow{2}{*}{$-1.9(7)$} & $4.09(8)$ & $3.86(3)$ & \multirow{2}{*}{$3.91$}  \\
& & $-10.3(3)$ & $-10.5$ & & $-2.47(17)$ & $-1.93$ & & $4.33(13)$ & $4.11(4)$ & \\
 
\multirow{2}{*}{$^{34}$Cl $\rightarrow$ $^{34}$S} & &
$-12.8(1)$  & $-12.8$ & \multirow{2}{*}{$-13.2(8)$} & $-1.43(11)$ & $-0.80$ & \multirow{2}{*}{$-2.2(1.0)$} & $4.40(4)$ & $4.14(2)$ & \multirow{2}{*}{$4.24(1)$} \\ 
& & $-13.6(4)$ & $-13.8$ & & $-3.05(16)$ & $-2.45$ &  & $4.66(14)$ & $4.45(3)$ & \\
 
\multirow{2}{*}{$^{38}$K $\rightarrow$ $^{38}$Ar} & & 
$-14.3(2)$ & $-14.2$ & \multirow{2}{*}{$-14.8(8)$} & $-1.45(15)$ & $-0.67$ & \multirow{2}{*}{$-2.4(1.1)$}& $4.53(5)$ & $4.27(2)$ & \multirow{2}{*}{$4.39(1)$} 
\\
& & $-15.3(3)$  & $-15.3$ & &$-3.34(17)$ & $-2.55$ & & $4.84(10)$ & $4.60(3)$ & \\
 
\multirow{2}{*}{$^{42}$Sc $\rightarrow$ $^{42}$Ca} & &
$-15.8(2)$ & $-15.9$ & \multirow{2}{*}{$-16.5(9)$} & $-1.46(16)$ &  $-0.86$ & \multirow{2}{*}{$-2.6(1.3)$} & $4.67(6)$ & $4.39(3)$ & \multirow{2}{*}{$4.53$} 
\\
& & $-17.2(2)$ & $-17.2$ & & $-3.74(14)$ & $-3.08$ & &$5.08(4)$ & $4.76(2)$ &
 \\

\multirow{2}{*}{$^{46}$V $\rightarrow$ $^{46}$Ti}
 & & $-17.3(2)$ & $-17.4$ & \multirow{2}{*}{$-18.1(9)$} &$-1.14(19)$ & $-0.71$ & \multirow{2}{*}{$-2.4(1.5)$} & $4.79(4)$ & $4.46(3)$ & \multirow{2}{*}{$4.66$} \\
& &  $-18.8(2)$ & $-18.9$ & &$-3.73(11)$ & $-3.30$ & &$5.20(4)$ & $4.85(2)$ & \\
 
 \multirow{2}{*}{$^{50}$Mn $\rightarrow$ $^{50}$Cr} 
& & $-18.5(1)$  & $-18.5$ & \multirow{2}{*}{$-19.3(1.0)$} &$-0.56(18)$ & $0.05$ & \multirow{2}{*}{$-2.0(1.6)$} & $4.82(3)$ & $4.49(2)$ & \multirow{2}{*}{$4.73$} 
 \\
& &  $-20.1(2)$ & $-20.1$ & &$-3.48(11)$ & $-2.85$ & & $5.23(5)$ & $4.90(2)$  &
 \\

 \multirow{2}{*}{$^{54}$Co $\rightarrow$ $^{54}$Fe} 
& & $-19.5(1)$  & $-19.4$ & \multirow{2}{*}{$-20.3(1.2)$} & $0.28(17)$ & $1.15$ & \multirow{2}{*}{$-1.3(1.8)$}  &$4.80(3)$ & $4.50(1)$ & \multirow{2}{*}{$4.76(1)$} \\
& & $-21.0(4)$  & $-21.2$ & & $-2.97(12)$ & $-2.06$ & & $5.19(9)$ & $4.92(2)$ & \\
\botrule
  \end{tabular}
  \caption{Fermi matrix elements from ab initio calculations and density estimates. The adopted values are determined as in Table~\ref{tab:LST}. The last three columns show the conversion to $\langle r_{jk}\rangle$ and $\langle \log r_{jk}\rangle$ according to Eq.~\eqref{MF_constant}, in comparison to $R=\sqrt{5/3}R_\text{ch}$ with the experimental charge radius of the final-state nucleus. A more detailed interpretation is given in the text.}\label{tab:Fermi}
\end{table*}

\subsection{\texorpdfstring{$\boldsymbol{\delta^E_{\text{NS}}}$}{} and \texorpdfstring{$\boldsymbol{\mathcal O(\alpha^2)}$}{} corrections to \texorpdfstring{$\boldsymbol{\delta^{(0)}_{\text{NS}}}$}{} }\label{sec:fermi} 

We next consider the corrections from the Fermi matrix elements $M_{\text{F}, p}^E$,
$M_{\text{F}, p}^+$, and $M_{\text{F}, p}^{+\,\text{3b}}$. 
The first matrix element contributes to $\delta_{\text{NS}}^E$ at $\mathcal O(\alpha E_e R)$, and its size grows with the nuclear charge $Z$. 
$M^E_{\text{F}, p}$ corresponds, in the standard formalism, to corrections to the phase-space factor that arise from the nucleus finite size~\cite{Cirigliano:2024msg}.
$M^+_{\text{F},p}$ has the same spin and isospin dependence as 
$M^E_{\text{F},p}$, with a different radial function, $r \rightarrow \log r$. Its role is to provide an operator definition of the scale that absorbs the renormalization scale dependence of the $\mathcal O(\alpha^2 Z)$ corrections to $\delta_R^\prime$. 
Similarly, the calculation of $M_{\text{F}, p}^{+\,\text{3b}}$ would allow one to replace terms in the traditional Fermi function
that scale as $\alpha^2 Z^2 \log R$ and depend on the nuclear radius $R$. The matrix element $M_{\text{F}, p}^{+\,\text{3b}}$ would provide a determination of the actual scale to be used in the Fermi function. 

As three-body matrix elements are difficult to compute in CC,
we will present here results for $M^E_{\text{F}, p}$
and $M^{+}_{\text{F}, p}$. 
We will then compare the results of the ab initio calculation to the approximation discussed in Sec.~\ref{sec:density}, which will give us confidence in adopting these approximations also for $M^{+\, \text{3b}}_{\text{F},p}$. We will further validate the three-body matrix elements via explicit QMC calculations.

The two-body Fermi matrix elements $M^E_{\text{F}, p}$ and $M^+_{\text{F}, p}$ calculated with the \intem~and \intsat~interactions are shown in Table~\ref{tab:Fermi},
for the choice $\Lambda = 1/R_A$, and $R_A = 1.2 A^{1/3}$ fm. This choice is arbitrary, and absorbed by the dependence in $C_{\delta}$ and $C_{\delta}^{\text{3b}}$.
We also show QMC results 
 in $^{10}$C~\cite{King:2025fph} and $^{14}$O~\cite{Cirigliano:2024msg}. The former were obtained using two interactions models  NV2+3Ia and NV2+3Ia$^{\star}$ in the Norfolk family~\cite{Piarulli:2014bda,Piarulli:2016vel,Piarulli:2017dwd,Baroni:2018fdn}.
These models differ in how the three-body force was determined, with Ia
using as input the trinucleon binding energies and the neutron--deuteron doublet scattering length~\cite{Piarulli:2017dwd}, while Ia$^{\star}$ the trinucleon binding energies and the tritium GT $\beta$ decay matrix element~\cite{Baroni:2018fdn}.
The calculation of $\bar\delta_\text{NS}$ in $^{14}$O used 
the NNLO local chiral potential of Ref.~\cite{Gezerlis:2014zia}, and value of the cutoff $R_0 =1\fm$.

\begin{table*}[t]
\centering
    \renewcommand{\arraystretch}{1.3}
\begin{tabular}{cc rrrrr rrrrr rr}
\toprule
& & \multicolumn{5}{c}{ \intem} & \multicolumn{5}{c}{\intsat}    & \multicolumn{2}{c}{$R_\text{ch}^\text{exp}$} \\
Nucleus & $Z$ & $R_p$ & $R_\text{ch}$ & $R_W^{(0)}$ & $R_W$ & $R_W^\text{anch}$  & $R_p$ & $R_{\text{ch}}$ &  $R_W^{(0)}$ & $R_W$ & $R_W^\text{anch}$& Ref.~\cite{Ohayon:2024dwt} & Ref.~\cite{Angeli:2013epw}  \\\colrule
$^{10}$B & 5 & $2.454 $ & $2.502 $ & \multirow{2}{*}{2.871} & \multirow{2}{*}{2.951} & \multirow{2}{*}{2.988} & $2.752$ & $2.772$ & \multirow{2}{*}{2.978} & \multirow{2}{*}{3.034} & \multirow{2}{*}{2.909} &   $2.531(38)$ & $2.428(50)$ \\
$^{10}$C & 6 & $2.527 $ & $2.581 $ & & & & $2.652$ & $2.679$ &  & & &  $2.638(36)$ & --\\
\colrule 
$^{14}$N & 7 & $2.409 $ & $2.479 $ & \multirow{2}{*}{2.927} & \multirow{2}{*}{3.024} & \multirow{2}{*}{3.138} & $2.682$ & $2.729$ & \multirow{2}{*}{3.115} & \multirow{2}{*}{3.192} & \multirow{2}{*}{3.141} & $2.623(9)$ & $2.558(7)$\\
$^{14}$O & 8 & $2.504 $ & $2.578 $ & & & & $2.667$ & $2.719$ &  & & &  $2.705(10)$ & --\\
\colrule
$^{26}$Mg & 12 & $2.850 $ & $2.929 $ & \multirow{2}{*}{3.379} & \multirow{2}{*}{3.481} & \multirow{2}{*}{3.582} & $3.072$ & $3.138$ & \multirow{2}{*}{3.590} & \multirow{2}{*}{3.681} & \multirow{2}{*}{3.564} &  $3.030(3)$ & $3.0337(18)$\\
\multirow{2}{*}{$^{26}$Al} & \multirow{2}{*}{13} & $2.870 $ & $2.952 $ & & & & $3.184$  & $3.250$ & & & & \multirow{2}{*}{$3.088(3)$} & \multirow{2}{*}{--}\\
&& $2.889 $ & $2.970 $ & \multirow{2}{*}{3.388} & \multirow{2}{*}{3.491} & \multirow{2}{*}{3.586} & $3.235$  & $3.330$ & \multirow{2}{*}{3.604} & \multirow{2}{*}{3.694} & \multirow{2}{*}{3.552} &  & \\
$^{26}$Si & 14 & $2.952$ & $3.034$ & & & & $3.177$  & $3.245$ & & & & $3.136(4)$ & --\\
\colrule
$^{34}$S & 16 & $3.031 $ & $3.112 $ & \multirow{2}{*}{3.601} & \multirow{2}{*}{3.702} & \multirow{2}{*}{3.862}& $3.299$ & $3.368$ & \multirow{2}{*}{3.863} & \multirow{2}{*}{3.953} & \multirow{2}{*}{3.850} & $3.284(4)$ & $3.2847(21)$\\
\multirow{2}{*}{$^{34}$Cl} & \multirow{2}{*}{17} & $3.042$ & $3.124$ & & & & $3.413$ &$3.481$ & & &  &\multirow{2}{*}{$3.327(4)$} & \multirow{2}{*}{--}\\
& & $3.061 $ & $3.143  $ & \multirow{2}{*}{3.603} & \multirow{2}{*}{3.704} & \multirow{2}{*}{3.854}  & $3.458$ & $3.525$ & \multirow{2}{*}{3.867} & \multirow{2}{*}{3.956} & \multirow{2}{*}{3.828}  &  &\\
$^{34}$Ar & 18 & $3.118 $ & $3.200 $ & & & & $3.388$ & $3.459$ &  & & & $3.365(11)$ & $3.3654(40)$\\
\colrule
$^{38}$Ar & 18 & $3.144 $ & $3.224 $ & \multirow{2}{*}{3.677} & \multirow{2}{*}{3.778} & \multirow{2}{*}{3.947}& $3.414$ & $3.483$ & \multirow{2}{*}{3.955} & \multirow{2}{*}{4.045} & \multirow{2}{*}{3.944}&  $3.402(6)$ & $3.4028(19)$\\
\multirow{2}{*}{$^{38}$K} & \multirow{2}{*}{19} & $3.142 $ & $3.224 $ & & & & $3.522$ & $3.591$ & & & & \multirow{2}{*}{$3.440(6)$} & \multirow{2}{*}{$3.4264(51)$}\\
& & $3.160 $ & $3.242 $ & \multirow{2}{*}{3.680} & \multirow{2}{*}{3.781} & \multirow{2}{*}{3.941}   & $3.568$ & $3.636$ & \multirow{2}{*}{3.964} & \multirow{2}{*}{4.053} & \multirow{2}{*}{3.924}  & & \\
$^{38}$Ca & 20 & $3.218 $ & $3.300 $ & & & & $3.494$ & $3.565$ & & & & $3.469(3)$ & --\\
\colrule 
$^{42}$Ca & 20 & $3.238 $ & $3.318 $ & \multirow{2}{*}{3.827} & \multirow{2}{*}{3.926}& \multirow{2}{*}{4.104}& $3.515$ & $3.584$ & \multirow{2}{*}{4.144} & \multirow{2}{*}{4.232} & \multirow{2}{*}{4.136} & $3.510(3)$ & $3.5081(21)$\\
$^{42}$Sc & 21 & $3.237 $ & $3.318 $ & & & & $3.626$ & $3.695$ & & & & $3.545(3)$ & $3.570(24)$\\
\colrule 
$^{46}$Ti & 22 & $3.349 $ & $3.428 $ & \multirow{2}{*}{3.891} & \multirow{2}{*}{3.988}& \multirow{2}{*}{4.162}& $3.647$ &$3.716$ & \multirow{2}{*}{4.232} & \multirow{2}{*}{4.318}& \multirow{2}{*}{4.192} & $3.610(3)$ & $3.6070(22)$\\
$^{46}$V & 23 & $3.344 $ & $3.424 $ & & & &  $3.759$ & $3.827$ & & & & $3.642(5)$ & --\\
\colrule
$^{50}$Cr & 24 & $3.402 $ & $3.480 $ & \multirow{2}{*}{3.867} & \multirow{2}{*}{3.966} & \multirow{2}{*}{4.145}& $3.716$ & $3.784$  & \multirow{2}{*}{4.213} & \multirow{2}{*}{4.301}& \multirow{2}{*}{4.161} & $3.664(5)$ & $3.6588(65)$\\
$^{50}$Mn & 25 & $3.391 $ & $3.471 $ & & & & $3.820$ & $3.887$  & & & & $3.692(4)$ & $3.712(20)$\\
\colrule
$^{54}$Fe & 26 & $3.408 $ & $3.487 $ & \multirow{2}{*}{3.831} & \multirow{2}{*}{3.932} & \multirow{2}{*}{4.129}& $3.731$ & $3.800$  &\multirow{2}{*}{4.181} & \multirow{2}{*}{4.271}& \multirow{2}{*}{4.143} & $3.688(4)$ & $3.6933(19)$\\
$^{54}$Co & 27 & $3.392 $ & $3.473 $ & & & & $3.820$ & $3.888$  & & & & $3.714(4)$ &--\\
\botrule
\end{tabular}
\caption{Point-proton, charge, and weak radii from CC calculations, in comparison to experiment. For $^{26}$Al, $^{34}$Cl, and $^{38}$K we show the results of two different implementations, based on the two reference states in the same triplet. $R_p$ and $R_W^{(0)}$ are evaluated directly from the sum rules over $\rho_p(r)$ and $\rho_W(r)$, respectively, while $R_\text{ch}$ and $R_W$ include finite-size, relativistic, and center-of-mass (CM) corrections as detailed in App.~\ref{app:CCdensities} (using $r_p^2=0.7071\,\text{fm}^2$~\cite{Antognini:2013txn}, $r_n^2=-0.1155\,\text{fm}^2$~\cite{ParticleDataGroup:2026mpi,Koester:1995nx,Kopecky:1997rw}), see also Fig.~\ref{fig:charge_radii}. 
The final result for $R_W$ is anchored to the experimental value of the charge radius from Ref.~\cite{Ohayon:2024dwt}, $\big(R_W^\text{anch}\big)^2 = \big(R_\text{ch}^\text{exp}\big)^2 + \big[R_W^2-R_\text{ch}^2\big]_\text{CC}$, using the average of initial- and final-state squared charge radii as calibration. 
All numbers in fm.
}\label{tab:charge_radii}
\end{table*}

\begin{table}[t]
\centering
    \renewcommand{\arraystretch}{1.3}
\begin{tabular}{cc rrrrr}
\toprule
Nucleus & $Z$ & $R_p$ & $R_\text{ch}$ & $R_W^{(0)}$ & $R_W$ & $R_W^\text{anch}$\\
\colrule
& &  \multicolumn{5}{c}{ NV2+3-Ia} \\
$^{10}$B & 5 &       $2.32$ & $2.46$ & \multirow{2}{*}{2.78} & \multirow{2}{*}{2.93} & \multirow{2}{*}{2.99}  \\ 
$^{10}$C & 6 &    $2.43$ & $2.57$  & & \\ 
\colrule
& &  \multicolumn{5}{c}{NV2+3-Ia$^*$}  \\
$^{10}$B & 5 & $2.48$ & $2.61$ & \multirow{2}{*}{2.94} & \multirow{2}{*}{3.08} & \multirow{2}{*}{3.03}    \\ 
$^{10}$C & 6 &    $2.56$ & $2.69$ & & &    \\ 
\colrule
& &   \multicolumn{5}{c}{NNLO} \\
$^{14}$N & 7      &  $2.531$ & $2.651$ & \multirow{2}{*}{2.730} & \multirow{2}{*}{2.882} & \multirow{2}{*}{2.881}  \\ 
$^{14}$O & 8 & $2.556$ & $2.681$ &  & \\
\botrule
\end{tabular}
\caption{Charge and weak transition radii from QMC. All numbers in fm.  
}\label{tab:charge_radii_qmc}
\end{table}

\begin{figure*}[t]
 \centering
\includegraphics[width=0.9\linewidth]{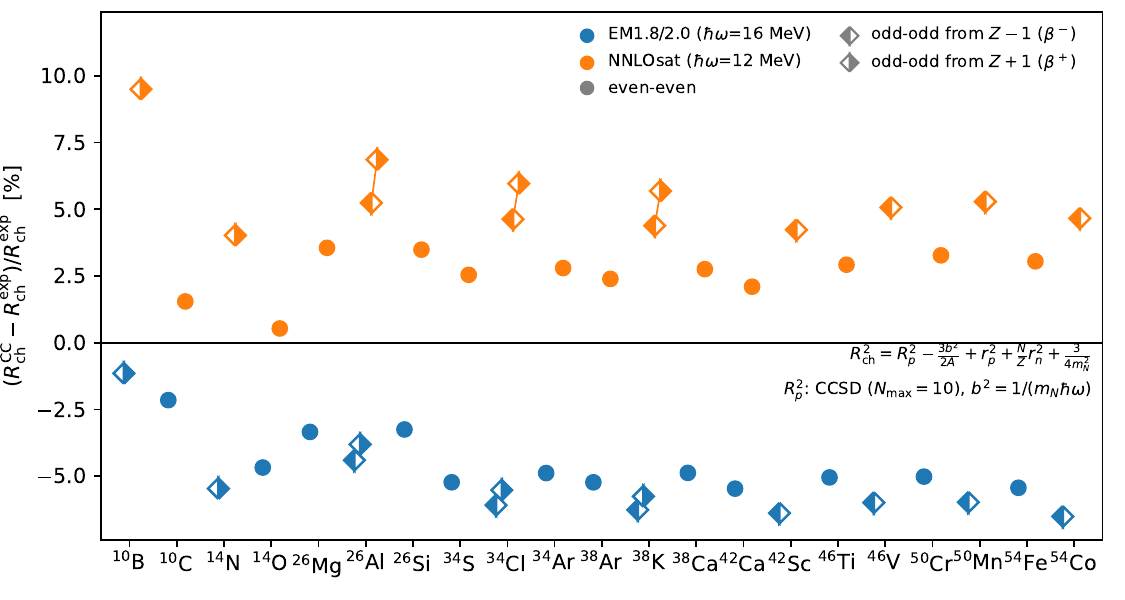}
\caption{Charge radii from CC calculations in comparison to experiment, see Table~\ref{tab:charge_radii}.
}\label{fig:charge_radii}
\end{figure*}

The same table shows an estimate of the distance between the decaying nucleon and a proton in the nucleus.
These estimates are obtained by equating the matrix elements
$M_{\text{F}, p}^E$ and $M_{\text{F}, p}^+$
 to the values one would obtain if $r_{jk}$ or $\log r_{jk}$ in Eqs.~\eqref{eq:hE} and \eqref{eq:hplus} were replaced by a constant, i.e., 
\begin{align}
    M_{\text{F}, p}^E &\equiv -  Z M_\text{F}^{(0)} \frac{\langle r_{jk} \rangle}{2 R_A}\,, \notag\\
    M_{\text{F}, p}^+ & \equiv - Z  M_\text{F}^{(0)} \left( \langle \log r_{jk} \rangle - \log R_A   \right)\,.
    \label{MF_constant}
\end{align}
These moments are the EFT counterparts of the nuclear-size parameters that enter the traditional treatment of finite-size effects in the phase space, which we discuss in Sec.~\ref{sec:phase_space} and App.~\ref{app:subtractions}. In that treatment, both the nuclear charge and the decaying nucleons are assumed to be distributed uniformly within a sphere. For two points distributed independently and uniformly within a sphere of a generic radius $R_s$, we have
\begin{equation}
\langle r_{12}\rangle = \frac{36}{35}R_s\,,\quad e^{\langle\log r_{12}\rangle} = 2e^{-3/4}R_s\simeq 0.945\,R_s\,.
\label{eq:uniformmoments}
\end{equation}
These expressions can be used in two different ways. First, if the decaying nucleon and the protons are distributed within the same sphere that is used in the traditional finite-size and shape corrections, that is setting $R_s= R \equiv \sqrt{5/3}\,R_\text{ch}$, with $R_\text{ch}$ the experimental charge radius of the final-state nucleus from Table~\ref{tab:charge_radii}, then $\langle r_{jk}\rangle = \frac{36}{35}R$ in $M^E_{\text{F},p}$ reproduces the $\mathcal O(\alpha Z R)$ terms of these corrections, see App.~\ref{app:subtractions}. Second, the expressions help to interpret the ab initio matrix elements. Since these are computed with point nucleons, we fill in the point radii of Table~\ref{tab:charge_radii}. Taking the same sphere for the decaying nucleon and the protons, $R_s=\sqrt{5/3}\,R_p$, Eq.~\eqref{eq:uniformmoments} underestimates the ab initio $\langle r_{jk}\rangle$ and $e^{\langle\log r_{jk}\rangle}$ by (3--13)\% for all transitions and both interactions. If instead the decaying nucleon is assigned its own sphere, with radius $\sqrt{5/3}\,R_W^{(0)}$ ($R_W^{(0)}$ being the weak radius for point nucleons), while the protons remain in the sphere with radius $\sqrt{5/3}\,R_p$, both moments are reproduced to within about 3\%, see Fig.~\ref{fig:twosphere}. The ab initio matrix elements are thus largely determined by the sizes of the weak transition and proton densities, and reflect the larger extent of the former, which is the effect described by the isovector correction in the traditional approach.

Table~\ref{tab:Fermi}
 shows that $M^E_{\text{F},p}$
differ by about 5\% to 10\% between different interactions. For \intem, the difference between CCSDT-1, CCSD(N10), and CCSD(N8) is small. \intsat~shows a larger variation, similar to the Hamiltonian variation. The logarithmic matrix elements $M_{\text{F},p}^+$ show a much larger variation. 
This indicates cancellations that are sensitive to the proton distribution in nuclei. 
The extracted average distances, however, are equally stable for the linear and logarithmic matrix elements. Their ratio, $e^{\langle\log r_{jk}\rangle}/\langle r_{jk}\rangle\simeq 0.93$--$0.95$, is close to the value $2e^{-3/4}\times 35/36\simeq 0.92$ for a uniform sphere, Eq.~\eqref{eq:uniformmoments}, showing that the difference between $e^{\langle\log r_{jk}\rangle}$ and $\langle r_{jk}\rangle$ is essentially geometric.

Table \ref{tab:Fermi} also shows the results for the two-body matrix elements in the representation in terms of EM and EW transition densities,  Eq.~\eqref{eq:densities2b}, 
truncated to the first intermediate state. 
We see that for $M_{\text{F},p}^E$ 
this approximation works extremely well. 
The density approximation does not agree as precisely with the full calculation for the logarithmic matrix element $M_{\text{F},p}^+$. This seems to be due to a partial cancellation between the long- and short-distance parts of the potential, which flips sign due to the logarithm, and at the same time avoids a coherent scaling with $Z$, as would be expected absent such cancellations. 

In practice, the density estimates are obtained equating the EM densities with the point-proton ones, given that the derivation presented in detail in App.~\ref{app:density} relies on the same assumption. Finite-size corrections could in principle be included along similar lines, but such corrections are beyond the accuracy that can be expected from the density approximation. As a validation of the CC EM and EW densities, see App.~\ref{app:CCdensities}, in Tables~\ref{tab:charge_radii} and~\ref{tab:charge_radii_qmc} we show the extracted charge and weak transition radii, compared to the available experimental information. The latter comparison is made more explicit in Fig.~\ref{fig:charge_radii}, showing that while charge radii deviate from their experimental values by typically $5\%$, the empirical values are at least covered by the span of the two interactions, indicating that the comparison of \intem~and \intsat~results should provide a realistic estimate of the Hamiltonian uncertainties in our calculation. For the evaluation of the phase-space factor, see Table~\ref{tab:phase_space}, we also provide versions of the weak radius anchored by the experimental value of the charge radius, which reduces the variation between \intem~and \intsat~significantly. This behavior is expected given that bulk properties of nuclei are strongly correlated~\cite{Hagen:2015yea,Payne:2019wvy,Heinz:2024cwg,Noel:2026byc}, which also points a path towards future precision calculations of weak and charge densities as input for superallowed NMEs, calibrated against empirical values for charge radii.

\begin{table*}[t]
\centering
    \renewcommand{\arraystretch}{1.3}
\begin{tabular}{ccrrrr}
\toprule 
Transition  & Interaction   &  \multicolumn{3}{c}{ $M^{+\, \text{3b}}_{\text{F}, p}$} & \multirow{2}{*}{$\frac{2}{3}\langle  r_{ijk}\rangle $ (fm) }\\
 & & density v1 & density v2 & adopted & \\
\colrule
\multirow{2}{*}{$^{10}$C $\rightarrow$ $^{10}$B} &  \intem   & $-12.1$ & $-10.4$ & \multirow{3}{*}{$-11.0(6)$} 
& \multirow{3}{*}{$3.74(15)$}
\\ 
& \intsat & $-12.9$ & $-11.5$ &  &\\
& NV2+3-Ia$^*$ & $-12.4$ & $-10.7$ & & 
\\
\multirow{2}{*}{$^{14}$O $\rightarrow$ $^{14}$N}  & \intem & $-20.3$ & $-17.7$   & \multirow{3}{*}{$-19.0(1.3)$}
& \multirow{3}{*}{$3.65(15)$}
\\
& \intsat & $-22.3$ & $-20.2$ & &\\
& NNLO & $-20.4$ & $-18.1$  & & \\
\multirow{2}{*}{$^{26}$Si $\rightarrow$ $^{26}$Al}  & & $-62.5$ & $-57.7$  & \multirow{2}{*}{$-62.7(5.0)$} & \multirow{2}{*}{$4.18(19)$}\\
& & $-71.1$ & $-67.7$  & &\\
\multirow{2}{*}{$^{34}$Ar $\rightarrow$ $^{34}$Cl}  & & $-101.0$ & $-94.6$ & \multirow{2}{*}{$-104(10)$} & \multirow{2}{*}{$4.45(22)$}\\
& & $-117.7$ & $-113.6$ & &\\
\multirow{2}{*}{$^{38}$Ca $\rightarrow$ $^{38}$K} & & $-123.6$ & $-116.4$ & \multirow{2}{*}{$-128(12)$} & \multirow{2}{*}{$4.57(23)$}\\
 & &  $-144.9$ & $-140.5$ & &\\
\colrule
\multirow{2}{*}{$^{26}$Al $\rightarrow$ $^{26}$Mg}  &  & $-52.3$ & $-48.4$ & \multirow{2}{*}{$-51.6(3.2)$} & \multirow{2}{*}{$4.12(14)$}\\
&  & $-59.6$ & $-54.7$ & & \\
\multirow{2}{*}{$^{34}$Cl $\rightarrow$ $^{34}$S} & & $-87.8$ & $-82.8$ & \multirow{2}{*}{$-89(6)$} & \multirow{2}{*}{$4.39(17)$} \\ 
& & $-102.5$ & $-95.8$ & &\\
\multirow{2}{*}{$^{38}$K $\rightarrow$ $^{38}$Ar} &  & $-109.1$ & $-103.7$  & \multirow{2}{*}{$-112(8)$} & \multirow{2}{*}{$4.51(17)$}
\\
& & $-127.7$ & $-120.1$ & & \\
\multirow{2}{*}{$^{42}$Sc $\rightarrow$ $^{42}$Ca} & & $-134.7$ & $-128.7$ & \multirow{2}{*}{$-139(11)$} & \multirow{2}{*}{$4.67(19)$} 
\\
& &  $-158.7$ & $-150.0$ & &
 \\
\multirow{2}{*}{$^{46}$V $\rightarrow$ $^{46}$Ti}
 &  & $-161.0$ & $-154.5$ & \multirow{2}{*}{$-168(14)$} & \multirow{2}{*}{$4.79(20)$}\\
&  & $-191.5$ & $-181.6$ & &\\
 \multirow{2}{*}{$^{50}$Mn $\rightarrow$ $^{50}$Cr} 
& & $-183.3$ & $-176.3$ & \multirow{2}{*}{$-193(17)$} & \multirow{2}{*}{$4.83(21)$} \\
& & $-220.6$ & $-209.6$ & &
 \\
 \multirow{2}{*}{$^{54}$Co $\rightarrow$ $^{54}$Fe} 
&  & $-202.0$ & $-194.7$ & \multirow{2}{*}{$-215(20)$} & \multirow{2}{*}{$4.82(21)$} \\
&  & $-246.5$ & $-234.8$ & & \\
\botrule
  \end{tabular}
  \caption{Three-body Fermi matrix elements in the density approximation. The two versions represent the two variants in Eq.~\eqref{eq:densities3b}. In cases in which we have two different proton distributions, based on two possible reference states, the one corresponding to the transition in question is adopted. The last column refers to $\langle r_{ijk}\rangle\simeq \frac{3}{2}R_s$ as defined in Eq.~\eqref{rijk}.}
  \label{tab:Fermi_3b}
\end{table*}

\begin{figure}[t]
    \centering
    \includegraphics[width=\linewidth]{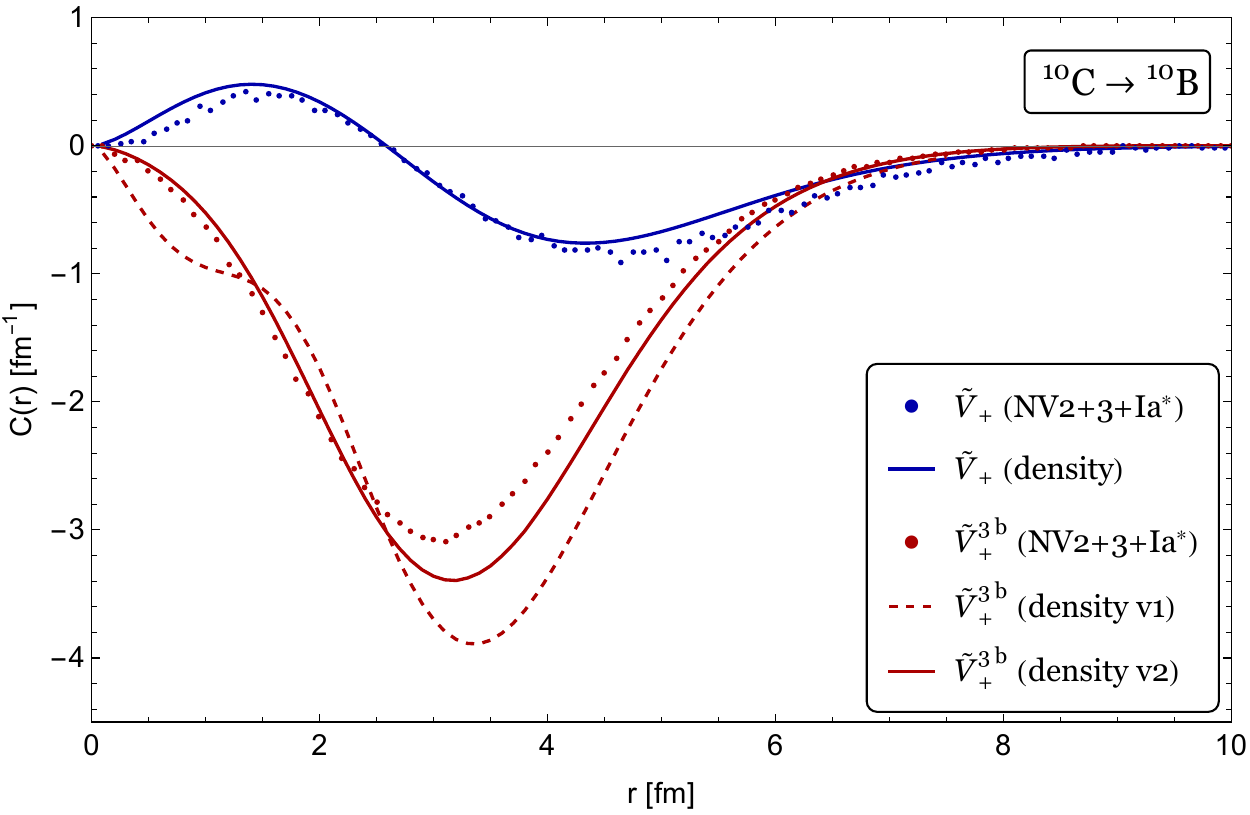}
    \caption{ The blue dots (line) shows the density for $M_{{\rm F}, p}^{+}$ as a function of the distance between the two nucleons, from an explicit QMC calculation (density estimate). The red dots (lines) show similar densities, from an explicit QMC calculation (density estimates),    
    for $M_{{\rm F}, p}^{+\, {\rm 3b}}$ as a function of the distance of the decaying nucleon to the center of mass of the other two.
    }
    \label{fig:alpha2densities}
\end{figure}

Having validated the density approximation for the two-nucleon matrix element, we can proceed to the three-nucleon sector. 
The values in the two versions of the density approximation are given in Table~\ref{tab:Fermi_3b}.

The explicit ab initio calculations yield, for $^{10}$C~\cite{Graham3N}
\begin{align}\label{eq:QMC3b}
&\text{NV2+3-Ia}: & \quad  M_{\text{F},p}^{+\, \text{3b}} &= -9.10(1)\,, \notag\\
& \text{NV2+3-Ia}^*: & \quad  M_{\text{F},p}^{+\, \text{3b}} &= -9.87(2)\,,
\end{align}
while for $^{14}$O
\begin{equation}
\label{eq:QMC3b14O}
    M_{\text{F},p}^{+\, \text{3b}} = -18.6\,,
\end{equation}
in good agreement with the density estimates obtained in Table~\ref{tab:Fermi_3b}.\footnote{We note that the results in Eq.~\eqref{eq:QMC3b} include the nonlogarithmic terms in ${\cal V}_+^{\rm 3b}$ in Eq.~\eqref{eq:V3plusMain}, while the density estimates in Table \ref{tab:Fermi_3b} do not. However, the difference between the two is small, for example, the result for the NV2+3-Ia* interaction would change $-9.87\to -9.98$.} In particular, we observe that the second variant (v2) leads to slightly better agreement with the explicit QMC calculations, and we therefore adopt this estimate, with an uncertainty derived from the comparison of the \intem~and \intsat~interactions. 

The density results can be expressed as
\begin{equation}
\label{rijk}
    M_{\text{F},p}^{+\, \text{3b}} \equiv - \frac{Z (Z-1)}{2} M_\text{F}^{(0)} \log \frac{\langle r_{ijk}\rangle}{R_A}\,,
\end{equation}
which defines an effective three-body distance $\langle r_{ijk}\rangle$. For a uniform sphere, $\langle r_{ijk}\rangle\simeq 1.50\,R_s$, and thus $\langle r_{ijk}\rangle \simeq 1.58\, e^{\langle\log r_{12}\rangle}$. The values in Tables~\ref{tab:Fermi} and~\ref{tab:Fermi_3b} give ratios between $1.52$ and $1.58$, consistent with this geometric picture.

The three-body matrix element plays a special role: it determines the dependence of the rate on the nuclear size at $\mathcal O(\alpha^2Z^2)$. In the traditional approach this dependence is carried by the factor $(2|{\mathbf p}_e|R)^{2(\eta-1)}$ of the Fermi function, see Eq.~\eqref{eq:FF}, which depends on the assumed nuclear radius $R$. In the EFT, the corresponding logarithm of the nuclear size enters through $\log\langle r_{ijk}\rangle$ in Eq.~\eqref{rijk}, while the dependence on the matching scale cancels against the running of $C^{(g_V)}_\text{eff}$ and of the Fermi function $\bar F$. Hence, $M_{\text{F},p}^{+\, \text{3b}}$ determines the actual scale that replaces $R$ in the traditional Fermi function.

In summary, the density approximation reproduces the ab initio values of $M^E_{\text F,p}$ very well, and $M^+_{\text F,p}$ at a level consistent with its larger intrinsic uncertainty, while for $M_{\text{F},p}^{+\, \text{3b}}$ it is validated by the QMC calculations for $^{10}$C and $^{14}$O. 
The qualitative agreement between ab initio calculations and the density approximation can also be seen from Fig.\ \ref{fig:alpha2densities}, which shows the densities of the two- and three-body matrix elements as a function of $r$.
For the heavier emitters, the uncertainties of $M^+_{\text F,p}$ and especially of $M_{\text{F},p}^{+\, \text{3b}}$ are among the largest in $\bar\delta^{(0)}_\text{NS}$. Dedicated ab initio calculations of the three-body NMEs, beyond the density approximation, are therefore an important target for future work.

\begin{table}[t]
    \centering
    \renewcommand{\arraystretch}{1.3}
    \begin{tabular}{crrr}
\toprule 
       Transition  & $\mathcal Q_\text{EC}$ [keV] & $t$ [ms]     &   $\bar{\delta}_C$ [\%] \cite{Hardy:2020qwl}\\
       \colrule
$^{10}$C $\rightarrow$ $^{10}$B &  $1907.994(67)$ & $1321800(1800)$ 
  &  0.175(18) \\
$^{14}$O $\rightarrow$ $^{14}$N  & 
  $2831.543(76)$ & $71075(15)$ & 0.330(25) \\
$^{26}$Si $\rightarrow$ $^{26}$Al & 4840.85(10) 
& 2969.0(5.4) & 0.435(27)\\
$^{34}$Ar $\rightarrow$ $^{34}$Cl  & $6061.83 (8)$
  & $896.55(0.81) $  & 0.659(40)   \\
$^{38}$Ca $\rightarrow$ $^{38}$K  & 
  $6612.12(7) $ & $574.8(1.1)$  & 0.745(47) \\
\hline 
$^{26}$Al $\rightarrow$ $^{26}$Mg  & 
 $4232.72(15)$ & $6351.24(55)$   & 0.310(18) \\
$^{34}$Cl $\rightarrow$ $^{34}$S  & 
 $5491.662(46)$ & $1527.77^{+0.47}_{
-0.44}$  &  0.613(49) \\
$^{38}$K $\rightarrow$ $^{38}$Ar  & 
  $6044.240(48)$ & $925.42(28)$  &  0.628(54)  \\  
$^{42}$Sc $\rightarrow$ $^{42}$Ca  &
 $6426.34(12)$ & $681.44(26)$  & 0.690(46)\\
$^{46}$V $\rightarrow$ $^{46}$Ti  & $7052.45(10)$ & $423.114^{+0.068}_{-0.053 }$  &  0.620(63)	\\
$^{50}$Mn $ \rightarrow$ $^{50}$Cr &
$7634.453(66)$ & $283.68(11)$   & 0.660(32) \\
$^{54}$Co $ \rightarrow$ $^{54}$Fe &
$8244.38(26)$ & $193.495^{+0.086}_{-0.063}$   & 0.770(67) \\
\botrule
           \end{tabular}
    \caption{Experimental input and $\bar\delta_C$.}
    \label{tab:exp}
\end{table}

\subsection{\texorpdfstring{$\boldsymbol{\bar\delta_C}$}{}}

Potential and hard photon modes also contribute to the isospin-breaking correction $\bar\delta_C$, by inducing isospin-breaking $N\!N$ (and multi-nucleon) potentials. These  
induce differences in the initial- and final-state wave functions and cause the Fermi matrix element to deviate from $\sqrt{2}$. In principle, $\bar\delta_C$
needs to be computed with the same ab initio Hamiltonian and many-body method used in the calculation of $\bar\delta_{\text{NS}}$.
Preliminary studies for the decays $^{10}$C and $^{14}$O have been reported in 
Refs.~\cite{Farren:2024spl,Piarulli:2026abq}. A direct calculation of $\bar\delta_C$ with the CC method would require one to control the Fermi matrix element at the $(0.1\text{--}1)\%$ level, which is currently not feasible, so that a dedicated analysis to determine fully consistent $\bar\delta_{\text{NS}}$ and $\bar\delta_{C}$ corrections  is required. 
For this reason, we use at the moment the shell-model values given in Ref.~\cite{Hardy:2020qwl}
and reported in the last column in Table~\ref{tab:exp}, but stress that this is a shortcoming that needs to be addressed in future work. 

\begin{table*}[t]
 \centering
    \renewcommand{\arraystretch}{1.3}
    \begin{tabular}{crrrrr}
    \toprule
       Transition  &  \multicolumn{4}{c}{ $\left[C_\text{eff}^{(g_V)}\right]^2\bar f (1 + \bar\delta_R^\prime)$ } & $\tilde f_E$ \\
    & \multicolumn{1}{c}{Ref.~\cite{Ohayon:2024dwt}}  & \multicolumn{1}{c}{$R_W$ \intem} & \multicolumn{1}{c}{$R_W$ \intsat}   & \multicolumn{1}{c}{$R_W^\text{anch}$} & \\
       \colrule
$^{10}$C $\rightarrow$ $^{10}$B   & $2.39666(57)_{g_V}(0)_{\mu}$ & $2.39668(57)_{g_V}(0)_{\mu}$ & $2.39667(57)_{g_V}(0)_{\mu}$ & $2.39668(57)_{g_V}(0)_{\mu}$ & 1.10\\
$^{14}$O $\rightarrow$ $^{14}$N   & $44.484(11)_{g_V}(0)_{\mu}$ & $44.484(11)_{g_V}(0)_{\mu}$ & $44.482(11)_{g_V}(0)_{\mu}$ & $44.483(11)_{g_V}(0)_{\mu}$ & 0.95\\
$^{26}$Si $\rightarrow$ $^{26}$Al & $1064.4(3)_{g_V}(0)_\mu$ & $1064.3(3)_{g_V}(0)_\mu$ & $1064.1(3)_{g_V}(0)_\mu$ & $1064.2(3)_{g_V}(0)_\mu$ & 0.87\\
$^{34}$Ar $\rightarrow$ $^{34}$Cl & $3520.4(8)_{g_V}(2)_{\mu}$ & $3520.6(8)_{g_V}(2)_{\mu}$ & $3519.3(8)_{g_V}(2)_{\mu}$ & $3519.9(8)_{g_V}(2)_{\mu}$ & 0.86\\
$^{38}$Ca $\rightarrow$ $^{38}$K  & $5489.6(1.3)_{g_V}(0.5)_{\mu}$ & $5489.8(1.3)_{g_V}(0.5)_{\mu}$ & $5487.2(1.3)_{g_V}(0.5)_{\mu}$ & $5488.4(1.3)_{g_V}(0.5)_{\mu}$ & 0.85\\
\colrule
$^{26}$Al $\rightarrow$ $^{26}$Mg & $495.85(12)_{g_V}(2)_{\mu}$ & $495.82(12)_{g_V}(2)_{\mu}$ & $495.76(12)_{g_V}(2)_{\mu}$ & $495.79(12)_{g_V}(2)_{\mu}$ & 0.89\\
$^{34}$Cl $\rightarrow$ $^{34}$S  & $2062.7(5)_{g_V}(1)_{\mu}$ & $2062.8(5)_{g_V}(1)_{\mu}$ & $2062.2(5)_{g_V}(1)_{\mu}$ & $2062.5(5)_{g_V}(1)_{\mu}$ & 0.86\\
$^{38}$K $\rightarrow$ $^{38}$Ar  & $3402.0(8)_{g_V}(3)_{\mu}$ & $3402.2(8)_{g_V}(3)_{\mu}$ & $3400.9(8)_{g_V}(3)_{\mu}$ & $3401.4(8)_{g_V}(3)_{\mu}$ & 0.86\\
$^{42}$Sc $\rightarrow$ $^{42}$Ca & $4610.9(1.1)_{g_V}(0.4)_{\mu}$ & $4611.7(1.1)_{g_V}(0.4)_{\mu}$ & $4609.2(1.1)_{g_V}(0.4)_{\mu}$ & $4610.1(1.1)_{g_V}(0.4)_{\mu}$ & 0.86\\
$^{46}$V $\rightarrow$ $^{46}$Ti  & $7415.2(1.8)_{g_V}(0.9)_{\mu}$ & $7416.4(1.8)_{g_V}(0.9)_{\mu}$ & $7411.1(1.8)_{g_V}(0.9)_{\mu}$ & $7413.4(1.8)_{g_V}(0.9)_{\mu}$ & 0.85\\
$^{50}$Mn $\rightarrow$ $^{50}$Cr & $11025.5(2.6)_{g_V}(1.5)_{\mu}$ & $11028.4(2.6)_{g_V}(1.5)_{\mu}$ & $11019.0(2.6)_{g_V}(1.5)_{\mu}$ & $11023.3(2.6)_{g_V}(1.5)_{\mu}$ & 0.85\\
$^{54}$Co $\rightarrow$ $^{54}$Fe & $16132.9(3.8)_{g_V}(2.6)_{\mu}$ & $16139.1(3.8)_{g_V}(2.6)_{\mu}$ & $16122.8(3.8)_{g_V}(2.6)_{\mu}$ & $16129.5(3.8)_{g_V}(2.6)_{\mu}$ & 0.84\\
\botrule
\end{tabular}

    \caption{Phase-space factors for all isotopes of interest: 
    The second column evaluates the form factor correction at $R_{\text{ch}}$ and includes the isospin-breaking corrections of Ref.~\cite{Hayen:2017pwg}. The third and fourth columns evaluate the form factor at the ab initio weak radius $R_W$ of \intem~and \intsat, respectively (including the nucleon radii, see Table~\ref{tab:charge_radii}), 
    removing the isospin-breaking corrections. The fifth column gives our preferred evaluation in which $R_W$ is anchored to the experimental charge radius via $\big(R_W^\text{anch}\big)^2 = \big(R_\text{ch}^\text{exp}\big)^2 + \big[R_W^2-R_\text{ch}^2\big]_\text{CC}$, with the CC numbers taken as the mean of \intem~and \intsat.  $\tilde f_E$ is defined in Eq.~\eqref{tildefE}. All evaluations use $L^\text{sub}_0$, see App.~\ref{app:subtractions}.}
    \label{tab:phase_space}
\end{table*}

\section{Phase space}
\label{sec:phase_space}

In Table~\ref{tab:phase_space} we provide results for the combination
\begin{equation}\label{combscale}
\overline{P} =   \left[C_\text{eff}^{(g_V)}(\mu_{\text{ext}})\right]^2\bar f(\mu_{\text{ext}}) \Big[1 + \bar\delta_R^\prime(\mu_\text{ext})\Big]\,.
\end{equation}
While the individual ingredients in Eq.~\eqref{combscale} are separately dependent on the renormalization scale $\mu_{\text{ext}}$, $\overline{P}$ is $\mu$ independent up to higher-order corrections, so that its sensitivity to the renormalization scale can be used to assess the importance of missing orders.
The matching coefficients $C_\text{eff}^{(g_V)}$ are evaluated at the matching scale
\begin{equation}
\label{mupi}
    \mu_\pi = R^{-1} \exp\left(\frac{1}{2}-\gamma_E\right)\,, 
\end{equation}
with $R = \sqrt{5/3} R_{\text{ch}}$, and they are evolved down to the scale $\mu_{\text{ext}}$ using the anomalous dimension in Eq.~\eqref{eq:CeffRG}.
The phase-space factor $\bar f$ and the outer corrections $\bar\delta_R^\prime$ are evaluated at the scale $\mu_{\text{ext}}$.
We choose as central value $\mu_{\text{ext}} = 2 E_0$, and vary it between $E_0$ and $4 E_0$ to assess the impact of missing orders in the $\alpha$ expansion. 
Our evaluation uses an updated value of the single-nucleon vector coupling $g_V$, which includes $\mathcal O(\alpha \alpha_s)$ corrections~\cite{Moretti:2025ghw,Cirigliano:2026ios}, 
the evaluation of $\mathcal O(\alpha^2 Z)$ corrections derived in Ref.~\cite{Crosas:2025xyv}, and the corrected $C_\delta$ and $C^{\text{3b}}_\delta$ given in Eqs.~\eqref{eq:Cdelta} and~\eqref{eq:Cdelta3b}.
We see that for most isotopes the error on $\overline{P}$ is dominated by the error on $g_V$, which is in turn determined by the nonperturbative input to the $\gamma W$ box. The scale variation should capture $\mathcal O(\alpha^3 Z^2)$ terms, which start to become relevant for $^{50}$Mn and $^{54}$Co.

In addition to the Fermi function, the phase-space factor $\bar f$ contains the factor $\tilde C(E_e)$, Eq.~\eqref{eq:Ctildedecomp}, which collects corrections taken over from the traditional superallowed literature~\cite{Hayen:2017pwg,Wilkinson:1993he,Hardy:2004id,Hardy:2020qwl}. They consist of the finite-size correction to the Fermi function, the correction for the shape of the nuclear charge distribution, the shape factor and its isovector correction, atomic screening and overlap, and nuclear recoil. Several of these encode physics that our EFT approach computes explicitly through the matrix elements of $\bar\delta_\text{NS}$, in particular through $\delta^E_\text{NS}$ and the $\mathcal O(\alpha^2)$ matrix elements $M^+_{\text{F},p}$ and $M^{+\,\text{3b}}_{\text{F},p}$. In App.~\ref{app:subtractions} we identify the overlapping terms and define the subtractions needed to avoid double counting. 

With respect to App.~G of Ref.~\cite{Cirigliano:2024msg}, we make three changes. First, in the finite-size correction $L_0^\text{sub}$ we subtract only the genuine $\mathcal O(\alpha ZR)$ and $\mathcal O(\alpha^2Z^2)$ terms, Eq.~\eqref{eq:L0subpure}, and not the coefficients of the fit to the remainder. The additional subtraction did not impact transitions of light nuclei, the focus of Ref.~\cite{Cirigliano:2024msg}, but leads to 
a $Z$-dependent trend in the values of $V_{ud}$ extracted from the individual transitions, see Fig.~\ref{fig:vud_subtractions}. The modified subtraction scheme avoids this.
Second, the shape factor is evaluated with the weak radius $R_W$ of the CC calculation, including the nucleon radii, Eq.~\eqref{eq:Rw}, and the isovector correction $C_I$ is not applied. Third, we include the recoil correction. The individual contributions of all factors are listed in Table~\ref{tab:phase3}.

The columns of Table~\ref{tab:phase_space} illustrate the treatment of the weak radius. The second column uses the traditional evaluation of the shape factor, with the charge radius $R_{\text{ch}}$ of Ref.~\cite{Ohayon:2024dwt} and the isovector correction of Ref.~\cite{Wilkinson:1993he} (see also Ref.~\cite{Hayen:2017pwg}). The third and fourth columns use the weak radius calculated with the \intem\ and \intsat\ interactions, respectively. The difference between them, up to $1\times 10^{-3}$ in $\overline P$ for the heaviest emitters, mainly reflects the different nuclear radii predicted by the two Hamiltonians, see Table~\ref{tab:charge_radii}, rather than the difference between the weak and charge distributions, which is much less sensitive to the interaction. In our preferred evaluation, in the fifth column, we therefore anchor the computed weak radius to the measured charge radii, Eq.~\eqref{eq:RWanchored}.  

Table~\ref{tab:phase3} in App.~\ref{app:subtractions}
shows that, after removing  the terms in the phase-space factor that are replaced by the ab initio calculation of the two- and three-body Fermi matrix elements, of the weak and charge densities, and of the weak radii, the residual corrections, not fully controlled  in the EFT formalism, are  small. The largest remaining corrections are the  screening correction $S$, an atomic effect,  and the correction $U$
that, in the standard formalism, parameterizes deviations of the nuclear charge distribution from a uniform sphere. While $S$ has no EFT counterpart, parts of $U$
should already be captured in the EFT formalism. As $U$
is only provided as a fit to numerical solutions of the Dirac equation \cite{Wilkinson:1993he}, it is not clear how to define a proper subtraction scheme. We therefore keep $U$ in its traditional form, and we will assess the impact on $V_{ud}$ by performing fits with and without this correction.
Progress in ab initio calculations of nuclear charge and weak transition densities, e.g., along the lines of Refs.~\cite{Hagen:2015yea,Payne:2019wvy,Heinz:2024cwg,Noel:2026byc,He:2026zie}, as well as further data input~\cite{Seng:2022inj,Gorchtein:2025wli}, would allow one to remove the remaining model dependence in the phase-space factor.

\begin{table*}[t]
\centering
    \renewcommand{\arraystretch}{1.3}
\begin{tabular}{cc rr}
\toprule
Fit & Interaction & \multicolumn{1}{c}{$V_{ud}$} & $\chi^2/\text{dof}$ \\\colrule
 \multirow{2}{*}{EFT $(0\text{A})$}
 &   \intem &  $0.97409(12)_{\text{stat}}
(12)_{g_V}(1)_{\mu} (3)_{\text{CC}}$   & 0.90\\
 & \intsat & $0.97400(12)_{\text{stat}}
(12)_{g_V}(1)_{\mu} (4)_{\text{CC}}$  & 0.63
\\

\multirow{2}{*}{EFT $(0\text{B})$} & \intem &
$0.97411(12)_{\text{stat}}
(12)_{g_V}(1)_{\mu} (2)_{\text{CC}}$   & 0.62 \\
 & \intsat & $0.97403(12)_{\text{stat}}
(12)_{g_V}(1)_{\mu} (3)_{\text{CC}}$  & 0.25
 \\ 
\multirow{2}{*}{EFT $(0\text{C})$} & \intem &
$0.97413(13)_{\text{stat}}
(12)_{g_V}(1)_{\mu} (2)_{\text{CC}}$   & 0.60 \\
 & \intsat & $0.97405(13)_{\text{stat}}
(12)_{g_V}(1)_{\mu} (4)_{\text{CC}}$  & 0.22
\\ 
\colrule
\multirow{2}{*}{HT $(0\text{A})$} & \intem &  $0.97367(9)_{\text{stat}}(12)_{g_V} (4)_{\text{CC}}$  & 1.50 \\
     & \intsat & $0.97366(9)_{\text{stat}}(12)_{g_V} (2)_{\text{CC}}$ &  1.20 \\ 
   \multirow{2}{*}{HT $(0\text{B})$} & \intem &  $0.97369(9)_{\text{stat}}(12)_{g_V} (5)_{\text{CC}}$  & 1.29 \\
     & \intsat & $0.97369(9)_{\text{stat}}(12)_{g_V} (2)_{\text{CC}}$ &  0.88 \\
   \multirow{2}{*}{HT $(0\text{C})$} & \intem &  $0.97369(10)_{\text{stat}}(12)_{g_V} (5)_{\text{CC}}$  & 1.40 \\
     & \intsat & $0.97369(10)_{\text{stat}}(12)_{g_V} (2)_{\text{CC}}$ &  0.95 \\
\botrule
\end{tabular}
\caption{Fits to $V_{ud}$ with the LECs $\tilde{g}^{{N\!N}}_{V p}$ and $\tilde{g}^{{N\!N}}_{V n}$ set to zero. 
}\label{tab:fitVudLEC0}
\end{table*}

\begin{table*}[t]
\centering
    \renewcommand{\arraystretch}{1.3}
\begin{tabular}{cc rrrr}
\toprule
Fit & Interaction & \multicolumn{1}{c}{$V_{ud}$} & \multicolumn{1}{c}{$\tilde g_{Vp}^{{N\!N}}$} & \multicolumn{1}{c}{$\tilde g_{Vn}^{{N\!N}}$} & $\chi^2/\text{dof}$ \\\colrule
\multirow{2}{*}{EFT $(2\text{A})$} &   \intem &  $0.97340(127)_{\text{stat}}
(12)_{g_V}(10)_{\mu} (74)_{\text{CC}}$ & $2.5(2.8)_{\text{stat}}(0.2)_\mu (1.3)_{\text{CC}} $ & 
$-0.8(8)_{\text{stat}}(1)_{\mu}(3)_{\text{CC}}$
& 0.77\\
& \intsat & $0.97356(171)_{\text{stat}}
(12)_{g_V}(14)_{\mu} (96)_{\text{CC}}$ & $1.5(4.8)_{\text{stat}}(0.3)_{\mu}(1.9)_{\text{CC}}$ & $-0.1(1.3)_{\text{stat}}(0.2)_{\mu}(0.8)_{\text{CC}}$ & 0.75
\\
\multirow{2}{*}{EFT $(2\text{C})$} & \intem &
$0.97361(127)_{\text{stat}}
(12)_{g_V}(11)_{\mu} (24)_{\text{CC}}$ & $2.4(2.8)_{\text{stat}}
(0.2)_{\mu} (0.9)_{\text{CC}}$ & $-1.1(9)_{\text{stat}}
(1)_{\mu} (2)_{\text{CC}}$  & 0.24 \\
& \intsat & $0.97357(171)_{\text{stat}}
(12)_{g_V}(14)_{\mu} (105)_{\text{CC}}$ & $2.0(4.8)_{\text{stat}}
(0.3)_{\mu} (3.8)_{\text{CC}}$ & $-0.5(1.3)_{\text{stat}}
(0.2)_{\mu} (0.5)_{\text{CC}}$ & 0.21  \\
\multirow{2}{*}{EFT $(2\text{D})$} & \intem & $0.97430(47)_{\text{stat}}
(12)_{g_V}(1)_{\mu} (2)_{\text{CC}}$ & $0.4(9)_{\text{stat}}(1)_{\mu} (3)_{\text{CC}}$ & $-0.9(6)_{\text{stat}}(1)_{\mu} (3)_{\text{CC}}$  & $0.47$ \\
& \intsat & $0.97409(41)_{\text{stat}}
(12)_{g_V}(0)_{\mu} (7)_{\text{CC}}$ & $0.1(1.0)_{\text{stat}}(0)_{\mu} (0.1)_{\text{CC}}$ & $-0.3(8)_{\text{stat}}(1)_{\mu} (3)_{\text{CC}}$  & $0.25$ 
\\ \colrule
\multirow{2}{*}{EFT $(1\text{A})$} &   \intem &  $0.97386(18)_{\text{stat}}
(12)_{g_V}(3)_{\mu} (2)_{\text{CC}}$ & -- & 
$-1.0(6)_{\text{stat}}(1)_{\mu}(3)_{\text{CC}}$
& 0.70\\
& \intsat & $0.97394(21)_{\text{stat}}
(12)_{g_V}(3)_{\mu} (11)_{\text{CC}}$ & -- & $-0.3(1.0)_{\text{stat}}(0.1)_{\mu}(0.8)_{\text{CC}}$ & 0.68
 \\
\multirow{2}{*}{EFT $(1\text{C})$} & \intem &
$0.97387(18)_{\text{stat}}
(12)_{g_V}(3)_{\mu} (1)_{\text{CC}}$ & -- & $-1.2(6)_{\text{stat}}
(1)_{\mu} (3)_{\text{CC}}$  & 0.22 \\
& \intsat & $0.97393(21)_{\text{stat}}
(12)_{g_V}(3)_{\mu} (10)_{\text{CC}}$ & -- & $-0.7(1.0)_{\text{stat}}
(0.1)_{\mu} (0.8)_{\text{CC}}$ & 0.19
\\ 
\colrule
\multirow{2}{*}{HT $(2\text{A})$} & \intem &  $0.97265(65)_{\text{stat}}(12)_{g_V} (24)_{\text{CC}}$ &
$3.3(1.5)_{\text{stat}} (0.3)_{\text{CC}}$
& $-0.7(6)_{\text{stat}} (1)_{\text{CC}}$ & 1.16 \\
     & \intsat & $0.97241(83)_{\text{stat}}(12)_{g_V} (7)_{\text{CC}}$ &
   $3.8(2.4)_{\text{stat}} (0.6)_{\text{CC}}$& 
   $0.1(1.0)_{\text{stat}} (0.5)_{\text{CC}}$& 1.18 \\

\multirow{2}{*}{HT $(2\text{C})$}  & \intem &  $0.97280(65)_{\text{stat}}(12)_{g_V} (5)_{\text{CC}}$ &
$3.4(1.5)_{\text{stat}} (0.1)_{\text{CC}}$
& $-1.0(7)_{\text{stat}} (1)_{\text{CC}}$ & 0.85 \\
    & \intsat & $0.97255(83)_{\text{stat}}(12)_{g_V} (40)_{\text{CC}}$ &
   $4.1(2.4)_{\text{stat}} (1.7)_{\text{CC}}$& 
   $-0.4(1.1)_{\text{stat}} (0.3)_{\text{CC}}$& 0.82 
   \\
   \multirow{2}{*}{HT $(2\text{D})$}  & \intem &  $0.97356(43)_{\text{stat}}(12)_{g_V} (12)_{\text{CC}}$ &
$1.1(8)_{\text{stat}} (5)_{\text{CC}}$
& $-0.6(6)_{\text{stat}} (2)_{\text{CC}}$ & 1.39 \\
    & \intsat & $0.97355(40)_{\text{stat}}(12)_{g_V} (11)_{\text{CC}}$ &
   $0.6(9)_{\text{stat}} (4)_{\text{CC}}$& 
   $-0.1(7)_{\text{stat}} (2)_{\text{CC}}$& 1.13 
   \\
   \colrule
\multirow{2}{*}{HT $(1\text{A})$}  & \intem &  $0.97353(12)_{\text{stat}}(12)_{g_V} (4)_{\text{CC}}$ &
--
& $-1.2(6)_{\text{stat}} (3)_{\text{CC}}$ & 1.24 \\
     & \intsat & $0.97359(14)_{\text{stat}}(12)_{g_V} (13)_{\text{CC}}$ &
   --& 
   $-0.6(9)_{\text{stat}} (9)_{\text{CC}}$& 1.27 \\
\multirow{2}{*}{HT $(1\text{C})$} & \intem &  $0.97353(12)_{\text{stat}}(12)_{g_V} (1)_{\text{CC}}$ &
--
& $-1.4(6)_{\text{stat}} (2)_{\text{CC}}$ & 0.90 \\
     & \intsat & $0.97357(14)_{\text{stat}}(12)_{g_V} (10)_{\text{CC}}$ &
   --& 
   $-1.1(9)_{\text{stat}} (8)_{\text{CC}}$& 0.91 \\
\botrule
\end{tabular}
\caption{Fits to $V_{ud}$ and LECs. The correlation matrices are documented in Table~\ref{tab:fitVudcorr}.
}\label{tab:fitVudLEC2}
\end{table*}

\begin{table*}[t]
\centering
    \renewcommand{\arraystretch}{1.3}
\begin{tabular}{c ccccc}
\toprule
& $(2\text{A})$ & $(2\text{C})$ & $(2\text{D})$ & $(1\text{A})$ & $(1\text{C})$\\\colrule
EFT & 
$\begin{pmatrix}
1 & -0.971 & -0.607\\
-0.977 & 1 & 0.422\\
-0.598 & 0.424 & 1\\
\end{pmatrix}
$
&
$\begin{pmatrix}
1 & -0.970 & -0.610\\
-0.976 & 1 & 0.420\\
-0.594 & 0.415 & 1\\
\end{pmatrix}
$
&
$\begin{pmatrix}
1 & -0.842 & -0.573\\
-0.780 & 1 & 0.122\\
-0.594 & 0.058 & 1\\
\end{pmatrix}
$
&
$\begin{pmatrix}
   1 & 0.742\\
   0.813 & 1
\end{pmatrix}
$
&
$\begin{pmatrix}
   1 & 0.732\\
   0.806 & 1
\end{pmatrix}
$
\\
HT & 
$\begin{pmatrix}
1 & -0.870 & -0.380\\
-0.888 & 1 & -0.103\\
-0.357 & -0.097 & 1\\
\end{pmatrix}
$
&
$\begin{pmatrix}
1 & -0.860 & -0.394\\
-0.880 & 1 & -0.106\\
-0.364 & -0.107 & 1\\
\end{pmatrix}
$
&
$\begin{pmatrix}
1 & -0.754 & -0.580\\
-0.719 & 1 & -0.051\\
-0.634 & -0.029 & 1\\
\end{pmatrix}
$
& 
$\begin{pmatrix}
   1 & 0.602\\
   0.739 & 1
\end{pmatrix}
$
&
$\begin{pmatrix}
   1 & 0.591\\
   0.734 & 1
\end{pmatrix}
$
\\
\botrule
\end{tabular}
\caption{Correlation matrices corresponding to the fits in Table~\ref{tab:fitVudLEC2}. In each case, the upper triangle refers to \intem, the lower triangle to \intsat.
}\label{tab:fitVudcorr}
\end{table*}

\section{Fits to \texorpdfstring{$\boldsymbol{V_{ud}}$}{}}\label{sec:fits}

We extract $V_{ud}$ by fitting the experimental half-lives, tabulated in Table~\ref{tab:exp}, to the theoretical expression in Eq.~\eqref{eq:master2}.
In constructing the $\chi^2$ function, we treat the errors on $\bar\delta_C$, and on the tensor, spin-orbit, $M^{E}_{\text{F},p}$, $M^{+}_{\text{F},p}$, and $M^{+\, \text{3b}}_{\text{F},p}$ matrix elements as statistical, uncorrelated errors, which are summed in quadrature with the experimental errors on the half-lives. The errors on the phase-space factors are on the other hand correlated. We thus will perform different fits for the three choices of renormalization scales, $\mu_{\text{ext}} = \{E_0, 2 E_0, 4 E_0\}$ and for the three values of $g_V = g_V \pm \delta g_V$.
Finally, as the LECs $\tilde g^{{N\!N}}_{Vp}$
and $\tilde g^{{N\!N}}_{Vn}$ depend on the Hamiltonian, we perform separate fits for the \intem~and \intsat~interactions. To assess the theoretical error due to the CC method, we perform fits using the CCSDT-1, CCSD(N10), and CCSD(N8) matrix elements. 

We will thus present results in the form 
\begin{equation}
    X = X_{\text{ave}} (\sigma _X)_{\text{stat}}   (\sigma _X)_{g_V}  (\sigma_X)_{\mu} (\sigma_X)_{\text{CC}}\,,    
\end{equation}
where  $X_{\text{ave}}$ 
and $(\sigma_X)_{\text{stat}}$
are the results of the CCSDT-1 fits, with $g_V$ and the renormalization scales set to their central values. The ``statistical'' error thus includes the experimental errors and the theoretical errors arising from $\bar\delta_C$ and from the NMEs in Tables~\ref{tab:LST},
\ref{tab:Fermi},
and~\ref{tab:Fermi_3b},
which should not depend too strongly on the regulator used in the chiral interaction.
The $g_V$, scale variation, and CC errors are defined as 
\begin{align}
    (\sigma_X)_{g_V} &=\text{max} \left( |X_{g_V + \delta g_V} - X_{\text{ave}} |\,,
| X_{g_V - \delta g_V} - X_{\text{ave}} | \right)\,,\notag \\
    (\sigma_X)_{\mu} &=\text{max} \left( |X_{\mu = E_0} - X_{\text{ave}} |,
| X_{\mu = 4 E_0} - X_{\text{ave}} | \right),\notag \\
(\sigma_X)_{\text{CC}} &=
\sqrt{ \big( X_{\text{N10}} - X_{\text{ave}} \big)^2
+ \big( X_{\text{N10}} - X_{\text{N8}}\big)^2   }\,. 
\end{align}
These errors currently do not include the theoretical uncertainties arising from higher orders in the EW operators (work along these lines is in progress~\cite{Graham3N}), as these are currently negligible compared to the dominant uncertainties from the leading-order LECs, see below.

Given that due to $\bar\delta_C$, the three-body ${\mathcal O}(\alpha^2)$ matrix elements, and the phase-space factor we do not have a fully consistent EFT evaluation yet, we study several fit variants to assess the sensitivity to the inputs not (fully) controlled by the EFT. 
Fits labeled by ``EFT'' use the EFT formalism as far as possible at this point, i.e., EFT input is used for the $\mathcal O(\alpha)$ corrections, 
as well as for the $\mathcal O(\alpha^2 Z^2)$, 
$\mathcal O(\alpha^2 Z)$, 
and $\mathcal O(\alpha Z E_{e} R)$ contributions that are traditionally included in the phase-space factor, 
which in turn uses the subtractions of App.~\ref{app:subtractions} and the anchored weak radius, see Sec.~\ref{sec:phase_space}.

On the other hand, fits labeled by ``HT'' use the Hardy and Towner formalism, with ab initio matrix elements only employed for $\bar\delta_{\text{NS}}$, as detailed in App.~\ref{app:HT}, but the remainder taken from Ref.~\cite{Hardy:2020qwl}. For each set-up, we consider the following variants, starting with fits to $V_{ud}$ only, in which the LECs are set to zero:
\begin{enumerate}
\item $(0\text{A})$: a fit that uses all the isotopes 
listed in Table~\ref{tab:1} and the CC NMEs.
\item $(0\text{B})$: as $(0\text{A})$, but using QMC NMEs for $^{10}$C, see App.~\ref{app:NME10C}.
\item $(0\text{C})$: as $(0\text{A})$, but excluding $^{10}$C. 
\end{enumerate}
Variant $(0\text{B})$ is motivated by the fact that 
the CC method might not adequately describe $^{10}$C. 
Since the absence of LECs implies the assumption that the NMEs should be regulator independent, we can use the QMC calculation of $^{10}$C in this case,
which were also performed with an NNLO chiral interaction, albeit a different implementation than either of the interactions used for the CC calculations.

\begin{figure*}[t]
\centering
\includegraphics[width=0.85\textwidth]{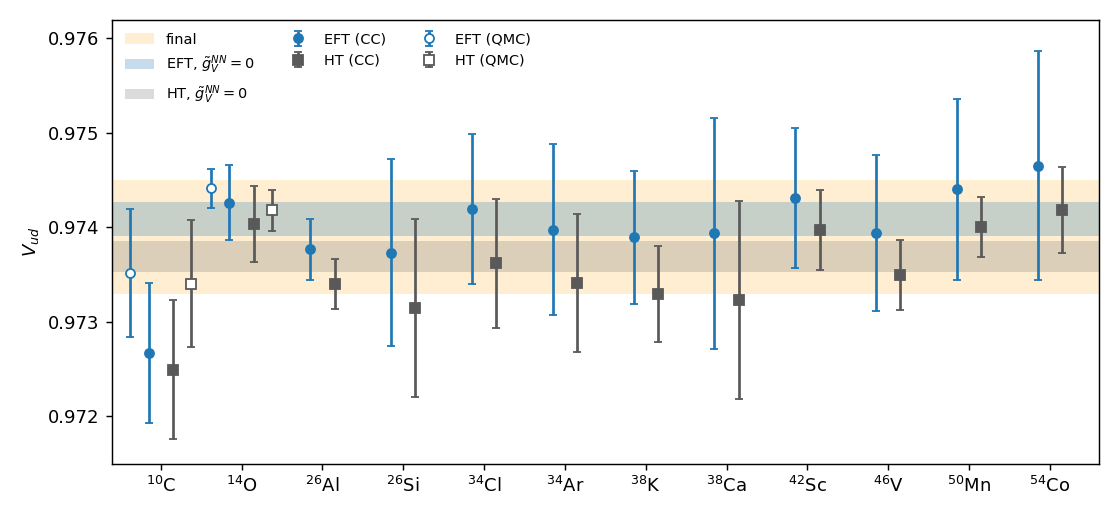}
\caption{$V_{ud}$ extracted from the individual transitions with vanishing LECs, for the EFT (blue circles) and HT (gray squares) set-ups, averaged over the \intem\ and \intsat\ interactions. Filled symbols use CC NMEs, open symbols QMC NMEs for $^{10}$C and $^{14}$O. The blue and gray bands show the corresponding fits (0C), the orange band the final result~\eqref{Vud_final}.}
\label{fig:vud_EFT_HT}
\end{figure*}

Finally, when we turn on the LECs, it is no longer consistent to ignore the cutoff dependence of different interactions, motivating the consideration of scenario $(0\text{C})$ in which we simply exclude $^{10}$C.
The results of these fits without LECs are shown in Table~\ref{tab:fitVudLEC0}. Figure~\ref{fig:vud_EFT_HT} shows the corresponding values of $V_{ud}$ extracted from the individual transitions. The transitions are consistent with each other in both set-ups, with the exception of $^{10}$C when CC NMEs are employed; with QMC NMEs, the $^{10}$C value moves into agreement with the other transitions, while for $^{14}$O, where CC and QMC agree well, the change is small. This supports the conclusion that the CC NMEs for $^{10}$C are less reliable. The EFT values lie systematically above the HT ones, reflecting the different treatment of the $\mathcal O(\alpha^2)$ and energy-dependent contributions and of the phase space, as discussed in App.~\ref{app:HT}.

\begin{figure*}[t]
\centering
\includegraphics[width=\textwidth]{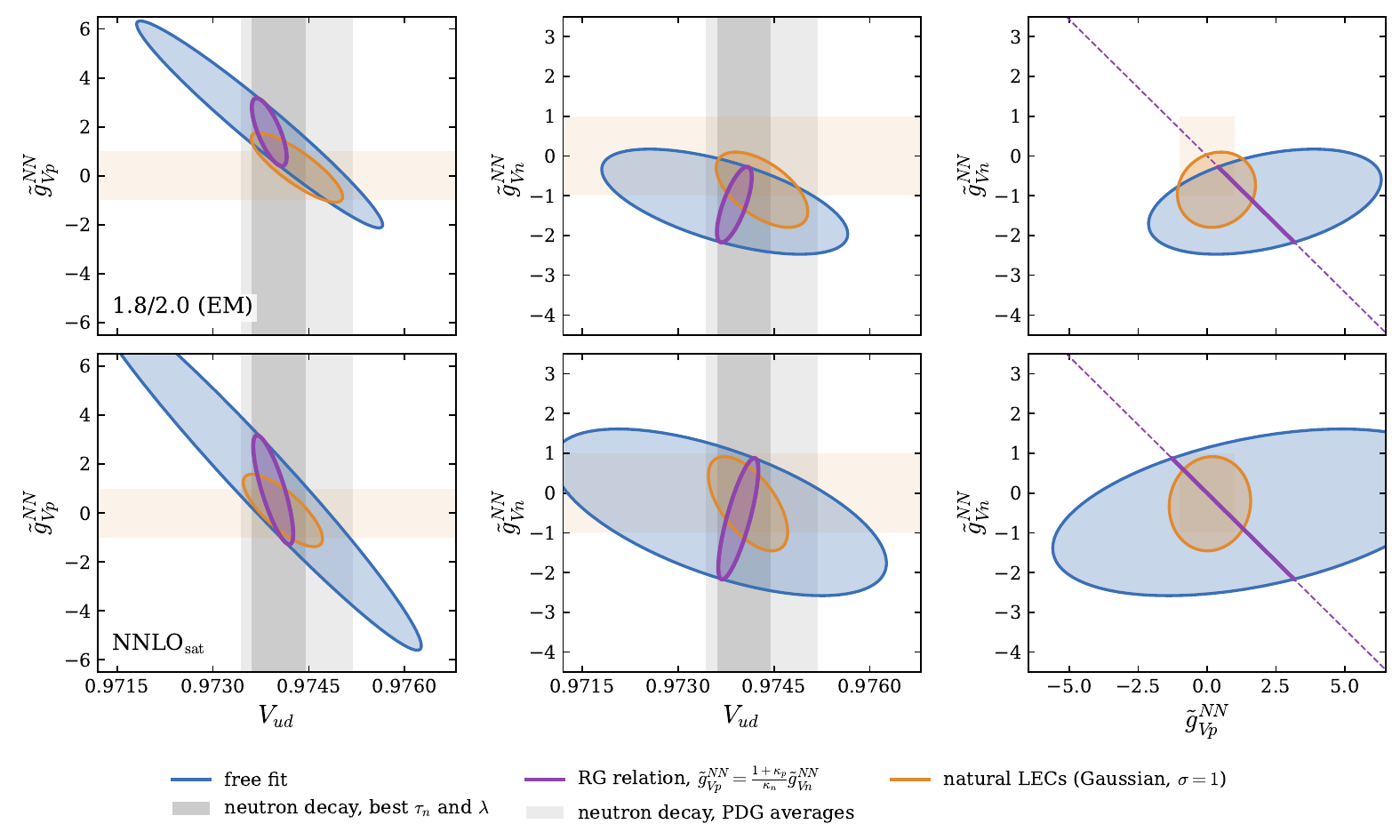}
\caption{Correlations between $V_{ud}$ and the LECs $\tilde g^{NN}_{Vp}$, $\tilde g^{NN}_{Vn}$ for \intem\ (top) and \intsat\ (bottom), with CCSDT-1 NMEs and $^{10}$C excluded. Shown are the 68\% regions ($\Delta\chi^2=2.30$) of the unconstrained fit (2C) (blue), of the fit with Gaussian priors of width 1 on both LECs (2D) (orange), and of the fit with the RG relation~\eqref{gVp_RG}, (1C) (purple), which in the right panels reduces to a segment on the dashed line. The light orange regions indicate $|\tilde g^{NN}_{Vp,n}|\leq 1$. We stress that the LECs can take different values for \intem\ and \intsat, since they absorb the dependence of $\bar\delta_\text{NS}$ on the regulator and other short-distance details of the chiral interactions. The gray bands show $V_{ud}$ from neutron decay, using the most precise measurements of $\tau_n$~\cite{UCNt:2021pcg} and $\lambda$~\cite{Markisch:2018ndu} (dark), $0.97403(42)$, and the PDG average including scale factors (light), $0.97431(88)$~\cite{ParticleDataGroup:2026mpi}.}
\label{fig:corner}
\end{figure*}

In addition, we performed fits including LECs in the following variants:
\begin{enumerate}
\setcounter{enumi}{3}
\item $(2\text{A})$: a fit that uses all the isotopes 
listed in Table~\ref{tab:1} and the CC NMEs, with both $\tilde g_{Vp}^{{N\!N}}$ and $\tilde g_{Vn}^{{N\!N}}$ as free fit parameters. 
\item $(2\text{C})$: as $(2\text{A})$, but excluding $^{10}$C.
\item $(2\text{D})$: as $(2\text{C})$, 
but with a normally distributed constraint on $\tilde g_{Vp}^{NN}$ and
$\tilde g_{Vn}^{NN}$, with width $\sigma = 1$.
\item $(1\text{A})$: as $(2\text{A})$, but constraining $\tilde g_{Vp}^{{N\!N}}=\frac{1+\kappa_p}{\kappa_n}\tilde g_{Vn}^{{N\!N}}$, and only including $\tilde g_{Vn}^{{N\!N}}$ as fit parameter.
\item $(1\text{C})$: as $(1\text{A})$, but excluding $^{10}$C.
\end{enumerate}
The fits leaving only one LEC as fit parameter are motivated by the 
relation between the anomalous dimensions, which suggests
\begin{equation}
    g_{V1}^{{N\!N}}
    = \frac{1+\kappa_p + \kappa_n}{1+\kappa_p - \kappa_n} g_{V2}^{{N\!N}}\,,
    \end{equation}
and thus
\begin{equation}
\label{gVp_RG}
    \tilde g_{Vp}^{{N\!N}} = \frac{1+\kappa_p}{\kappa_n} \tilde g_{Vn}^{{N\!N}}\,.
\end{equation}
The results of these fit variants are given in Table~\ref{tab:fitVudLEC2}, together with the correlation matrices in Table~\ref{tab:fitVudcorr}. We show the corner plots of the fits in Fig.~\ref{fig:corner}. 

The main conclusion from the fits is that $V_{ud}$ remains stable across the variants with at most one free LEC, but that the uncertainties significantly change once both LECs are included. This behavior emerges because $V_{ud}$ and $\tilde g_{Vp}^{N\!N}$ are highly correlated, with a correlation coefficient $\simeq -0.97$, see Fig.~\ref{fig:corner}. Accordingly, it is not possible to extract both quantities from superallowed data alone without incurring a significant loss in precision. This motivates analyses combining data fits with additional information on the short-distance two-nucleon physics, similarly to the
contact-term contribution to neutrinoless double $\beta$ decay~\cite{Cirigliano:2020dmx,Cirigliano:2021qko}.

Interestingly, values of $V_{ud}$ as determined from fits without LECs and using the constraint~\eqref{gVp_RG} agree at the level of $\simeq 2\times 10^{-4}$, with LECs that are of natural size ${\mathcal O}(1)$, and likewise fixing $V_{ud}$ from neutron decay would lead to natural LECs. Accordingly, we do not see indications that the EFT treatment would lead to a material shift in central value, but at the same time the EFT treatment reveals that control over the short-range two-nucleon physics is critical to control the uncertainty in $V_{ud}$.

We first assess the LEC uncertainty entirely within the EFT fits. The remaining non-EFT inputs are then related to $\bar\delta_C$, the residual uncertainty in the phase-space subtraction due to the charge-distribution correction $U$, and the three-body ${\mathcal O}(\alpha^2)$ NMEs, which are evaluated in the density approximation. Constraining the LECs with Gaussian priors of width 1, see fit (2D) in Table~\ref{tab:fitVudLEC2} and Fig.~\ref{fig:corner}, also accounts for the constraints the data impose on the LECs, while amounting to slightly stronger assumptions on the LECs than a scan over the entire range $[-1,1]$. Averaging the results for the two interactions, and adding an uncertainty from the charge-distribution correction ($U$), defined as the change when $U$ is removed, see App.~\ref{app:subtractions}, and the Hamiltonian uncertainty ($\chi$), defined as half the difference between \intem\ and \intsat, we obtain
\begin{align}
\label{Vud_EFT}
V_{ud} &= 0.97420(44)_\text{stat}(12)_{g_V}(1)_\mu(5)_\text{CC}(9)_U(11)_\chi \nonumber\\
       &= 0.9742(5)\,.
\end{align}
Varying the width of the Gaussian priors between 0.5 and 2 changes the central value by less than $1\times 10^{-4}$, while the total uncertainty changes from $3\times 10^{-4}$ to $8\times 10^{-4}$.

To assess the impact of the parts of the calculation not yet fully controlled by the EFT, it is instructive to compare the EFT and HT versions of the fit. The values of $V_{ud}$ from the HT fits lie systematically lower by $\simeq 4\times 10^{-4}$, see Fig.~\ref{fig:vud_EFT_HT}, and comparing central values for the zero- and one-LEC fits thus suggests a resulting uncertainty of $\simeq 2\times 10^{-4}$.
To account for those aspects of the EFT version of the fit that are not yet fully controlled within the EFT, we 
define as our final result the average of the EFT and HT (0C) fits, with the uncertainty from scanning the LECs over their natural range $[-1,1]$, $\delta V_{ud}\simeq 6\times 10^{-4}$. The LEC uncertainty then completely dominates over the other sources of uncertainty as given in Eq.~\eqref{Vud_EFT}, and we obtain
\beq
\label{Vud_final}
V_{ud}=0.9739(6)\,.
\eeq
This value is consistent with the EFT-only result~\eqref{Vud_EFT}, and reflects the uncertainty that we believe can currently be justified from an EFT treatment of NMEs for superallowed $\beta$ decays. It is also consistent with Eq.~\eqref{Vud_HT}, but our uncertainty is larger because it includes the contribution of the short-distance two-nucleon physics, which is not accounted for in the traditional treatment.

\section{Summary and conclusions}
\label{sec:summary}

In this paper we presented the details of our ab initio calculations of NMEs for superallowed $\beta$ decays~\cite{letter} and the consequences for the determination of $V_{ud}$. As key results, we performed CC calculations of the NMEs entering the nuclear correction $\bar\delta_\text{NS}$ in the EFT approach, across 12 transitions ranging from $^{10}$C $\rightarrow$ $^{10}$B to $^{54}$Co $\rightarrow$ $^{54}$Fe. We benchmarked these calculations against QMC results for light nuclei and provided detailed uncertainty estimates both regarding truncation errors in the many-body method and Hamiltonian uncertainties in the chiral interaction. We found that besides the NMEs for which we have explicit ab initio results also a class of three-body NMEs at ${\mathcal O}(\alpha^2)$ becomes relevant, due to a combination of large numerical prefactors and coherent scaling with $Z^2$. We developed a formalism based on nuclear densities to estimate these matrix elements, validated against the ab initio results in the two-body sector. 
These densities were derived from the same CC calculations, allowing for another consistency check against experimental charge radii. In addition, we identified the corrections in the traditional phase-space factor that are captured by the EFT matrix elements, including finite-size and isovector effects, and removed them consistently, which renders the values of $V_{ud}$ extracted from the individual transitions mutually consistent.

We found that, using a recent calculation of ${\mathcal O}(\alpha^2 Z)$ ultrasoft photons, the remaining scale uncertainty in the EFT is now negligible, and the final uncertainty arises overwhelmingly from contact-term contributions that are necessary to renormalize the GT matrix elements. One combination of the two LECs can be extracted from a fit to data, but the other proves strongly correlated with $V_{ud}$, so that additional input is required to separate $V_{ud}$ from nuclear structure effects. The results presented in this work constitute substantial progress towards a full EFT extraction of $V_{ud}$ from superallowed $\beta$ decays, to be followed along the lines of the roadmap formulated in Ref.~\cite{letter}, to obtain a robust $V_{ud}$ determination beyond the level of precision quoted in Eq.~\eqref{Vud_final} in the future.

\begin{acknowledgments}
We thank Heiko Hergert, Garrett King, and Saori Pastore for helpful discussions. 
 Financial support by the  SNSF (Project No.\ TMCG-2\_213690) 
and the U.S.\ DOE (Grant No.\ DE-FG02-00ER41132) is gratefully acknowledged.
We acknowledge support from the DOE Topical Collaboration ``Nuclear Theory
for New Physics,'' Award No.\ DE-SC0023663. 
 Computer time was provided by the Innovative and Novel Computational Impact on Theory and Experiment (INCITE) program. This research used resources of the Oak Ridge Leadership Computing Facility located at ORNL, which is supported by the Office of Science of the Department of Energy under Contract No.~DE-AC05-00OR22725.  J.~d.~V.\ is funded by the Dutch Research Council (NWO) in the form of a VIDI grant and by the European Union (ERC, CRUNS, 101230525). Views and opinions expressed are, however,  those of the author(s) only and do not necessarily reflect those of the European Union or the European Research Council. Neither the European Union nor the granting authority can be held responsible for them.
 The work of S.G.\ and E.M.\ is supported by the U.S.\ Department of Energy through
the Los Alamos National Laboratory. Los Alamos National Laboratory is
operated by Triad National Security, LLC, for the National Nuclear
Security Administration of U.S.\ Department of Energy (Contract No.\
89233218CNA000001), by the Office of Advanced Scientific Computing
Research, Scientific Discovery through Advanced Computing (SciDAC)
NUCLEI program, and by the LANL LDRD program, under projects
20260246ER and 20270410ER.
This research used resources provided by the Los Alamos National
Laboratory Institutional Computing Program, which is supported by the
U.S.\ Department of Energy National Nuclear Security Administration under
Contract No.\ 89233218CNA000001. The work of G.C.\ was supported by the U.S.\ Department of Energy (No.\ DE-SC0021027). QMC calculations were performed by the National Energy Research Scientific Computing Center under Contract No.\ DE-AC02-05CH11231
using NERSC award NP-ERCAP0036336 and NP-ERCAP0032885.
\end{acknowledgments}

\appendix

\section{Regularization of the two-body operators}
\label{app:regulator}

The EW two-body operators used in this work are derived in coordinate space in the EFT formulation of Refs.~\cite{Cirigliano:2024rfk,Cirigliano:2024msg}. To evaluate their matrix elements consistently with the nuclear interactions, we apply the same class of nonlocal regulator used in the two-nucleon interaction when constructing oscillator-basis matrix elements~\cite{Wirth:2021pij}. In terms of the relative momentum $p=|\pp_1-\pp_2|/2$, the regulator is
\begin{equation}
    f_{\Lambda_\text{reg}}(p) =
    \exp\left[-\left(\frac{p}{\Lambda_\text{reg}}\right)^{2n}\right]\,,
\label{eq:sm_regulator}
\end{equation}
where $\Lambda_\text{reg}$ is the momentum cutoff and $n=3$ for both interactions used in this work~\cite{Entem:2003ft, Ekstrom:2015rta}. For the scaled relative momentum used in the oscillator basis,
$k=|\mathbf p_1-\mathbf p_2|/\sqrt{2}$, so that $p=k/\sqrt{2}$, we define
\begin{equation}
    F_{\Lambda_\text{reg}}(k) \equiv
    f_{\Lambda_\text{reg}}\left(\frac{k}{\sqrt{2}}\right)\,.
\label{eq:sm_regulator_k}
\end{equation}
The partial-wave momentum-space operator between relative angular momenta $\lambda',\lambda$ is then regulated according to
\begin{equation}
    \widetilde {\mathcal V}_{\lambda'\lambda}(k',k) \rightarrow
    F_{\Lambda_\text{reg}}(k')\,
    \widetilde {\mathcal V}_{\lambda'\lambda}(k',k)\,
    F_{\Lambda_\text{reg}}(k)\,.
\label{eq:sm_regulated_V}
\end{equation}
The cutoff scale $\Lambda_\text{reg}$ is chosen consistently with the underlying interaction. We use $\Lambda_\text{reg}=500\MeV\ (2.534\,\text{fm}^{-1})$ for \intem~and $\Lambda_\text{reg}=450\MeV\ (2.280\,\text{fm}^{-1})$ for \intsat.

The corresponding radial component of the regulated matrix element between relative oscillator states is
\begin{align}
    \langle n'\lambda'|\widetilde {\mathcal V}|n\lambda\rangle_{\Lambda_\text{reg}} &=
    \int_0^\infty dk'\,k'^2 \int_0^\infty dk\,k^2 \nonumber\\
    &\quad\times \widetilde\phi^{*}_{n'\lambda'}(k')\, F_{\Lambda_\text{reg}}(k')\, \widetilde {\mathcal V}_{\lambda'\lambda}(k',k) \nonumber\\
    &\quad\times F_{\Lambda_\text{reg}}(k)\, \widetilde\phi_{n\lambda}(k)\,.
\label{eq:sm_me_k}
\end{align}
Here $\widetilde\phi_{n\lambda}(k)$ denotes the radial oscillator function in momentum space.  The coordinate- and momentum-space radial wave functions are related by the Hankel transform
\begin{equation}
    \widetilde\phi_{n\lambda}(k) = \sqrt{\frac{2}{\pi}}
    \int_0^\infty dr\,r^2\, j_\lambda(kr)\, \phi_{n\lambda}(r)\,.
\label{eq:sm_hankel}
\end{equation}
It is convenient to absorb the regulator into a regulated coordinate-space oscillator function,
\begin{equation}
    \phi_{n\lambda:\Lambda_\text{reg}}(r) =   \sqrt{\frac{2}{\pi}} \int_0^\infty dk\,k^2\, j_\lambda(kr)  F_{\Lambda_\text{reg}}(k)\, \widetilde\phi_{n\lambda}(k)\,.
\label{eq:sm_regulated_basis}
\end{equation}
Then the regulated momentum-space matrix element in Eq.~\eqref{eq:sm_me_k} can be expressed in coordinate space as
\begin{align}
\label{eq:sm_me_r}
    \langle n'\lambda'|{\mathcal V}|n\lambda\rangle_{\Lambda_\text{reg}} & =
    \int_0^\infty dr'\,r'^2 \int_0^\infty dr\,r^2 \\
    &\quad \times \phi^{*}_{n'\lambda':\Lambda_\text{reg}}(r')\, {\mathcal V}_{\lambda'\lambda}(r',r)\, \phi_{n\lambda:\Lambda_\text{reg}}(r)\,,\notag
\end{align}
where the partial-wave coordinate-space kernel is
\begin{align}
\label{eq:sm_V_coordinate}
    {\mathcal V}_{\lambda'\lambda}(r',r) &= \frac{2}{\pi}
    \int_0^\infty dk'\,k'^2 \int_0^\infty dk\,k^2 \nonumber\\
    &\quad\times j_{\lambda'}(k'r')\, \widetilde {\mathcal V}_{\lambda'\lambda}(k',k)\, j_\lambda(kr)\,.
\end{align}
Thus, for coordinate-space EW potentials, the nonlocal momentum-space regulator is incorporated through the regulated oscillator functions $\phi_{n\lambda:\Lambda_\text{reg}}(r)$, and their derivatives where required.

For a momentum-independent contact interaction in the $S$ wave,
\begin{equation}
    \widetilde {\mathcal V}_{00}(k',k)=C\,,
\label{eq:sm_contact_kernel}
\end{equation}
the regulated matrix element becomes
\begin{equation}
    \langle n'0|{\mathcal V}_{\rm CT}|n0\rangle_{\Lambda_\text{reg}} =
    C\,\frac{\pi}{2}\, \phi^{*}_{n'0:\Lambda_\text{reg}}(0)\, \phi_{n0:\Lambda_\text{reg}}(0)\,.
\label{eq:sm_contact_me}
\end{equation}

\section{Corrections to the phase space and their subtraction}
\label{app:subtractions}

The phase-space factor $\bar f$, Eq.~\eqref{eq:PS}, contains the function $\tilde C(E_e)$, which collects corrections taken over from the traditional formalism~\cite{Hayen:2017pwg,Wilkinson:1993he,Hardy:2004id,Hardy:2020qwl},
\begin{equation}
\tilde C = L_0\times U\times C\times C_I\times S\times r\times R(E_0)\,,
\label{eq:Ctildedecomp}
\end{equation}
where all factors, except for $R(E_0)$, depend on the positron energy $E_e$. $L_0$ is the finite-size correction to the Fermi function for a uniformly charged sphere, $U$ corrects $L_0$ for the difference between a realistic charge distribution and the uniform sphere, $C$ is the shape factor and arises from the integration of the lepton wave functions over the nuclear volume, and $C_I$ is the isovector correction
that accounts for the difference between the weak and the charge distributions. $S$ and $r$ are the atomic screening and overlap corrections, and $R(E_0)$ is the recoil correction.

Several of these encode physics that, in our EFT approach, is computed explicitly through the nuclear matrix elements of $\bar\delta_\text{NS}$, most notably through $\delta^E_\text{NS}$ at $\mathcal O(\alpha Z E_e R)$ and through the $\mathcal O(\alpha^2)$ matrix elements $M^+_{\text{F},p}$ and $M^{+\,\text{3b}}_{\text{F},p}$. To avoid double counting, the overlapping terms have to be removed from $\tilde C$. In this appendix we identify these overlapping terms, and specify the subtractions used in this work. 

All expressions are given for $\beta^+$ decay. We use the radii of Table~\ref{tab:charge_radii}: 
$R_p$ and $R_W^{(0)}$ refer to the radii for point-like nucleons as defined in Eq.~\eqref{radii}, while finite-size and relativistic corrections are required to obtain the physical charge and weak radii, see Eqs.~\eqref{eq:Rch} and~\eqref{eq:Rw}. A uniformly charged sphere with root-mean-square radius $R_X$ has radius $\sqrt{5/3}\,R_X$. In the traditional factors below, $R=\sqrt{5/3}\,R_\text{ch}$ is the radius of the uniform sphere with the experimental charge radius of the final-state nucleus, as in the main text.

\begin{table}[t]
\centering
\renewcommand{\arraystretch}{1.3}
\begin{tabular}{lrrr}
\toprule
 & $E_e$ & $E_0$ & $m_e^2/E_e$ \\
\colrule
$L_0$ & $1$ & -- & $\frac{1}{2}$ \\
$C$ & $\frac{13}{35}$ & $\frac{6}{35}$ & $-\frac{1}{70}$ \\
$L_0\times C$ & $\frac{48}{35}$ & $\frac{6}{35}$ & $\frac{17}{35}$ \\
\colrule
$\delta^E_\text{NS}$, $\langle r_{12}\rangle=\frac{36}{35}R$ & $\frac{48}{35}$ & $\frac{6}{35}$ & $\frac{18}{35}$ \\
\botrule
\end{tabular}
\caption{Coefficients of the $\mathcal O(\alpha ZR)$ terms in the traditional factors $L_0$ and $C$ for a uniformly charged sphere, compared to $\delta^E_\text{NS}$ evaluated in the same model. All coefficients are given in units of $\alpha ZR$.}
\label{tab:aZRcomparison}
\end{table}

\begin{figure*}[t]
\centering
\includegraphics[width=\textwidth]{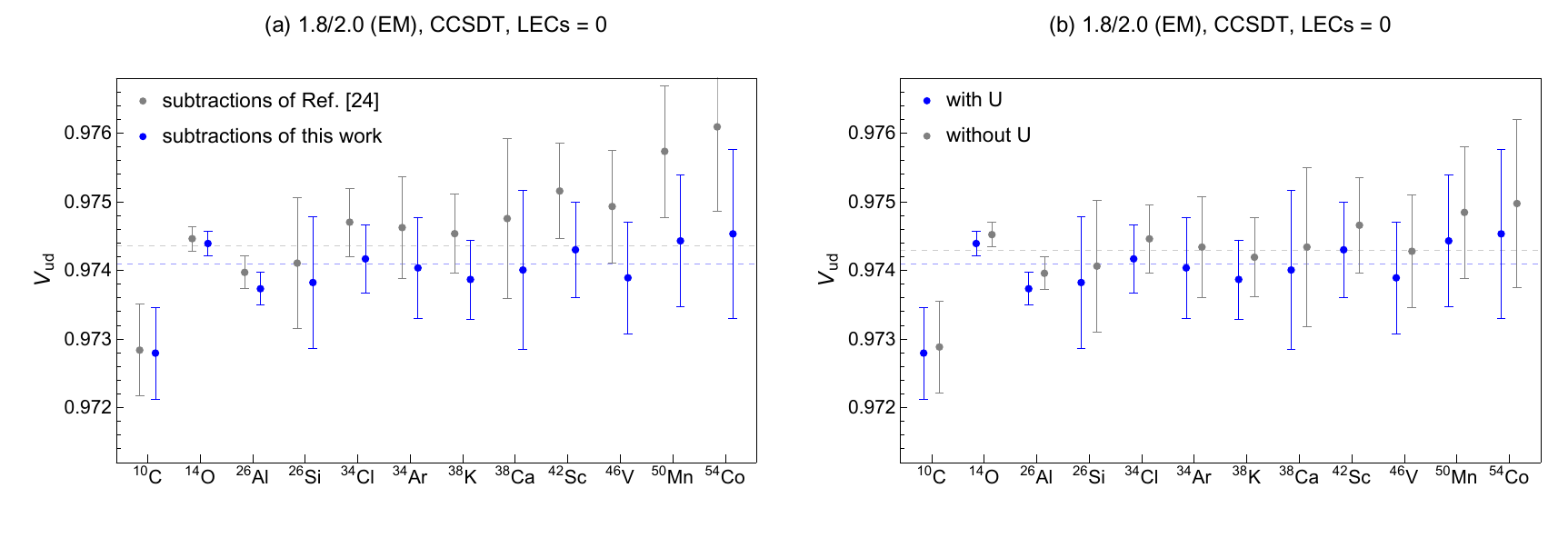}
\caption{$V_{ud}$ extracted from the individual transitions with $\tilde g^{N\!N}_{Vp}=\tilde g^{N\!N}_{Vn}=0$, for \intem\ and CCSDT-1 matrix elements. (a) Subtractions of Ref.~\cite{Cirigliano:2024msg} and the traditional shape factor $C(R_\text{ch})$ with $C_I$ (gray), compared to the subtractions of this work, Eq.~\eqref{eq:Ctildefinal} (blue). (b) The subtractions of this work with (blue) and without (gray) the charge-distribution correction $U$.}
\label{fig:vud_subtractions}
\end{figure*}

\subsection{Atomic corrections}
The atomic corrections $S$ and $r$ concern the electron cloud and do not overlap with $\bar\delta_\text{NS}$. They are mostly kept the same as in App.~G of Ref.~\cite{Cirigliano:2024msg}, however, some of the expressions there require generalization as we consider a wider range of transitions in this work. In particular, the atomic screening factor, $S$, depends on the function $N(Z+1)$. Here we use the  values given in Ref.~\cite{Matese:1966zz} and find that the following functional form fits well for $x\lesssim 50$,
\begin{align}
N(x) &= 1.03 -2.60\times 10^{-2} x+5.40\times 10^{-4}x^2\nonumber\\
&-4.13\times 10^{-6}x^3+0.276 \log x\,.
\end{align}

In addition, the atomic overlap factor, $r$, depends on the electron binding energy, $B(G)$, which can be generalized from the form given in Ref.~\cite{Cirigliano:2024msg} by
\begin{align}
B(G) = 
\begin{cases}
13.080(Z+1)^{2.42}\, {\rm eV} & 5\leq Z\leq 9\\
14.9545(Z+1)^{2.37}\, {\rm eV} & 10\leq Z\leq 29\\
11.435(Z+1)^{2.45}\, {\rm eV} & 30\leq Z\leq 38
\end{cases}\,.
\end{align}

\subsection{The \texorpdfstring{$\boldsymbol{\mathcal O(\alpha ZR)}$}{} terms of \texorpdfstring{$\boldsymbol{L_0}$}{} and \texorpdfstring{$\boldsymbol{C}$}{}}

The finite-size correction $L_0$ and the shape factor $C$ are computed in the traditional formalism for a uniformly charged sphere~\cite{Behrens:1982,Hayen:2017pwg}. Their terms linear in $\alpha Z R$ are, for $\gamma=\sqrt{1-(\alpha Z)^2}\to 1$,
\begin{align}
L_0 &\supset \alpha Z R\left(E_e+\frac{m_e^2}{2E_e}\right)\,, \label{eq:L0aZR}\\
C &\supset \alpha Z R\left(\frac{13}{35}E_e+\frac{6}{35}E_0-\frac{1}{70}\frac{m_e^2}{E_e}\right)\,. \label{eq:CaZR}
\end{align}
In the EFT, the corresponding physics is contained in $\delta^E_\text{NS}$, see Eq.~\eqref{deltaNSE}. With the density approximation $M^E_{\text{F},p}= -Z M_\text{F}^{(0)}\langle r_{jk}\rangle/(2R_A)$, Eq.~\eqref{MF_constant}, the spectral form of $\delta^E_\text{NS}$ before phase-space averaging is
\begin{equation}
\delta^E_\text{NS}(E_e) = \alpha Z \langle r_{jk}\rangle\left(\frac{4}{3}E_e+\frac{1}{6}E_0+\frac{1}{2}\frac{m_e^2}{E_e}\right)\,,
\label{eq:deltaEspectral}
\end{equation}
where $\langle r_{jk}\rangle$ is the mean distance between the decaying nucleon and a proton. If we evaluate Eq.~\eqref{eq:deltaEspectral} in the model that underlies $L_0$ and $C$ (both nucleon distributions are uniform spheres of radius $R$) we obtain $\langle r_{12}\rangle = \frac{36}{35}R$. Table~\ref{tab:aZRcomparison} compares the resulting coefficients. The terms proportional to $E_e$ and $E_0$ agree exactly, while the terms proportional to $m_e^2/E_e$ differ only by $\frac{1}{35}\alpha Z R\, m_e^2/E_e$, a numerically negligible effect. Hence, $\delta^E_\text{NS}$ captures the $\mathcal O(\alpha ZR)$ content of $L_0\times C$ in the uniform-sphere limit and generalizes it to realistic densities. We therefore have to perform subtractions on the traditional expressions to avoid double counting.

\subsection{\texorpdfstring{$\boldsymbol{L_0^\text{sub}}$}{}}

The $\mathcal O(\alpha ZR)$ terms of $L_0$ are contained in $\delta^E_\text{NS}$, and its $R$-independent $\mathcal O(\alpha^2Z^2)$ terms in $\bar F$, $C^{(g_V)}_\text{eff}$, and the $\mathcal O(\alpha^2)$ matrix elements, so both have to be removed from $L_0$. The small-$Z$ expansion of $L_0$ for a uniform sphere is~\cite{Behrens:1982}
\begin{equation}
L_0 = 1 + \alpha Z R E_e + \frac{13}{60}(\alpha Z)^2 + \frac{\alpha Z R\, m_e^2}{2E_e} + \dots\,,
\label{eq:L0Taylor}
\end{equation}
which agrees with the analytic terms of the parametrization of Refs.~\cite{Wilkinson:1993he,Hayen:2017pwg} used in the numerics. That parametrization adds to the analytic terms a series $a_{-1}R\, m_e^2/E_e+\sum_{n} a_n (E_eR)^n$ with $a_n=\sum_{x=1}^6 b_{x,n}(\alpha Z)^x$. Although this has the form of an expansion in $\alpha Z$, the coefficients $b_{x,n}$ are obtained by fitting the numerical solution of the Dirac equation over a wide range of $Z$, $R$ and $E_e$, so that contributions of different orders are redistributed among them. As noted in Ref.~\cite{Behrens:1982}, ``the coefficients of an economized power series are different from the Taylor coefficients of the analytic function.'' This can also be seen from the fact that the electron and positron fits are not related by $Z\to -Z$, as genuine expansion coefficients would be: $b_{1,1}=0.0248$ and $0.0794$, and $b_{2,0}=0.0072$ and $0.0665$, for electrons and positrons, respectively~\cite{Hayen:2017pwg}. The terms $\alpha Z\, b_{1,n}$ and $(\alpha Z)^2 b_{2,0}$ are therefore not the $\mathcal O(\alpha Z)$ and $\mathcal O(\alpha^2Z^2)$ parts of $L_0$ and if they were, the agreement in Table~\ref{tab:aZRcomparison}, which relies on the analytic coefficients, would be spoiled. In Ref.~\cite{Cirigliano:2024msg} we subtracted these terms together with the analytic ones, which had a marginal impact on the transition of ${}^{14}$O, the main focus of that work. Here we subtract only the Taylor coefficients of Eq.~\eqref{eq:L0Taylor},
\begin{equation}
L_0^\text{sub} = L_0 - \alpha Z R E_e\left(1+\frac{m_e^2}{2E_e^2}\right) - \frac{13}{60}(\alpha Z)^2\,,
\label{eq:L0subpure}
\end{equation}
and keep the $a_n$ series fit as the higher-order remainder. The difference between the two prescriptions grows smoothly with $Z$, from $6\times 10^{-5}$ in $\bar P$ for $^{10}$C to $3.4\times 10^{-3}$ for $^{54}$Co, and is the largest change in the phase-space factors with respect to Ref.~\cite{Cirigliano:2024msg}. With Eq.~\eqref{eq:L0subpure}, the remainder $L_0^\text{sub}-1$ is at most $4\times 10^{-4}$, and the $Z$-dependent trend of the values of $V_{ud}$ extracted from the individual transitions, found with the previous prescription, disappears. We show this in Fig.~\ref{fig:vud_subtractions}. 

\subsection{\texorpdfstring{$\boldsymbol{C^\text{sub}}$}{} and the isovector correction}

In the traditional formalism, the shape factor $C$ is evaluated with the charge distribution, and the isovector correction $C_I$ accounts for the difference between the weak (transition) and charge distributions~\cite{Wilkinson:1993he,Hayen:2017pwg}. To first order in $ R_W^2-R_\text{ch}^2$,
\begin{align}
   C_I - 1 &= -\frac{R_W^2-R_\text{ch}^2}{3}\left[(E_0-E_e)^2+(E_e+V_0)^2-m_e^2\right]\,, \nonumber\\
V_0 &= -\frac{3\alpha Z}{2R}\,.
\label{eq:CI}
\end{align}
This expression contains terms of three orders.  The terms $(E_0-E_e)^2$ and $E_e^2$ scale as $\mathcal O((E R)^2)$, the cross term $2E_eV_0$ as $\mathcal O(\alpha Z E_e R)$, and the $V_0^2$ term as $\mathcal O(\alpha^2Z^2)$.

The last two terms are corrections to the Coulomb interaction of the lepton with the nuclear charge, evaluated where the decay takes place (over the weak rather than the charge distribution). In the EFT, this physics is part of the matrix elements of $\delta^E_\text{NS}$ and of the $\mathcal O(\alpha^2)$ terms, which involve the relative distances between the decaying nucleon and the protons and are computed with the actual weak and proton distributions. That the ab initio matrix elements indeed reflect the larger extent of the weak distribution can be checked with simple models of the densities. Since the matrix elements are computed with point nucleons, we use the point radii $R_p$ and $R_W^{(0)}$ of Table~\ref{tab:charge_radii} in these models.

\begin{figure}[t]
\centering
\includegraphics[width=\columnwidth]{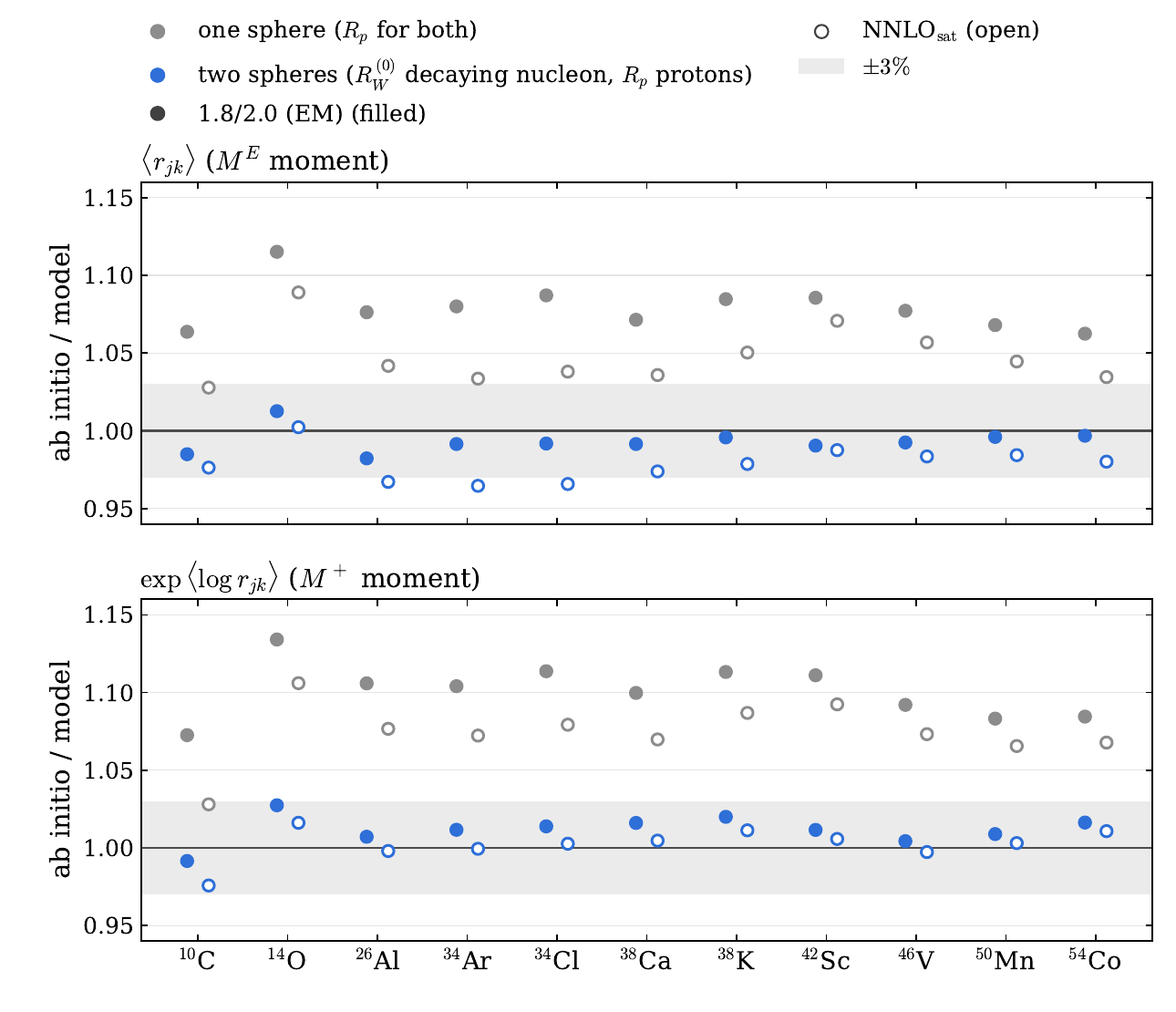}
\caption{Ratio of the ab initio moments $\langle r_{jk}\rangle$ (top) and $e^{\langle \log r_{jk}\rangle}$ (bottom) of Table~\ref{tab:Fermi} to uniform-sphere models. The one-sphere model uses the point-proton radius $R_p$ for both the decaying nucleon and the protons (gray). The two-sphere model uses the point weak radius $R^{(0)}_W$ for the decaying nucleon and $R_p$ for the protons (blue). Radii are taken from Table~\ref{tab:charge_radii}.}
\label{fig:twosphere}
\end{figure}

In a first model, which corresponds to the assumptions underlying $L_0$ and $C$, the decaying nucleon and the protons are distributed uniformly within the same sphere of radius $\sqrt{5/3}\,R_p$. The moments entering $M^E_{\text{F},p}$ and $M^+_{\text{F},p}$ are then
\begin{align}
\langle r_{12}\rangle &= \frac{36}{35}\sqrt{5/3}\,R_p\,,\nonumber\\
e^{\langle \log r_{12}\rangle} &= 2e^{-3/4}\sqrt{5/3}\,R_p\,,
\end{align}
obtained by averaging $r_{12}$ and $\log r_{12}$ over the positions of two independent points in the sphere. In a second model, the decaying nucleon is distributed uniformly within a sphere of radius $\sqrt{5/3}\,R_W^{(0)}$ and the protons within a sphere of radius $\sqrt{5/3}\,R_p$. The moments are then evaluated numerically, by first averaging over the relative angle, which can be done analytically, and then integrating over the two radial coordinates. As shown in Fig.~\ref{fig:twosphere}, the one-sphere model already works well but underestimates the ab initio $\langle r_{jk}\rangle$ and $e^{\langle\log r_{jk}\rangle}$ by (3--13)\% for all transitions and both interactions. The two-sphere model is a significant improvement and reproduces the ab initio $\langle r_{12}\rangle $ and $e^{\langle \log r_{12}\rangle} $ to within 3\%. The ab initio matrix elements thus already account for the difference between the weak and the charge distributions, which is the effect described by the $\mathcal O(\alpha Z E_e R)$ and $\mathcal O(\alpha^2 Z^2)$ terms of $C_I$. Within the two-sphere model, one can also verify that the dependence of $M^E_{\text{F},p}$ and $M^{+\,\text{3b}}_{\text{F},p}$ on $R_W^{(0)}$ has the same structure and sign as these terms of Eq.~\eqref{eq:CI}, with coefficients of similar size. Including $C_I$ in full would therefore double count these contributions.

The remaining $\mathcal O((ER)^2)$ part of $C_I$ is included by evaluating the $\mathcal O(\alpha^0)$ shape factor with the weak radius, as in Ref.~\cite{Cirigliano:2024msg},
\begin{align}
\label{eq:Csub}
C^\text{sub}&=1+\frac{5}{3}(E_0R_W)^2\\
&\times\left(-\frac{1}{5}+\frac{4}{15}\frac{E_e}{E_0}+\frac{2}{15}\frac{m_e^2}{E_eE_0}-\frac{4}{15}\frac{E_e^2}{E_0^2}\right)\,.\notag
\end{align}
Here the lepton probes the full weak form factor, so $R_W$ is the weak radius including finite-size and relativistic corrections, as given in Eq.~\eqref{eq:Rw}. Since at this order only the (root-mean-square) radius $R_W$ enters, Eq.~\eqref{eq:Csub} holds for any weak density. 
 Evaluating Eq.~\eqref{eq:Csub} with $R_W$ instead of $R_\text{ch}$ reproduces the $\mathcal O((ER)^2)$ part of Eq.~\eqref{eq:CI} to within about 25\%. We suspect that the difference arises because Eq.~\eqref{eq:CI} keeps only the large components of the lepton wave functions, while Eq.~\eqref{eq:Csub} has the full structure of Ref.~\cite{Behrens:1982}. We therefore use $C^\text{sub}$ with the weak radius and, unlike the traditional evaluation, do not apply $C_I$.

The absolute value of the computed $R_W$ depends on the Hamiltonian, see Table~\ref{tab:charge_radii}, while the difference between the computed weak and charge radii is much less sensitive to the interaction. We therefore anchor the weak radius to the measured charge radii
\begin{equation}
\big(R_W^\text{anch}\big)^2 = \big(R_\text{ch}^\text{exp}\big)^2 + \big[R_W^2-R_\text{ch}^2\big]_\text{CC}\,,
\label{eq:RWanchored}
\end{equation}
where the squared charge radii are averaged over the initial- and final-state nuclei, and the CC difference over \intem\ and \intsat. The resulting values of $R_W^\text{anch}$ are given in Table~\ref{tab:charge_radii}.

\begin{table*}[t]
    \centering
    \renewcommand{\arraystretch}{1.3}
    \begin{tabular}{c r r r r r r r}
    \toprule
       Transition
   &   $\overline{P}_0$   & $C_0(R_W^\text{anch})$ & $L^{\text{sub}}_0$ & $U$   & $S$ & $r$ & $R(E_0)$ \\
       \colrule
$^{10}$C $\rightarrow$ $^{10}$B   & $2.39308$ & $-0.00019$ & $-0.00002$ & $0.00044$ & $0.00444$ & $-0.00053$ & $-0.00054$ \\
$^{14}$O $\rightarrow$ $^{14}$N   & $44.4364$ & $-0.0129$ & $-0.0001$ & $0.0118$ & $0.0650$ & $-0.0057$ & $-0.0119$ \\
$^{26}$Si $\rightarrow$ $^{26}$Al & $1063.8$ & $-1.50$ & $0.024$ & $0.52$ & $1.73$ & $-0.085$ & $-0.29$ \\
$^{34}$Ar $\rightarrow$ $^{34}$Cl & $3522.0$ & $-9.6$ & $0.24$ & $2.2$ & $6.3$ & $-0.24$ & $-0.93$ \\
$^{38}$Ca $\rightarrow$ $^{38}$K  & $5494.5$ & $-19.0$ & $0.62$ & $3.8$ & $10.3$ & $-0.35$ & $-1.4$ \\
\colrule
$^{26}$Al $\rightarrow$ $^{26}$Mg & $495.37$ & $-0.51$ & $0.008$ & $0.23$ & $0.85$ & $-0.045$ & $-0.11$ \\
$^{34}$Cl $\rightarrow$ $^{34}$S  & $2062.5$ & $-4.5$ & $0.10$ & $1.2$ & $3.8$ & $-0.15$ & $-0.49$ \\
$^{38}$K $\rightarrow$ $^{38}$Ar  & $3403.1$ & $-9.7$ & $0.29$ & $2.3$ & $6.6$ & $-0.24$ & $-0.80$ \\
$^{42}$Sc $\rightarrow$ $^{42}$Ca & $4614.4$ & $-16.4$ & $0.60$ & $3.4$ & $9.5$ & $-0.31$ & $-1.0$ \\
$^{46}$V $\rightarrow$ $^{46}$Ti  & $7425.9$ & $-33.3$ & $1.5$ & $5.9$ & $15.7$ & $-0.47$ & $-1.7$ \\
$^{50}$Mn $\rightarrow$ $^{50}$Cr & $11048.2$ & $-58.2$ & $3.3$ & $9.4$ & $24.0$ & $-0.67$ & $-2.5$ \\
$^{54}$Co $\rightarrow$ $^{54}$Fe & $16176.8$ & $-99.9$ & $6.9$ & $14.7$ & $35.9$ & $-0.92$ & $-3.7$ \\
\botrule
\end{tabular}
    \caption{Non-EFT corrections to the phase-space factor. $\overline{P}_0$ is the phase-space combination of Eq.~\eqref{combscale} without the corrections in $\tilde C$. The other columns give the change of $\overline{P}$ due to each factor in Eq.~\eqref{eq:Ctildefinal} separately, $\overline{P}_0\,(\langle X\rangle-1)$, the shape factor $C_0$ evaluated with the anchored weak radius (averaged over \intem\ and \intsat, see Table~\ref{tab:charge_radii}), the finite-size correction $L^\text{sub}_0$, Eq.~\eqref{eq:L0subpure}, the charge-distribution correction $U$, the atomic screening and overlap corrections $S$ and $r$, and the recoil correction $R(E_0)$, Eq.~\eqref{eq:recoil}.}
    \label{tab:phase3}
\end{table*}

\subsection{The charge-distribution correction \texorpdfstring{$\boldsymbol{U}$}{}}

By the same argument as for $C_I$, the $\mathcal O(\alpha ZR)$ part of $U$ is already contained in $\delta^E_\text{NS}$. Unfortunately, $U$ is available only as a fit to numerical solutions of the Dirac equation~\cite{Wilkinson:1993he}, which, similarly to the coefficients $b_{x,n}$ of $L_0$, does not separate the orders. We therefore keep $U$ in its traditional form. Removing it entirely, an upper bound on the effect of the potential double counting, increases $V_{ud}$ in the fits with $\tilde g^{N\!N}_{Vp}=\tilde g^{N\!N}_{Vn}=0$ by $2\times 10^{-4}$, see Fig.~\ref{fig:vud_subtractions}. 
 When the LECs are included as free parameters, however, the effect is largely absorbed by $\tilde g^{N\!N}_{V1}$, see Eq.~\eqref{LECs_linear_combinations}, which remains of natural size, and $V_{ud}$ is essentially unchanged. This illustrates a more general point. The LECs are required by renormalization, but as a side effect they also make the analysis robust against residual uncertainties in the description of the phase space. The uncertainty on $V_{ud}$ that results from the LECs thus also covers effects of this kind.
 
 In conclusion, we keep $U$ as in Ref.~\cite{Cirigliano:2024msg} and note that a consistent subtraction would require isolating the term linear in $\alpha Z$ in a numerical calculation of $U$, which we leave for future work. 

\subsection{Recoil}

Finally, we include the recoil correction 
\begin{equation}\label{eq:recoil}
R(E_0) = 1-\frac{3E_0}{2M_A}
\end{equation}
in Eq.~\eqref{eq:Ctildedecomp}~\cite{Hardy:2004id}. It does not overlap with $\bar\delta_\text{NS}$ and no subtractions are needed. 

\subsection{Summary}

In summary, the phase-space factors in Table~\ref{tab:phase_space} are evaluated with
\begin{equation}\label{eq:Ctildefinal}
\tilde C(E_e) = L_0^\text{sub}\times U \times C^\text{sub}\times S\times r\times R(E_0)\,,
\end{equation}
with $L_0^\text{sub}$ from Eq.~\eqref{eq:L0subpure}, $C^\text{sub}$ from Eq.~\eqref{eq:Csub} with the anchored weak radius of Eq.~\eqref{eq:RWanchored}, and without $C_I$. The individual contributions of the factors in Eq.~\eqref{eq:Ctildefinal} are listed in Table~\ref{tab:phase3}.

Figure~\ref{fig:vud_subtractions} illustrates the impact of these choices on $V_{ud}$. For transparency, we show the values of $V_{ud}$ extracted from the individual transitions with the short-distance LECs, $\tilde g_{Vp}^{{N\!N}}$ and $\tilde g_{Vn}^{{N\!N}}$, set to zero, for a single interaction and CC truncation (\intem, CCSDT-1), but for other ab initio calculations the figures look similar. With the prescription of Ref.~\cite{Cirigliano:2024msg}, that is $L_0^\text{sub}$ including the subtraction of the fit coefficients and the traditional shape factor $C(R_\text{ch})$ combined with $C_I$, the extracted values increase with $Z$. With the subtractions of this work that avoid the double counting the trend disappears, and the transitions become consistent with each other. Almost all of this change is due to $L_0^\text{sub}$, while the treatment of the weak radius changes $V_{ud}$ by less than $5\times 10^{-5}$. The right panel shows the effect of removing $U$ altogether. All values move up by an amount that grows slowly with $Z$, changing the average $V_{ud}$ by about $2\times 10^{-4}$, without affecting the consistency of the transitions. This change could be interpreted as a rough estimate of the uncertainty from the phase-space treatment, but we emphasize that, as argued above, the uncertainty due to the LECs already covers similar effects.

\section{Nuclear densities}
\label{app:density}

Here we discuss the relation between two- and three-body matrix elements and nuclear densities, outlined in Sec.~\ref{sec:density}, in more detail.

The two- and three-body potentials can be written in terms of fields as 
\begin{align}
\hat V_\text{2b} &= \int_{\vx_i} N_{\alpha_1}^\dagger(\vx_1)N_{\alpha_2}^\dagger(\vx_2)\\
&\qquad\times V^{\alpha_1\alpha_2}_{\beta_1\beta_2}(\vx_1,\vx_2)N_{\beta_2}(\vx_2)N_{\beta_1}(\vx_1)\,,\nn\\
\hat V_\text{3b} &= \int_{\vx_i} N_{\alpha_1}^\dagger(\vx_1)N_{\alpha_2}^\dagger(\vx_2)N_{\alpha_3}^\dagger(\vx_3)\nn\\
&\qquad\times V^{\alpha_1\alpha_2\alpha_3}_{\beta_1\beta_2\beta_3}(\vx_1,\vx_2,\vx_3)N_{\beta_3}(\vx_3)N_{\beta_2}(\vx_2)N_{\beta_1}(\vx_1)\,,\notag
\end{align}
where the indices denote (iso)spins. The two- and three-body potentials corresponding to the ${\mathcal O}(\alpha^2)$ NMEs $M_\text{F}^+$ and $M_\text{F}^{+\,\text{3b}}$ take the form
\begin{align}
\label{eq:Visospin}
V^{\alpha_1\alpha_2}_{\beta_1\beta_2}(\vx_1,\vx_2)&=
\tau^+_{\alpha_1\beta_1} Q_{\alpha_2\beta_2} V(\vx_1,\vx_2)\,,\\
V^{\alpha_1\alpha_2\alpha_3}_{\beta_1\beta_2\beta_3}(\vx_1,\vx_2,\vx_3)&=
\tau^+_{\alpha_1\beta_1} Q_{\alpha_2\beta_2}Q_{\alpha_3\beta_3} V(\vx_1,\vx_2,\vx_3)\,,\nn
\end{align}
with 
\begin{align}\label{eq:pots}
V(\vx_1,\vx_2)&= -\log (r_{12} \Lambda)\,,\\
V(\vx_1,\vx_2,\vx_3)&= -\frac{1}{2}\log\left[ (r_{12}+r_{23}+r_{31})\Lambda/2\right]\,,\nn
\end{align}
and $r_{ij} = |\vx_i-\vx_j|$. The potential corresponding to $M_{\text{F},p}^{E}$ takes a similar form,
\begin{equation}\label{eq:potsVE}
V(\vx_1,\vx_2)= -\frac{r_{12}}{2R_A}\,.
\end{equation}

To connect the NMEs of these potentials with the density operator, we can use the equal-time anticommutation relation, $\{ N^\dagger_\alpha(\vx_1),\, N_\beta(\vx_2)\} = \delta^3(\vx_1-\vx_2)\delta_{\alpha\beta}$, to write the above potentials in terms of bilinears. One possible way to do so gives
\begin{align}\label{eq:Appdensities}
\hat V_\text{2b} &= \int_{\vx_i} (p^\dagger p)_{\vx_2}    (p^\dagger n)_{\vx_1} 
V(\vx_1,\vx_2)
-\int_{\vx} (p^\dagger n)_{\vx} V(\vx,\vx)\,,\nn\\
\hat V_\text{3b} &= \int_{\vx_i}  (p^\dagger p)_{\vx_3}  (p^\dagger p)_{\vx_2}
 (p^\dagger n)_{\vx_1}
V(\vx_1,\vx_2,\vx_3)\nn\\
&-\int_{\vx_i} \left[2(p^\dagger p)_{\vx_2} (p^\dagger n)_{\vx_1}
+(p^\dagger n)_{\vx_2}(p^\dagger p)_{\vx_1} 
\right]\notag\\
&\qquad\times V(\vx_1,\vx_2,\vx_2)\nn\\
&+\int_{\vx_1}(p^\dagger n)_{\vx_1}
V(\vx_1,\vx_1,\vx_1)\,.
\end{align}
We can now insert complete sets of states, $1=\sum_n \int_{\vq}|n(\vq)\rangle\langle n(\vq)|$, between each of the bilinears and write their matrix elements as,
\begin{equation}
\langle n(\vp)| N^\dagger \Gamma^\alpha N(\vx) | m(\vq)\rangle  = e^{i \vx \cdot (\vq-\vp)} \rho^{\alpha}_{mn}(\vq-\vp )\,,
\end{equation}
where $\Gamma^\text{EM}=Q$, $\Gamma^{W}=\tau^+$, and $\rho^{\text{EM}}$, $\rho^W$ are the charge and weak densities in momentum space. Inserting these relations into Eq.~\eqref{eq:Appdensities} and moving to coordinate space, we obtain the matrix elements, $\langle \vp_f|\hat V_{i}|\vp_i\rangle = (2\pi)^3\delta ^3(\vp_i - \vp_f)M_i$, with
\begin{align}
M_\text{2b}&=
 \sum_n \int_{\vx_i}\rho^\text{EM}_{fn}(\vx_2)\rho^{W}_{ni}(\vx_1)V(\vx_1,\vx_2)\nn\\
 &-\int_{\vx} \rho^W_{fi}(\vx)V(\vx,\vx)\,,\nn\\
M_\text{3b} &= 
\sum_{m,n}\int_{\vec x_i}\rho^\text{EM}_{fm}(\vx_3)\rho^\text{EM}_{mn}(\vx_2)\rho^{W}_{ni}(\vx_1)V(\vx_1,\vx_2,\vx_3)\nn\\
&-
\sum_n\int_{\vx_i}\left[2\rho^\text{EM}_{fn}(\vx_2)\rho^{W}_{ni}(\vx_1)+\rho^{W}_{fn}(\vx_2)\rho^\text{EM}_{ni}(\vx_1)\right]
\nn\\
&\qquad\times  V(\vx_1,\vx_2,\vx_2) \notag\\
&+\int_{\vx}\rho^{W}_{fn}(\vx) V(\vx,\vx,\vx)\,.
\label{eq:coordinate}
\end{align}

We could, in principle, use the above to evaluate the NMEs of the Fermi potentials in Eqs.~\eqref{eq:pots} and \eqref{eq:potsVE}. However, we would encounter apparent divergences in the one-body terms of Eq.~\eqref{eq:coordinate}, proportional to $V(\vx,\vx)$ or $V(\vx,\vx,\vx)$, when considering the logarithmic potentials in Eq.~\eqref{eq:pots}.
As the left-hand side of Eq.~\eqref{eq:coordinate} converges, this implies the appearing divergences should cancel between the one-, two-, and three-body terms, if one sums over a complete set of states. 
However, when approximating these expressions by restricting the sums to the initial or final states, these cancellations are no longer explicit. To estimate the ${\mathcal O}(\alpha^2)$ potentials in Eqs.~\eqref{eq:pots}, it is therefore useful to rewrite the one-body terms by again using the anticommutation relations,
\begin{align}
&\rho^W_{fi}(\vx_1)\delta^3(\mathbf{r}_{12}) \notag\\
&= \sum_n \left(\rho^\text{EM}_{fn}(\vx_1) \rho^{W}_{ni}(\vx_2)-\rho^{W}_{fn}(\vx_2) \rho^\text{EM}_{ni}(\vx_1)\right)\,.
\end{align}
This allows us to express Eq.~\eqref{eq:coordinate} as
\begin{align}
M_\text{2b}&=
 \sum_n \int_{\vx_i}\rho^{W}_{fn}(\vx_2)\rho^\text{EM}_{ni}(\vx_1)V(\vx_1,\vx_2)\,,\nn\\
M_\text{3b}&= 
\sum_{m,n}\int_{\vx_i}\rho^\text{EM}_{fm}(\vx_3)\rho^\text{EM}_{mn}(\vx_2)\rho^{W}_{ni}(\vx_1)V(\vx_1,\vx_2,\vx_3)\nn\\
&-
\sum_n\int_{\vx_i}\left[\rho^\text{EM}_{fn}(\vx_2)\rho^{W}_{ni}(\vx_1)+2\rho^{W}_{fn}(\vx_2)\rho^\text{EM}_{ni}(\vx_1)\right]
\nn\\
&\qquad\times  V(\vx_1,\vx_2,\vx_2)\,,
\label{eq:coordinate2}
\end{align}
which can be evaluated numerically when restricting the sums to a single state. 
Going through the same steps, but choosing different orderings of the charge and weak bilinears in Eq.~\eqref{eq:Appdensities}, then leads to the other two relations in Eq.~\eqref{eq:densities3b}. 
Finally, the matrix elements discussed above are relevant for $\beta^-$ decays. However, since the potentials describing $\beta^\pm$ decays are related by Hermitian conjugation, their matrix elements can be related by 
\begin{equation}
\langle f| \hat V_{\beta^-}|i\rangle  = \langle i| \hat V_{\beta^+}|f\rangle^* \,.
\end{equation}
The expressions for $\beta^+$ decays can therefore be obtained from those above by replacing $\rho_{ab}^W \to  (\rho_{ab}^W)^* $, 
$\rho_{ab}^\text{EM} \to \rho_{ba}^\text{EM}$,
and $f\leftrightarrow i$. 

As discussed in Sec.~\ref{sec:fermi} and can be seen from Table~\ref{tab:Fermi}, these density estimates agree rather well for the two-body matrix elements, especially in the case of $M^{E}_{\text{F}, p}$. The fact that the approximation does not work as well for $M^{+}_{\text{F}, p}$ seems to be due to the fact that the corresponding potential changes sign, which leads to a partial cancellation, making it more sensitive to the details of the charge distribution.

As a further consistency check, we discuss a simpler Fermi-gas estimate in App.~\ref{app:FermiGas}.

\section{Fermi-gas estimate}
\label{app:FermiGas}

Here we discuss Fermi-gas estimates of the three-body matrix elements. We start by rewriting the transition matrix elements in terms of matrix elements that only involve the initial or final state,
by making use of the fact that these are part of the same isospin triplet. There are two types of $\beta^+$  transitions, namely those with either $T_z=1$ or $T_z=0$ in the initial state, which can be rewritten as
\begin{align}
\langle 1,0| \hat V | 1,-1\rangle &= \frac{1}{\sqrt{2}}
 \langle 1,-1| [J_+,\hat V] |1, -1\rangle \,,\nonumber\\
 \langle 1,1| \hat V | 1, 0\rangle &= -\frac{1}{\sqrt{2}}
\langle 1,1| [J_+,\hat V] | 1, 1\rangle \,,
\end{align}
where $J_+ = \int_{\vx} N^\dagger \tau^+ N(\vx)$ is the isospin-raising operator and the states are labeled by their isospin, $|T, T_z\rangle$. The matrix elements of these commutators can now be evaluated using usual Hartree--Fock techniques by approximating the nuclear $T_z=\pm1$ states by Fermi-gas wave functions. The modified  potentials, $\hat V' = [J_+,\hat V]$, have the same form as the two- and three-body potentials in Eq.~\eqref{eq:Appdensities}, but with their isospin structure in Eq.~\eqref{eq:Visospin} replaced by
\begin{align}\label{eq:VprimeIsospin}
V_{\beta_1\beta_2}^{\alpha_1\alpha_2} &\to
(\tau^3_{\alpha_1\beta_1} Q_{\alpha_2\beta_2}-\tau^-_{\alpha_1\beta_1} \tau^+_{\alpha_2\beta_2}) V\,,\nonumber\\
V_{\beta_1\beta_2\beta_3}^{\alpha_1\alpha_2\alpha_3} &\to 
(\tau^3_{\alpha_1\beta_1} Q_{\alpha_2\beta_2}-2\tau^-_{\alpha_1\beta_1} \tau^+_{\alpha_2\beta_2}) Q_{\alpha_3\beta_3} V\,.
\end{align}
The matrix elements of the $\hat V'$ potentials in the Fermi gas state are obtained by contracting the appearing nucleon fields in all possible ways and using the finite density propagator in momentum or coordinate space,\footnote{These propagators can be derived from noninteracting nucleons confined to a box of radius $R_A$. In the limit where the number of nucleons tends to infinity, the resulting constant momentum distribution can be taken to be continuous and boundary effects (which would lead to propagators that depend on both the relative and CM coordinates) can be neglected.}
\begin{align}
D_N(q) &= \frac{i}{v\cdot q+i\epsilon}-2\pi \delta (v\cdot q)\theta(k_N-|\vq|)\,,\notag\\
D_N(r_0,\mathbf{r}) &= \theta(r_0)\delta^3(\mathbf{r})\nonumber\\
&+\frac{1}{2\pi^2 r^3}\left(r k_N\cos (k_N r)-\sin (k_N r)\right)\,,
\end{align}
where $r=|\mathbf{r}|$ and $k_N$ is the Fermi momentum for neutrons ($N=n$) or protons ($N=p$), which is related to the density by $\frac{k_N^3}{6\pi^2} = \rho_N$. 
In practice, only the nonzero density pieces involving $k_N$ contribute, and, 
for potentials with isospin structure as in Eq.~\eqref{eq:VprimeIsospin}, one obtains
\begin{align}\label{eq:FermiGasApprox}
\langle \hat V_\text{2b}'\rangle & = \int_{\vx_1 \vx_2}\bigg[
4\Big(D_p(0)-D_n(0)\Big)D_p(0)\\
&\qquad-2\Big(D_p(r_{12})
-D_n(r_{12})\Big)D_p(r_{12})
\bigg]V_\text{2b}(r_{12})\,,\nonumber\\
\langle \hat V_\text{3b}'\rangle & = \int_{\vx_1 \vx_2,\vx_3}\bigg[
-8\Big(D_p(0)-D_n(0)\Big)\big[D_p(0)\big]^2
\nonumber\\
&
\qquad+4\Big(3D_p(0)-D_n(0)\Big)\big[D_p(r_{12})\big]^2\nonumber\\
&\qquad-4 D_p(r_{12})D_p(r_{23})D_p(r_{31})
\nonumber\\
&\qquad-8 D_p(0)D_p(r_{12})D_n(r_{12})\notag\\
&\qquad+4 D_n(r_{12})D_p(r_{23})D_p(r_{31})
\bigg]V_\text{3b}(r_{12},r_{23})\,,\notag
\end{align}
where $r_{ij} = |\vx_i-\vx_j|$ and the integrals are restricted to $|\vx_i|\leq R_A$.
As the integrands only depend on the relative coordinates, the integral over the CM coordinate can be performed analytically, and is related to the overlap of two or three spheres. 

The resulting matrix elements for the $V_{+,E}$ and $V_+^\text{3b}$ potentials are shown in Table~\ref{tab:FermiGas}. Comparing with Tables~\ref{tab:Fermi} and \ref{tab:Fermi_3b}, we see that the estimates for the $V_{E}$ and $V_+^\text{3b}$ potentials agree within $(10\text{--}20)\%$ with the full many-body calculations and density estimates, respectively. In contrast, the matrix elements related to $V_{+}$ do not reproduce the correct sign. This is related to the partial cancellation between contributions from shorter and longer distances, which occur because the $V_+$ potential changes sign. This avoids the coherent enhancement as observed for $V_+^\text{3b}$, but also renders the result of the cancellation more sensitive to the nuclear model.

\begin{table}
\centering
\renewcommand{\arraystretch}{1.3}
\begin{tabular}{c rrr}
\toprule
Transition  &   { $M^{E}_{\text{F}, p}$} &$M^{+}_{\text{F}, p}$ & $M^{+\, \text{3b}}_{\text{F}, p}$
\\\colrule
{$^{10}$C $\rightarrow$ $^{10}$B}     & $-4.3 $&$ 0.11 $&$ -9.4$\\ 
{$^{14}$O $\rightarrow$ $^{14}$N}  
  &$ -5.7 $&$  0.22 $&$-17.1
$\\
$^{26}$Si $\rightarrow$ $^{26}$Al&$ -10.1$&$ 0.60 $&$ -54.1$\\
{$^{34}$Ar $\rightarrow$ $^{34}$Cl} &$ 
 -13.0 $&$  0.87 $&$-90.2$\\

{$^{38}$Ca $\rightarrow$ $^{38}$K}  &$ -14.5  $&$ 1.0 $&$ -112$\\

\colrule
{$^{26}$Al $\rightarrow$ $^{26}$Mg} 
&$  -8.7 $&$  0.50 $&$-39.7 $\\

{$^{34}$Cl $\rightarrow$ $^{34}$S} &$
-11.6 $&$  0.76 $&$ -71.3 $\\ 

{$^{38}$K $\rightarrow$ $^{38}$Ar} &$ 
 -13.0   $&$ 0.90 $&$-90.5
$\\ 
{$^{42}$Sc $\rightarrow$ $^{42}$Ca} &$
 -14.5 $&$  1.0 $&$ -112
$\\

{$^{46}$V $\rightarrow$ $^{46}$Ti}
 &$ -15.9 $&$ 1.2 $&$-136 $\\

 {$^{50}$Mn $\rightarrow$ $^{50}$Cr} 
  &$ -17.4 $&$1.3 $&$ -162
 $\\
 {$^{54}$Co $\rightarrow$ $^{54}$Fe} 
  &$ -18.8  $&$1.5  $&$-190$ \\
\botrule
  \end{tabular}
  \caption{Fermi matrix elements computed using the Fermi-gas approximation in Eq.~\eqref{eq:FermiGasApprox}, see main text for details.}
  \label{tab:FermiGas}
\end{table}

\section{Higher-body potentials at \texorpdfstring{$\boldsymbol{{\mathcal O}(\alpha^{4,6})}$}{}}
\label{app:higher-body}

Additional $n$-body potentials, with $n>3$, are induced beyond ${\mathcal O}(\alpha^2)$. Although the exact form of such potentials will require new explicit calculations, here we discuss the coefficients of certain logarithmic terms that can be predicted from the RG equations alone.  
These arguments allow for an order-of-magnitude estimate of the higher-body terms that are currently not included in our calculations. This is  especially useful for large-$Z$ nuclei as the $M_{\text{F},p}^{+ \,\text{3b}}$ matrix elements scale as $Z^2$, while higher-body terms are expected to grow with even larger powers of $Z$. 

The RG arguments start from the fact that the effective coupling, $C_\text{eff}^{(g_V)}$, in Eq.~\eqref{eq:Ceff}
should follow the RG equation in Eq.~\eqref{eq:CeffRG}, which is known to all orders in $(\alpha Z)^n$ and $\alpha^n Z^{n-1}$. The known ${\mathcal O}(\alpha^2)$ terms in $C_\text{eff}^{(g_V)}$,  together with the RG equation, therefore imply $\log \mu_\pi$-dependent terms at higher orders in $\alpha$ and $Z$. Each logarithm $\propto \log \mu_\pi$ requires a  compensating scale, which can only be provided by higher-body analogs of ${\mathcal V}^{}_{ +}$ and ${\mathcal V}^\text{3b}_{ +}$, implying that the $\log\mu_\pi$ dependence determines terms of the form $\propto \log r_{ij}$ in these potentials.

A generalized form of Eq.~\eqref{eq:Ceff} can be written as\footnote{Note that  the $n$-body ${\mathcal V}^{n\text{b}}_{\delta}$ and ${\mathcal V}^{n\text{b}}_{+}$ potentials contribute to the amplitude as $\propto\sum_n \big(C^{n\text{b}}_\delta {\mathcal V}^{n\text{b}}_{\delta}+C_+^{n\text{b}}{\mathcal V}^{n\text{b}}_{+}\big)$. To ${\mathcal O}(\alpha^2)$, these terms can be factorized between $C_\text{eff}^{(g_V)}$ (containing ${\mathcal V}_{\delta}$) and $\bar\delta_\text{NS}$ (which includes ${\mathcal V}_{+}$), as done in the main text. However, beyond this order it is more natural to use the nonfactorized form of Eq.~\eqref{eq:Ceffapp}, in which both types of potentials enter $\bar C^{(g_V)}_{\text{eff}} = g_V\left[1+\sum_n \big(C^{n\text{b}}_\delta {\mathcal V}^{n\text{b}}_{\delta}+C_+^{n\text{b}}{\mathcal V}^{n\text{b}}_{+}\big)+\dots\right]$ additively.}
\bea\label{eq:Ceffapp}
\bar C^{(g_V)}_{\text{eff}}(\mu_\pi) &=& g_V(\mu_\pi)\Bigg[1+ \sum_{n=1}\sum_{m\leq n} c_{n}^{(m)}(\mu_\pi)\alpha^nZ^{n-m}
\Bigg]\,.\nonumber\\
\eea
The first few ${\mathcal O}(\alpha^nZ^n)$ and ${\mathcal O}(\alpha^nZ^{n-1})$ terms are given by 
\begin{align} \label{eq:cijCoeff}
c_{1}^{(0)} &=0 \,,\nonumber\\ 
 \alpha^2Z^2 c_2^{(0)} &=Z^2 C_\delta^\text{3b} +C_+^\text{3b}\frac{M_{\text{F},p}^{+\,\text{3b}}}{M_\text{F}^{(0)}}\equiv -\frac{\alpha^2Z^2}{2}\log \frac{\mu_\pi}{\Lambda_2^{(0)}}\,,\nonumber\\
\alpha^2 Z c_2^{(1)} &= 
ZC_{\delta}+\frac{M_{\text{F},p}^+}{M_\text{F}^{(0)}}-(2\mp 1)\bigg(Z C_{\delta}^\text{3b}+ \frac{M_{\text{F},p}^{+\,\text{3b}}}{M_\text{F}^{(0)}}\bigg)\nonumber\\
&\equiv\mp \frac{\alpha^2Z}{2}\log \frac{\mu_\pi}{\Lambda_2^{(1)}}\,,
\end{align}
where the $\Lambda_n^{(m)}$ are defined in such a way that they absorb all $\mu_\pi$-independent terms.\footnote{Although we only introduce a single parameter at each order, this definition strictly speaking implies that the $\Lambda_i^{(j)}$ can be different for $\beta^\pm$ decays.} With the above coefficients, $\bar C_\text{eff}^{(g_V)}$ reproduces the RG equation in Eq.~\eqref{eq:CeffRG} up to ${\mathcal O}(\alpha^2Z^2)$ and ${\mathcal O}(\alpha^2Z)$, but  requires additional $\mu_\pi$-dependent terms at higher orders in $\alpha$ and $Z$. Equating terms of the same order in $\alpha$  and $Z$ in Eq.~\eqref{eq:CeffRG}, one finds that the $c_n^{(m)}$ coefficients can generally be written as polynomials in $L_{i,j}\equiv \log\frac{\mu_\pi}{\Lambda_i^{(j)}}$.
Approximating $\gamma^{(g_V)} = \sqrt{1-\alpha^2 Z(Z\pm1)}-1$,  this leads to the following forms for the leading and next-to-leading terms in $Z$ at ${\mathcal O}(\alpha^4)$ and ${\mathcal O}(\alpha^6)$,
\begin{align}\label{eq:Ceffcoeffs}
c_4^{(0)}(\mu_\pi)&=\frac{1}{8}\left(L_{2,0}^2-L_{4,0}\right)\,,\notag\\
c_4^{(1)}(\mu_\pi)&=\pm\frac{1}{8}\left(L_{2,0}^2+L_{2,1}^2-2L_{4,1}\right)\,,\nonumber\\
c_6^{(0)}(\mu_\pi)&=
\frac{1}{16}\left(-\frac{1}{3}L_{2,0}^3+\frac{1}{2}L_{2,0}^2+\frac{1}{2}L_{4,0}^2-L_{6,0}\right)
\,,\nonumber\\
c_6^{(1)}(\mu_\pi)&=\pm\frac{1}{16}\Big(-\frac{2}{3}L_{2,0}^3+L_{2,0}^2-\frac{1}{3}L_{2,1}^3+\frac{1}{2}L_{2,1}^2\nonumber\\
&+\frac{1}{2}L_{4,0}^2+L_{4,1}^2-3 L_{6,1}\Big)
\,.
\end{align}
The coefficient at each order involves an integration constant not determined by the RG, which can be written in terms of $\log \Lambda_n^{(m)}$.
These $\Lambda_n^{(m)}$ are  fixed by the requirement that they absorb the $\mu_\pi$-independent terms at the corresponding order, as in Eq.~\eqref{eq:cijCoeff} for $\Lambda_2^{(0,1)}$. As they have mass dimension $1$ and are related to terms with a factor of $Z^{n-m}$, dimensional analysis implies these should be related to logarithmic terms in the $(n-m+1)$-body potential.
Similar to ${\mathcal V}_+$ and ${\mathcal V}^\text{3b}_+$, such a term for the $N$-body potential would take the form 
\begin{align}\label{eq:VplusPots}
C_+^{N\text{b}}{\mathcal V}_+^{N\text{b}}& = C_+^{N\text{b}} \sum_{i_1\neq i_2\dots \neq i_N}\tau^{\mp (i_1)}P_p^{(i_2)}\dots P_p^{(i_N)}\nonumber\\
&\qquad\times \left[v_N\log r_N+f(r_{i_1},\dots, r_{i_N})\right]\,,
\end{align} 
where $r_N$ is a function (with mass dimension $-1$) of the distances between the $N$ interacting nucleons, $v_N$ is a coefficient that can be read off from Eq.~\eqref{eq:Ceffcoeffs}, and $f(r_{i_1},\dots, r_{i_N})$ captures nonlogarithmic terms not determined by the RG. 
In particular, the ${\mathcal O}(\alpha^4)$ and ${\mathcal O}(\alpha^6)$ logarithmic terms in the $4\text{--}7$-body potentials are related to the $c_n^{(m)}$ coefficients by 
\begin{align}
\alpha^4\dot c_4^{(0)}&=-C_+^\text{5b}  \,,  &\alpha^4\dot c_4^{(1)}&=-\left(C_+^\text{4b} -(8\mp 2)C_+^\text{5b}\right)\,,\nonumber\\
\alpha^6\dot c_6^{(0)}&=-C_+^\text{7b}  \,,&\alpha^6 \dot c_6^{(1)}&=-\left(C_+^\text{6b} -(18\mp 3)C_+^\text{7b}\right)\,,
\end{align}
where $\dot c_i^{(j)} = d c_i^{(j)}/d\log\Lambda_i^{(j)}$. Note that the expressions for the $c_n^{(1)}$ coefficients are more complicated as they receive contributions from both the $(n+1)$- and $n$-body potentials. This then leads to
\begin{align}
C_+^\text{5b}&= -\frac{1}{8}\alpha^4 \,, 
&C_+^\text{4b}&= -\alpha^4\,,\nonumber\\
C_+^\text{7b}&=  -\frac{1}{16}\alpha^6\,,\qquad 
&C_+^\text{6b}&= -\frac{9}{8}\alpha^6\,,
\end{align}
where the first line is determined by $c_{4}^{(0)}$ and $c_{4}^{(1)}$, while $c_{6}^{(0)}$ and $c_{6}^{(1)}$ determine the second. 

Although the above arguments only determine part of the higher-body potentials, they allow for a crude estimate of the logarithmic terms. Here we focus on the $\alpha^4 Z^4$ contribution and assume the dependence of the radii in Eq.~\eqref{eq:VplusPots} to take the form 
$\log r_N+ f\to \log \frac{r_N}{ R_A}$,
where $r_N = (r_{12}+r_{13}+r_{14}+r_{15}+r_{23}+r_{24}+r_{25}+r_{34}+r_{35}+r_{45})/10$, in analogy to the ${\mathcal V}_+^\text{3b}$ potential.
Evaluating this potential using the density approximation described in App.~\ref{app:density} leads to $\langle {\mathcal V}_+^\text{5b}\rangle\simeq 6.4\times 10^{4}$, so that $C_+^\text{5b} \langle {\mathcal V}_+^\text{5b}\rangle\simeq -2\times 10^{-5}$ for the largest $Z$ we consider, $^{54}$Co $\rightarrow$  $^{54}$Fe.

\section{Two-body operators in the harmonic-oscillator basis}
\label{app:TBME}

The reduced matrix element of a scalar operator between harmonic oscillator states can be written in terms of the relative and CM components~\cite{Mihaila:2011yi, Payne:2018},
\begin{align}
 \label{eq:tmbe_scalar}
    {}^{\mu}{\mathcal V}^{(\mathbf{pq}:J)'}_{\mathbf{\ pq}:J} &= \delta_{JJ'} \hspace{-4pt}\sum_{\substack{N\Lambda \\ N'\Lambda'}} \sum_{\substack{L'S' \\ n'\lambda'}}
    \sum_{\substack{LS \\ n\lambda}}{}^{\mu}\hat{\chi}^{S'}_{S}
    {}^{\mu}\hat{\mathcal{U}}^{(N\Lambda n\lambda:L)'}_{\ N\Lambda n\lambda:L} \\
    &\quad\times G^{(LSJ\mathbf{pq})'}_{(N\Lambda n\lambda)'} G^{N\Lambda n\lambda}_{LSJ\mathbf{pq}}\sixjj{L'}{S'}{J}{S}{L}{\mu}(-1)^{P}\,.\notag
\end{align}
Here, $L,S,J$ are the orbital angular momentum, total spin, and total angular momentum of the coupled two-body state; $N,\Lambda$ are the two-body CM radial and orbital angular momentum quantum numbers; and $n,\lambda$ are the two-body relative radial and orbital angular momentum quantum numbers, respectively. The oscillator states are denoted by $\mathbf{p},\mathbf{q}$, and primed (unprimed) values indicate final (initial) states. The spherical tensors, ${}^{\mu}\hat{\chi}$ and ${}^{\mu}\hat{\mathcal{U}}$, are the spin--isospin and spatial components of $\mathcal{V}$ both with rank $\mu$. The curly brackets are the Wigner 6j symbol~\cite{Varshalovich:1988ifq} and $P=S'+J+L$.

The coefficient $G$ combines the $jj\rightarrow LS$ recoupling and the transformation from single-particle oscillator states to relative and CM states~\cite{Miyagi:2023qce},
\begin{align}
    G^{LSJ\mathbf{pq}}_{N\Lambda n\lambda} &= [j_{p}][j_{q}][L][S]
    \ninejj{l_{p}}{\sfrac{1}{2}}{j_{p}}{l_{q}}{\sfrac{1}{2}}{j_{q}}{L}{S}{J}A^{\lambda S}_{\mathbf{pq}} \nonumber\\
    &\quad\times \inner{N\Lambda,n\lambda}{n_{p}l_{p},n_{q}l_{q}}_{1}\,.
\end{align}
Here, the curly brackets are the Wigner 9j symbol, the angled bracket is the Talmi--Moshinsky bracket for equal masses~\cite{Talmi:1952, Moshinsky:1959qbh, Kamuntavicius:2001pf}, the square brackets are shorthand for $[I]\equiv \sqrt{2I+1}$, and $A^{\lambda S}_{\mathbf{pq}} = 1 + \delta_{t_{z_{p}}t_{z_{q}}}(-1)^{\lambda+S}$ is the antisymmetrization operator. In the Talmi--Moshinsky transformation, we use the scaled relative and CM coordinates $\mathbf{r}=\mathbf{r}_{jk}/\sqrt{2}$ and $\mathbf{R}=(\mathbf{r}_j +\mathbf{r}_k)/\sqrt{2}$, such that radial functions defined in terms of the physical internucleon separation are evaluated at $r_{jk}=\sqrt{2}r$. The scaled coordinate $\rr$ is conjugate to the scaled relative momentum $k=\sqrt{2}\,p$ of App.~\ref{app:regulator}.

For operators that are independent of the CM coordinates, the spatial component reduces to
\begin{align}
    {}^{\mu}\hat{\mathcal{U}}^{(N\Lambda n\lambda:L)'}_{\ N\Lambda n\lambda:L} &= \delta_{NN'}\delta_{\Lambda\Lambda'}
    \langle n'\lambda'\vert h(r_{jk})\vert n\lambda\rangle \nonumber\\
    &\quad\times [L][L']\sixjj{\lambda'}{L'}{\Lambda}{L}{\lambda}{\mu}(-1)^{P'}\,,
    \label{eq:tmbe_scalar0}
\end{align}
with $P'=L'+\lambda+\Lambda+\mu$.

The full expressions for the different two-body matrix elements are written below. For readability, the isospin operators in Eqs.~\eqref{eq:OF}--\eqref{eq:Oso} are left implicit in the expressions below. Also, the transformation coefficient indices are suppressed and their product is simply written as $GG'$. Additionally, the $\delta_{JJ'}$ are suppressed, and the summations in Eq.~\eqref{eq:tmbe_scalar} are simply written as $\sum_{\cdots}$. Lastly, the operator $\mathcal{V}^{\mathcal X}_{i,{\rm N}}$ is simply written as $\mathcal{V}$, and the reduced, regulated matrix element is denoted as $\langle \mathbf{p'q'}\hspace{-3pt}:\hspace{-2pt}J \vert\vert \mathcal{V}^{\mathcal X}_{i,{\rm N}}(r_{jk}) \vert\vert \mathbf{pq}\hspace{-1pt}:\hspace{-2pt}J \rangle_{\Lambda_\text{reg}}\equiv {\mathcal V}^{\mathbf{p'q'}:J}_{\mathbf{\ p\ q}:J}$.

The two-body Fermi transition has $\mu=0$, $\hat{\chi}=\mathds{1}$, $\hat{\mathcal{U}}=h(r_{jk})$, and simplifies to
\begin{align}
    \text{F}:\hspace{5pt}{\mathcal V}^{\mathbf{p'q'}:J}_{\mathbf{\ p\ q}:J} &= \sum_{\cdots}GG' \times\delta_{NN'}\delta_{\Lambda\Lambda'}\delta_{LL'}\delta_{SS'}\delta_{\lambda\lambda'} \nonumber\\
    &\times \langle n'\lambda' \vert h^{\mathcal X}_{\rm F,N}(r_{jk})\vert n\lambda\rangle_{\Lambda_\text{reg}}\,.
\end{align}

The two-body GT transition has $\mu=0$, $\hat{\chi}=\boldsymbol{\sigma}_{j}\cdot\boldsymbol{\sigma}_{k}$, $\hat{\mathcal{U}}=h(r_{jk})$, and with the spin function $\mathcal{S}(S) = 4S-3$, simplifies to
\begin{align}
    \text{GT}:\hspace{5pt}{\mathcal V}^{\mathbf{p'q'}:J}_{\mathbf{\ p\ q}:J} &=
    \sum_{\cdots}GG' \times \delta_{NN'}\delta_{\Lambda\Lambda'}\delta_{LL'}\delta_{SS'}\delta_{\lambda\lambda'} \nonumber\\
    &\times [S]\mathcal{S}(S) \langle n'\lambda' \vert h^{\mathcal X}_{\rm GT,N}(r_{jk})\vert n\lambda\rangle_{\Lambda_\text{reg}}\,.
\end{align}

The two-body tensor transition has $\mu=2$, $\hat{\chi}={}^{2}[\boldsymbol{\sigma}_{j}\otimes\boldsymbol{\sigma}_{k}]$, and the first nontrivial tensor coupling in the spatial component with $\hat{\mathcal{U}}={}^{2}[\hat{\boldsymbol{r}} \otimes \hat{\boldsymbol{r}}]\ h(r_{jk})$ simplifies to
\begin{align}
    {\text{T}}:\hspace{5pt}{\mathcal V}^{\mathbf{p'q'}:J}_{\mathbf{\ p\ q}:J} &=
    2\sqrt{30}\sum_{\cdots}GG' \times \delta_{NN'}\delta_{\Lambda\Lambda'}\delta_{S1} \delta_{S'1} \nonumber\\
    \times &[\lambda'][L][L']
    \sixjj{\lambda'}{L'}{\Lambda}{L}{\lambda}{2}
    \sixjj{L'}{1}{J}{1}{L}{2}(-1)^{P_T}\nonumber\\
    \times &\inner{\lambda'_{0}2_{0}}{\lambda_{0}} \times \langle n'\lambda' \vert h^{\mathcal X}_{\text{T},N}(r_{jk})\vert n\lambda\rangle_{\Lambda_\text{reg}}\,,
\end{align}
with $P_T = 1+J+L+L^{'}+\lambda+\Lambda$.

Both the relative and CM two-body spin-orbit transitions have $\mu=1$ and $\hat{\chi}=\boldsymbol{\sigma}^{(j)}P_{N}^{(k)} \pm P_{N}^{(j)}\boldsymbol{\sigma}^{(k)}$. The spin--isospin components use the matrix elements
\begin{equation}
    \mathcal{S}_{a}(S',S)\equiv \langle S'\vert\vert \boldsymbol{\sigma}^{(a)}\vert\vert S\rangle=\sqrt{3(S'+S)}(-1)^{S_{a}+1}\,,
\end{equation}
where $a=j,k$ and $S_{j}=S',\ S_{k}=S$. Then these are combined to give, 
\begin{equation}
    \mathcal{S}^{\pm}_{N}(S',S)\equiv \mathcal{S}_{j}(S',S)P_{N}^{(k)} \pm P_{N}^{(j)}\mathcal{S}_{k}(S',S)\,,
\end{equation}
where $+$ ($-$) refers to the relative (CM) component.

The spatial component of the two-body relative spin-orbit operator is simply the relative angular momentum with respect to the scaled relative coordinate, $\mathbf{L}_{jk} = -i(\mathbf{r}\times\boldsymbol{\nabla}_{r})$. Then the full expression reduces to
\begin{align}
    \text{LS}:\hspace{5pt}{\mathcal V}^{\mathbf{p'q'}:J}_{\mathbf{\ p\ q}:J} &=
    \frac{1}{2}\sum_{\cdots}GG' \times \delta_{NN'}\delta_{\Lambda\Lambda'}\delta_{\lambda\lambda'}
    \mathcal{S}^{+}_{N}(S',S)\nonumber\\
    \times &\sqrt{\lambda(\lambda+1)}
    \sixjj{\lambda}{L'}{\Lambda}{L}{\lambda}{1}
    \sixjj{L'}{S'}{J}{S}{L}{1}(-1)^{P_{LS}}\nonumber\\
    \times &[L][L'][\lambda] \times \langle n'\lambda' \vert h_{\text{LS},N}(r_{jk})\vert n\lambda\rangle_{\Lambda_\text{reg}}\,,
\end{align}
with $P_{LS}=1+S'+J+L+L'+\lambda+\Lambda$.

\begin{figure*}[t]
\centering
\includegraphics[width=0.45\textwidth]{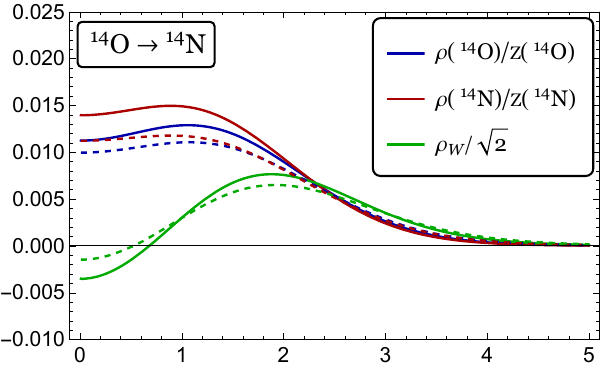}\quad
\includegraphics[width=0.45\textwidth]{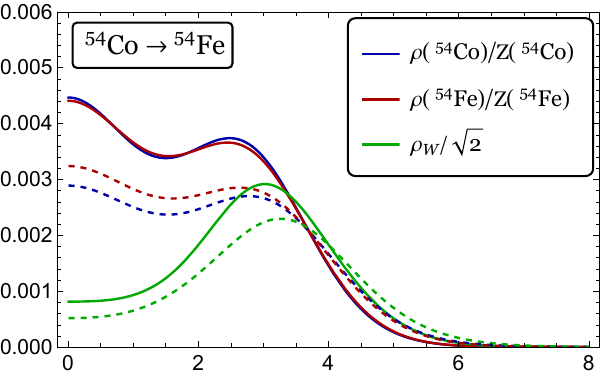}
\caption{Proton densities and weak transition density from CC calculations for \intem~(solid) and \intsat~(dashed), for the examples of $^{14}$O $\rightarrow$ $^{14}$N (left) and $^{54}$Co $\rightarrow$ $^{54}$Fe (right).}
\label{fig:CC_distributions}
\end{figure*}

In the Talmi--Moshinsky coordinates, the CM component of the spin-orbit operator becomes $\mathbf{L}_{jk}^\text{CM} = -i(\mathbf{r}\times\boldsymbol{\nabla}_{R})$. This introduces the first nontrivial CM component~\cite{Varshalovich:1988ifq},
\begin{align}
    \langle N'\Lambda' \vert\vert {\boldsymbol{\nabla}}_{R} \vert\vert N\Lambda \rangle =&\ \delta_{\Lambda',\Lambda+1}\sqrt{\Lambda+1}\ \mathcal{A} \nonumber\\
    &- \delta_{\Lambda',\Lambda-1}\sqrt{\Lambda}\ \mathcal{B}\,,
\end{align}
where~\cite{Scalesi:2023qdh}
\begin{align}
    \mathcal{A} &\equiv \langle N'\Lambda'\vert \frac{\partial}{\partial R} - \frac{\Lambda}{R} \vert N\Lambda\rangle\,, \nonumber \\
    \mathcal{B} &\equiv \langle N'\Lambda'\vert \frac{\partial}{\partial R} + \frac{\Lambda+1}{R} \vert N\Lambda\rangle\,.
\end{align}
Then the full expression is
\begin{align}
    {\rm LS^{CM}}\hspace{-2pt}:\hspace{-1pt}{\mathcal V}^{\mathbf{p'q'}:J}_{\mathbf{\ p\ q}:J} &=
    \frac{\sqrt{6}}{2}\sum_{\cdots}GG'\ \langle N'\Lambda' \vert\vert \boldsymbol{\nabla}_{R} \vert\vert N\Lambda \rangle \nonumber\\
    \times &\inner{\lambda'_{0}1_{0}}{\lambda_{0}}\ \mathcal{S}^{-}_{N}(S',S) (-1)^{S'+J+L+1} \nonumber\\
    \times &[L][L'][\lambda'] \sixjj{L'}{S'}{J}{S}{L}{1}\ninejj{\Lambda'}{\lambda'}{L'}{\Lambda}{\lambda}{L}{1}{1}{1} \nonumber\\
    \times & \langle n'\lambda' \vert r h_{\text{LS},N}(r_{jk})\vert n\lambda\rangle_{\Lambda_\text{reg}}\,.
\end{align}
Here, the explicit $r$ is the scaled relative coordinate, while $h_{\text{LS},N}$ remains a function of the physical separation $r_{jk}$.

\section{Transition densities and radii from coupled cluster}
\label{app:CCdensities}

The point-proton and weak transition densities are one-body densities between the correlated CC states of Sec.~\ref{sec:NMEs}. Since the reference states are deformed, we use the angle-averaged densities, evaluated on a Gauss--Legendre grid in $r$. With $\hat\rho_p(r)$ and $\hat\rho_W(r)$ the angle-averaged operators $\sum_a P_p^{(a)}\delta^3(\rr-\rr_a)$ and $\sum_a \tau_a^{\pm}\delta^3(\rr-\rr_a)$, the proton density of the even--even nucleus is~\cite{Novario:2021low}
\begin{equation}
    \rho_p(r) = \Big\langle\Phi_0\Big|(1+\widetilde\Lambda)\,
    \overline{\hat\rho_p(r)}\Big|\Phi_0\Big\rangle\,,
\end{equation}
and that of the odd--odd nucleus is the analogous expectation value in the isobaric analog state, $\langle\Phi_0|\widetilde L_k\, \overline{\hat\rho_p(r)}\,R_k|\Phi_0\rangle$. For $^{26}$Al, $^{34}$Cl, and $^{38}$K, this state is reached from both even--even neighbors, giving the two results listed in Table~\ref{tab:charge_radii}.

The weak transition density is evaluated at each $r$ as in Eq.~\eqref{eq:CC_ME},
\begin{align}
    \rho_W(r) &= \Big|
    \Big\langle\Phi_0\Big|(1+\widetilde\Lambda)\,
    \overline{\hat\rho_W^\dagger(r)}\,R_k\Big|\Phi_0\Big\rangle \notag\\
    &\quad\times
    \Big\langle\Phi_0\Big|\widetilde L_k\,
    \overline{\hat\rho_W(r)}\Big|\Phi_0\Big\rangle
    \Big|^{1/2}\,,
\end{align}
with signs fixed as described in Sec.~\ref{sec:NMEs}; the operator densities in Fig.~\ref{fig:GTdensities} are obtained in the same way. Examples for the resulting distributions are shown in Fig.~\ref{fig:CC_distributions}.
As a check, $\rho_p$ integrates to $Z$ and $\rho_W$ to $M_\text{F}^{(0)}=\sqrt{2}$. The radii are
\begin{align}
\label{radii}
    R_p^2 &= \frac{4\pi}{Z}\int_0^\infty dr\, r^4 \rho_p(r)\,, \notag\\
    \big(R_W^{(0)}\big)^2 &= \frac{4\pi}{M_\text{F}^{(0)}}\int_0^\infty dr\, r^4 \rho_W(r)\,.
\end{align}

Since the densities are expectation values of one-body operators, $R_p$ is defined with respect to the laboratory origin rather than the nuclear CM. If the CC wave function factorizes into intrinsic and Gaussian CM parts~\cite{Hagen:2009pq}, the laboratory density is the intrinsic one folded with the CM Gaussian, and the charge radius follows as~\cite{Novario:2021low}
\begin{equation}
    R_\text{ch}^2 = R_p^2 + r_p^2 + \frac{N}{Z}r_n^2 + \frac{3}{4m_N^2} - \frac{3b^2}{2A}\,,
\label{eq:Rch}
\end{equation}
where $r_p^2$, $r_n^2$ are the squared proton and neutron charge radii~\cite{Antognini:2013txn, ParticleDataGroup:2026mpi, Koester:1995nx, Kopecky:1997rw} and the oscillator length is $b^2=1/(\mN\hbar\omega)$. As the factorization holds only approximately, the CM term is an estimate. Analogously, the weak radius is calculated as 
\begin{equation}
    R_W^2 = \big(R_W^{(0)}\big)^2 + r_p^2 - r_n^2 + \frac{3}{4m_N^2} - \frac{3b^2}{2A}\,.
\label{eq:Rw}
\end{equation}

\begin{table*}[t]
    \centering
    \renewcommand{\arraystretch}{1.3}
    \begin{tabular}{c r r r r}
    \toprule
       Transition
   & $f t$ $[s]$     & 
   $\overline{ft}$ $[s]$ & $\mathcal P t$ $[s]$
   &  $\overline{\mathcal P} t$ $[s]$      \\
       \colrule
$^{10}$C $\rightarrow$ $^{10}$B   &   $3042.2(4.1)$ &  $3040.0(4.1)$  & $3169.1(4.3)$ & $3168.2(4.3)$\\
$^{14}$O $\rightarrow$ $^{14}$N     & $3042.2(8)$ & $3040.0(8)$  & $3164.9(1.2)$ & $3163.4(8)$ \\
$^{26}$Si $\rightarrow$ $^{26}$Al  & $3052.2(5.6)$ & $3048.7(5.9)$  & $3172.0(5.9)$ &  $3168.4(6.1)$ \\
$^{34}$Ar $\rightarrow$ $^{34}$Cl  & $3058.0(2.8)$ &  $3054.6(4.4)$  & $3177.2(3.0)$ & $3173.6(4.6)$ \\
$^{38}$Ca $\rightarrow$ $^{38}$K   & $3062.8(6.0)$ &  $3058.7(7.2)$   &   $3182.2(6.3)$& $3177.6(7.5)$\\
\colrule
$^{26}$Al $\rightarrow$ $^{26}$Mg  & $3037.6(7)$ &  $3035.0(1.4)$  & $3158.1(1.0)$ & $3155.7(1.4)$\\
$^{34}$Cl $\rightarrow$ $^{34}$S  & $3049.4(9)$ & $3045.8(2.7)$   & $3169.3(1.3)$& $3165.5(2.8)$ \\
$^{38}$K $\rightarrow$ $^{38}$Ar    & $3051.5(9)$ & $3048.0(3.2)$  & $3171.3(1.2)$ & $3167.4(3.3)$ \\
$^{42}$Sc $\rightarrow$ $^{42}$Ca  & $3047.7(1.2)$ & $3046.1(4.0)$  & $3167.8(1.5)$& $3165.7(4.2)$ \\
$^{46}$V $\rightarrow$ $^{46}$Ti   & $3050.3(5)$ & $3048.2(4.7)$   & $3170.2(1.0)$& $3167.4(4.8)$ \\
$^{50}$Mn $\rightarrow$ $^{50}$Cr &  $3048.4(1.2)$ & $3046.7(5.8)$  & $3168.2(1.5)$& $3165.6(6.0)$ \\
$^{54}$Co $\rightarrow$ $^{54}$Fe & $3050.8(1.4)$ & $3048.7(7.1)$   & $3170.7(1.6)$ & $3167.6(7.4)$\\
\botrule
\end{tabular}
    \caption{$ft$ values 
    and the product $\mathcal P$ used in the HT fit variants. The $ft$ values and $\delta_R^\prime$ are taken from Ref.~\cite{Hardy:2020qwl}. The barred quantities are the analogous quantities defined in the EFT formalism, see Eqs.~\eqref{eq:fbarEFT} and \eqref{eq:PbarEFT}. 
    }
    \label{tab:phase4}
\end{table*}

\section{Fits using the Hardy and Towner formalism}
\label{app:HT}

The EFT master formula contains several new ingredients, both in $\bar\delta_{\text{NS}}$ and in the treatment of $\mathcal O(\alpha Z E_e R)$ and $\mathcal O(\alpha^2)$ corrections that are traditionally considered as part of the phase-space factor. To assess the impact of the ab initio calculation of the NME for $\bar\delta_{\text{NS}}$, we therefore  performed fits in which all other ingredients are taken from Ref.~\cite{Hardy:2020qwl}. 
In practice, we start from the traditional master formula 
\begin{equation}
    ft (1 + \delta_R^\prime) (1 + \delta_{\text{NS}} - \delta_C )= \frac{K}{2 G_F^2 V_{ud}^2 (1 + \Delta_R^V)}\,,
\end{equation}
with 
\begin{equation}
K=\frac{2\pi^3\log 2}{m_e^5}=8120.27618(1)\times 10^{-10}\GeV^{-4}\text{s}\,,
\end{equation}
and replace
\begin{equation}
    \delta_{\text{NS}} \rightarrow  \delta^{(0)}_{\text{NS}} \Big|_{\alpha^2 \rightarrow 0}\,,
\end{equation}
where $\delta^{(0)}_{\text{NS}}$ is defined in Eq.~\eqref{deltaNS0}, and by $\alpha^2 \rightarrow 0$ we imply that we are setting $M^+_{\text{F}, N}$ and $M_{\text{F}, N}^{+\,\text{3b}}$ to zero in Eq.~\eqref{deltaNS0}. This is necessary as the terms required to cancel the $\Lambda$ dependence in 
$M_{\text{F}, N}^{+}$ and $M_{\text{F}, N}^{+\,\text{3b}}$ are missing in the standard approach. 
$\delta_C$ is the same as given in Table~\ref{tab:exp}, while the values of $ft$ and $\delta_R^\prime$ are taken from Table XVI of Ref.~\cite{Hardy:2020qwl} (neglecting the uncertainty in $\delta_R^\prime$).
The universal correction $\Delta_R^V$, see Eq.~\eqref{DeltaRV}, is updated to be consistent with the treatment of $g_V$ as described in Sec.~\ref{sec:CgVeff}.
We report the $ft$ values and the combination 
\begin{align}
\mathcal P =  ft  \left( 1 + \Delta_R^V \right) (1 + \delta_R^\prime)\,, 
\end{align}
in Table~\ref{tab:phase4}.

As discussed in Sec.~\ref{sec:fits}, the EFT and HT fit variants with zero or one LEC  
yield values of $V_{ud}$ that differ by about $4 \times 10^{-4}$. 
These differences primarily arise from the treatment of 
$\alpha^2 Z^2 \log R$ and 
$\alpha^2 Z \log R$ corrections,
and corrections scaling as $\alpha^n Z^n (E_e R)^m$. In the HT formalism,  
these 
corrections are mostly captured in the phase-space factor, either in the Fermi function
or in finite-size corrections, while 
$\alpha^2 Z \log R$ corrections are booked in  $\delta_R^\prime$.
Large logarithms of $E_e/m_n$ are resummed in $\delta_R^\prime$, while the logarithms
of $E_e/k_F$
in the Fermi function and in the 
$\alpha^2 Z$ corrections are included at fixed order.
In the EFT approach, the dependence on $\log R$ and linear terms in $E_e R$
arise from integrating out physics at the nuclear physics scale, $\mu \simeq k_F$, and they
require new matching coefficients and nuclear matrix elements, 
which are calculated at the scale $\mu_\pi$. These corrections appear in the
$C^{(g_V)}_{\text{eff}}$,
$\delta^E_{\text{NS}}$,
and $\mathcal O(\alpha^2)$ corrections to $\delta^{(0)}_{\text{NS}}$. 
The matching coefficient is then evolved to the scale $\mu = 2 E_0$,  resumming large logarithms. 
The Fermi function, the Sirlin function, and higher-order corrections in $\alpha$ arising from ultrasoft photons are then calculated at this low-energy scale. 
Because of the different approaches, it is not immediately obvious how to identify the origin of the difference in the extracted values of $V_{ud}$.
To facilitate the comparison between the EFT and HT fits, we can define the EFT analogs of $ft$ and $\overline{\mathcal P}$  
\begin{align}
\overline{f t} &=    \frac{|C_{\text{eff}}(\mu_\pi)|^2}{ \left[g_V(m_n)\right]^2} \frac{1 + \frac{5\alpha(2E_0)}{8\pi}}{1 + \frac{5\alpha(m_n)}{8\pi}} \Big[ 1 + \overline{\delta^E_{\text{NS}}} + \delta_{\text{NS}}^{(0)}\Big|_{\alpha^2}\Big]\bar f(\mu_\pi) t\,, \label{eq:fbarEFT} \\
\overline{\mathcal P} &=  \overline{P}  \Big[1 + \overline{\delta^E_{\text{NS}}} + \delta_{\text{NS}}^{(0)}\Big|_{\alpha^2}\Big]\,, \label{eq:PbarEFT}
\end{align}
where $\delta^{(0)}_{\text{NS}}|_{\alpha^2}$ denotes only the $\alpha^2$ corrections to $\bar\delta_{\text{NS}}$, and $\overline{P}$, $\bar f(\mu_\pi)$ are as in Eq.~\eqref{combscale} and Table~\ref{tab:phase_space}.
In the first quantity, $\overline{ft}$, we turn off the resummation of $\log \frac{2 E_0}{k_F}$
and $\log \frac{2 E_0}{m_n}$. The former is not included in the HT formalism, while the latter is included in $\delta_R^\prime$.
We notice that $\overline{ft}$ is a scheme-dependent object. The factors of $1 + \frac{5\alpha}{8\pi}$ serve to convert to the scheme for the Sirlin function used in Ref.~\cite{Hardy:2020qwl}, while a residual dependence at $\mathcal O(\alpha^2 Z)$
is not accounted for in Eq.~\eqref{eq:fbarEFT}~\cite{Crosas:2025xyv}.
The quantity $\overline{\mathcal P}$ is scheme independent, and 
it thus allows for a better comparison.
Table~\ref{tab:phase4} shows that $\overline{ft}$
and $\overline{\mathcal P}t$ are about $0.1\%$ smaller than $ft$ and $\mathcal P$
for all the isotopes we consider, resulting in a larger value of $V_{ud}$ in the EFT fits.
The fact that the ratios $ft/\overline{ft}$ and $\mathcal P/\overline{\mathcal P}$ are basically the same indicates that the treatment of
$g_V$ and of $\mathcal O(\alpha)$
and $\mathcal O(\alpha^2 Z)$ ultrasoft corrections is not responsible for significant differences. 
It is thus likely that the main origin of the difference 
is to be traced to the $M_{\text{F},p}^E$
and $M^{+ \text{3b}}_{\text{F},p}$
matrix elements.
These matrix elements are also responsible for the larger error in the EFT values in Table~\ref{tab:phase4}. These observations reiterate the  
importance of having fully controlled calculations of 
$M_{\text{F},p}^E$
and especially $M^{+\,\text{3b}}_{\text{F},p}$, for which it will be desirable to go beyond the density approximation presented here. 
 We stress, on the other hand, that replacing these matrix elements
with integrals of the charge distributions, as effectively done in the 
standard approach, might be too simplistic, and likely underestimates the uncertainty in $ft$.

\section{QMC matrix elements for \texorpdfstring{$\boldsymbol{^{10}}$}{}C and \texorpdfstring{$\boldsymbol{^{14}}$}{}O }
\label{app:NME10C}

In the fit variant (0B),
for $^{10}$C we use the GFMC matrix elements obtained with the NV2+3-Ia interaction~\cite{King:2025fph}.
The long-range magnetic and spin-orbit matrix elements are
\begin{align}\label{eq:QMC10}
   M_{\text{GT,p}}^{\text{mag}} &= -0.515\,, &    M_{\text{GT,n}}^{\text{mag}} &= 0.082\,, \nonumber \\ 
   M_{\text{T,p}}^{\text{mag}} + M_{\text{T,n}}^{\text{mag}} &= 0.002\,, & M_{\text{LS}, p} &= -0.005\,.
\end{align}
The two-body Fermi matrix elements $M_{\text{F},p}^E$ and $M_{\text{F},p}^+$ are
\begin{equation}
    M_{\text{F},p}^E = -4.74\,, \qquad  M_{\text{F},p}^+ = -1.42\,.
\end{equation}
The three-body matrix element is given in Eq.~\eqref{eq:QMC3b}, $M_{\text{F},p}^{+\, \text{3b}} = -9.10(1)$. The three-body matrix elements was calculated in VMC, but Ref.~\cite{King:2025fph} shows that the GFMC evolution has a small effect. For QMC matrix elements, we neglect the small statistical errors, and do not account for model errors.

The NMEs for the decay of $^{14}$O were presented in Ref.~\cite{Cirigliano:2024msg}, using the NNLO interaction of Ref.~\cite{Gezerlis:2014zia}.  Their values are
\begin{align}
   M_{\text{GT,p}}^{\text{mag}} &= -0.480\,, &    M_{\text{GT,n}}^{\text{mag}} &= 0.180\,, \nonumber \\ 
   M_{\text{T,p}}^{\text{mag}} + M_{\text{T,n}}^{\text{mag}} &= 0.016\,, & M_{\text{LS}, p} &= -0.020\,.
\end{align}
The biggest difference with respect to Ref.~\cite{Cirigliano:2024msg} concerns the value of the spin-orbit matrix element. Reference~\cite{Cirigliano:2024msg}
had a sign mistake in the radial function $h_{\text{LS}, p}$
and neglected the component proportional to ${\mathbf L}^{\text{CM}}$ in Eq.~\eqref{eq:Oso}. An updated calculation gives
\begin{equation}
    M^{\text{rel}}_{\text{LS}, p} = -0.114\,, \quad
    M^{\text{CM}}_{\text{LS}, p} =  0.134\,,
\end{equation}
so that the ${\mathbf L}$ and ${\mathbf L}^{\text{CM}}$ components largely cancel. The same cancellation was observed in the QMC calculation of $^{10}$C~\cite{King:2025fph} and in the CC calculations presented in this work. 
The two-body Fermi matrix elements $M_{\text{F},p}^E$ and $M_{\text{F},p}^+$ are
\begin{equation}
    M_{\text{F},p}^E = -6.22\,, \qquad  M_{\text{F},p}^+ = -1.52\,,
\end{equation}
and the three-body matrix element is given in Eq.~\eqref{eq:QMC3b14O}, $M_{\text{F},p}^{+\, \text{3b}} = -18.6$.

\bibliography{delta}

\end{document}